\documentclass[a4paper,11pt]{article}
\pdfoutput=1 

\usepackage{jheppub} 

\usepackage[bottom]{footmisc}
\usepackage{amssymb}
\usepackage{amsmath}
\usepackage{amsthm}
\usepackage[usenames,dvipsnames]{xcolor}
\usepackage{epsfig}
\usepackage{dcolumn}
\usepackage{tikz}
\usetikzlibrary{shapes.geometric, arrows}
\usetikzlibrary{calc}
\usepackage{upgreek}
\usepackage{setspace}
\usepackage{enumerate}
\usepackage[inline]{enumitem}
\usepackage[makeroom]{cancel}

\usepackage{array,multirow,bigdelim,arydshln}
\usepackage{appendix}
\usepackage{xparse}
\usepackage{stmaryrd}
\usepackage[T1]{fontenc} 
\usepackage{mathtools}
\usepackage{physics} 
\usepackage{adjustbox}
\usepackage{multirow}
\usepackage{graphicx} 
\usepackage{float} 
\graphicspath{{./images/}}
\usepackage[nottoc]{tocbibind}
\usepackage{hyperref}
\usepackage[utf8]{inputenc}
\hypersetup{
	colorlinks,
	urlcolor=Maroon,
	linkcolor=Maroon,
	citecolor=Maroon
	}

\NewDocumentCommand{\binomial}{omm}
 {%
  \genfrac(){0pt}{}{#2}{#3}%
  \IfValueT{#1}{_{\!#1}}%
 }
\NewDocumentCommand{\eulerian}{omm}
 {%
  \genfrac<>{0pt}{}{#2}{#3}%
  \IfValueT{#1}{_{\!#1}}%
 }

\usepackage{latexsym}
\usepackage{tikz}

\theoremstyle{plain}

\theoremstyle{definition}

\def\bea#1\eea{\begin{eqnarray}#1\end{eqnarray}}
\def\be#1\ee{\begin{equation}#1\end{equation}}
\def\ba#1\ea{\begin{align}#1\end{align}}

\def\nl{\nonumber\\} 
\def\non{\nonumber}

\def\yz#1\yz {{\color{blue} [[YZ: #1]] }}

\usepackage{amsmath}
\usepackage{multicol}
\usepackage{bbm}
\usepackage{enumerate}

\usepackage{amsthm}
\usepackage{mathrsfs}
\usepackage{upgreek}
\usepackage{amssymb}
\usepackage{bm}
\usepackage{setspace}
\usepackage{array,multirow,arydshln}
\usepackage{bigdelim}
\usepackage{scalerel}
\usepackage{diagbox}
\usepackage{caption}
\usepackage{tabularx}

\usepackage{tikz}

\usetikzlibrary{shapes.geometric,arrows,arrows.meta,decorations.pathmorphing,decorations.markings,patterns}

\def\<{\langle}
\def\>{\rangle}
\def\a{\alpha}
\def\b{\beta}

\def\e{\epsilon}

\font\tenshuffle=shuffle10 \font\sevenshuffle=shuffle7 \font\fiveshuffle=shuffle7 at 5pt
\def\shuffle{{%
		\def\Dshuffle{\mathbin{\hbox{\tenshuffle\char'001}}}%
		\def\Sshuffle{\mathbin{\hbox{\sevenshuffle\char'001}}}%
		\def\SSshuffle{\mathbin{\hbox{\fiveshuffle\char'001}}}%
		\mathchoice{\Dshuffle}{\Dshuffle}{\Sshuffle}{\SSshuffle}}}

\def\beq{\begin{equation}}
\def\eeq{\end{equation}}

\let\Im\relax

\DeclareMathOperator{\Im}{Im}

\usepackage{enumitem}

\usepackage{enumerate}

\usepackage[percent]{overpic}
\usepackage{multirow} 
\usepackage{slashed}

\usepackage{cleveref}

\newcommand{\ap}{{\alpha'}}

\usepackage{subfig}
\usepackage{float}

\def \CC {{\bm C}}
\def \MM {{\bm M}}

\def \CCF {{\tilde {\bm C}}}
\def \MMF { {\tilde {\bm  M}}}

\def\beq{\begin{equation}}
\def\eeq{\end{equation}}

\let\Im\relax

\DeclareMathOperator{\Im}{Im}

\usepackage{enumitem}

\usepackage{enumerate}

\allowdisplaybreaks[1]

\DeclareFontFamily{U}{mathx}{\hyphenchar\font45}
\DeclareFontShape{U}{mathx}{m}{n}{
      <5> <6> <7> <8> <9> <10>
      <10.95> <12> <14.4> <17.28> <20.74> <24.88>
      mathx10
      }{}
\DeclareSymbolFont{mathx}{U}{mathx}{m}{n}
\DeclareFontSubstitution{U}{mathx}{m}{n}
\DeclareMathAccent{\widecheck}{0}{mathx}{"71}

\usepackage{mathrsfs}

\title{Basis decompositions of genus-one string integrals}

\author[a]{Carlos Rodriguez,}
\author[a]{Oliver Schlotterer}
\author[a,b,c]{and Yong Zhang}

\affiliation[a]{Department of Physics and Astronomy, Uppsala University, Box 516, 75120 Uppsala, Sweden}

\affiliation[b]{Perimeter Institute for Theoretical Physics, Waterloo, ON N2L 2Y5, Canada}

\affiliation[c]{
School of Physical Science and Technology, Ningbo University, Ningbo 315211, China}

\emailAdd{carlos.rodriguez@physics.uu.se}
\emailAdd{oliver.schlotterer@physics.uu.se}
\emailAdd{yzhang@perimeterinstitute.ca}

\date{\today}

\abstract{One-loop scattering amplitudes in string theories involve configuration-space integrals over genus-one surfaces with coefficients of Kronecker-Eisenstein series in the integrand. A conjectural genus-one basis of integrands under Fay identities and integration by parts was recently constructed out of chains of Kronecker-Eisenstein series. In this work, we decompose a variety of more general genus-one integrands into the conjectural chain basis. The explicit form of the expansion coefficients is worked out for infinite families of cases where the Kronecker-Eisenstein series form cycles. Our results can be used to simplify multiparticle amplitudes in supersymmetric, heterotic and bosonic string theories and to investigate loop-level echoes of the field-theory double-copy structures of string tree-level amplitudes. The multitude of basis reductions in this work strongly validate the recently proposed chain basis and stimulate mathematical follow-up studies of more general configuration-space integrals with additional marked points or at higher genus.}

\preprint{UUITP--26/23}  

\begin{document}

\maketitle{}
\addtocontents{toc}{\protect\setcounter{tocdepth}{2}}

\pagenumbering{roman}

\newpage

\setcounter{page}{1}
\pagenumbering{arabic}

\setcounter{tocdepth}{2}

\numberwithin{equation}{section}

 \newpage

\section{Introduction}
\label{sec:intro}

Scattering amplitudes in string theories are derived from moduli-space integrals over punctured worldsheets of various genera. The integrands are correlation functions in the conformal field theory on the worldsheet of vertex operators encoding the scattering data. Simplified representations of these integrated correlators became increasingly important, not only for the sake of computational efficiency but primarily to unravel elegant structures and relations of string amplitudes and their field-theory limit. It proved particularly rewarding to decompose these correlators into basis functions of the moduli using integration by parts in the punctures along with algebraic manipulations of the integrand.

In $n$-point tree-level amplitudes of various string theories, the basis of functions can be spanned by so-called Parke-Taylor factors \beq
 {\rm PT}(1,2,\ldots,n) \coloneqq \frac{1}{z_{1,2}z_{2,3} \ldots z_{n-1,n} z_{n,1}} \,, \ \ \ \ \ \ z_{i,j} = z_i - z_j   
 \label{intr.1}
\eeq
depending on $n$ punctures $z_1,z_2,\ldots,z_n$ on a sphere or disk. Permutations of (\ref{intr.1}) in the external-state labels $1,2,\ldots,n$ are related by integration-by-parts relations and, as firstly derived by Aomoto \cite{Aomoto87}, fall into $(n{-}3)!$-dimensional bases. The underlying integration-by-parts relations at genus zero are driven by the universal Koba-Nielsen factor
\beq
{\cal I}_n^{\rm tree} = \prod_{1\leq i<j}^n |z_{i,j}|^{\ap k_i \cdot k_j} \,,
\label{intr.2}
\eeq
which accompanies (\ref{intr.1}) and depends on the inverse string tension $\ap$ as well as the external momenta $k_i$. 
Moreover, the Koba-Nielsen factor aligns genus-zero string integrals into the setup of twisted (co)homologies, see \cite{Mizera:2017cqs, Mizera:2017rqa, Mizera:2019gea} and \cite{Aomoto:1975Vanishing,Aomoto87ComplexSelberg,KitaYoshida1994paperI,Mimachi2004IntNumbersSelbergInt} for references in the physics and mathematics literature.

The integration-by-parts relations among permutations of ${\cal I}_n^{\rm tree} {\rm PT}(1,2,\ldots,n)$ and related building blocks of genus-zero correlators have profound implications for field theory, string theory and mathematics, see \cite{Mafra:2022wml} for a review: First, they determine the Kleiss-Kuijf \cite{Kleiss:1988ne} and Bern-Carrasco-Johansson \cite{BCJ} relations among gauge-theory amplitudes \cite{Zfunctions, Stieberger:2014hba} and reduce Einstein-Yang-Mills tree amplitudes to gauge-theory ones \cite{Schlotterer:2016cxa}. Second, $(n{-}3)!$-term representations of genus-zero correlators reveal field-theory double-copy structures in open- and closed-string tree amplitudes \cite{Mafra:2011nv, Zfunctions, Carrasco:2016ldy, Azevedo:2018dgo} that resemble the Kawai-Lewellen-Tye relations between gravity amplitudes and squares of gauge-theory ones \cite{Kawai:1985xq}. Third, in deriving the $\ap$-expansion of string tree-level amplitudes from the Drinfeld associator \cite{Broedel:2013aza}, the constituting braid matrices \cite{Mizera:2019gea, Kaderli:2019dny} are obtained from integration-by-parts manipulations of genus-zero correlators with an additional unintegrated puncture.

The variety of tree-level insights resulting from Parke-Taylor bases motivate the quest for analogous integration-by-parts bases for loop-level correlators. In one-loop string amplitudes, the correlators on a genus-one surface such as the torus or the cylinder can be expressed in terms of Jacobi theta functions. In particular, genus-one correlators involve a ubiquitous uplift of the Koba-Nielsen factor (\ref{intr.2}) including $|\theta_1(z_{i,j},\tau)|^{\ap k_i \cdot k_j}$ in the place of $|z_{i,j}|^{\ap k_i \cdot k_j}$, where $\theta_1$ is the odd Jacobi theta function depending on the modular parameter~$\tau$ of the surface. The main focus of this work lies on the leftover functions of $z_j,\tau$ that accompany the one-loop Koba-Nielsen factor. These functions may be viewed as one-loop analogues of the Parke-Taylor factors (\ref{intr.1}) and will be systematically reduced to conjectural bases under integration by parts and so-called Fay relations among the theta functions. The entirety of Fay and integration-by-parts relations will be referred to as F-IBP.

Genus-one correlators for massless bosonic, heterotic and supersymmetric  string amplitudes can be expressed via coefficients $f^{(w)}(z_{i,j},\tau)$ of the Kronecker-Eisenstein series with modular weights $w \geq 0$ \cite{Dolan:2007eh, Broedel:2014vla, Gerken:2018jrq}. While specific string amplitudes impose multiplicity-dependent upper bounds on the weight $w$ of $f^{(w)}$, a recent proposal for F-IBP bases at genus one \cite{Mafra:2019ddf, Mafra:2019xms, Gerken:2019cxz} is built at the level of generating series, i.e.\ products of Kronecker-Eisenstein series. The necessity of generating series is plausible by the fact that $\tau$-derivatives effectively add two units of modular weight, so any bounded collection of $f^{(w)}(z_{i,j},\tau)$ cannot close under moduli derivatives.

More specifically, the conjectural genus-one bases of \cite{Mafra:2019ddf, Mafra:2019xms, Gerken:2019cxz} are built from chains
\beq
\Omega(z_{1,2},\eta_2{+}\eta_3{+}\ldots{+}\eta_n,\tau)
\Omega(z_{2,3},\eta_3{+}\ldots{+}\eta_n,\tau)
\ldots
\Omega(z_{n-1,n},\eta_n,\tau)
\label{intr.3}
\eeq
of Kronecker-Eisenstein series $\Omega(z,\eta,\tau)=\sum_{w=0}^{\infty}\eta^{w-1}f^{(w)}(z,\tau)$,
whose permutations in the labels $j=2,3,\ldots,n$ of $z_j,\eta_j$ yield $(n{-}1)!$-component vectors. By isolating different orders in the Laurent expansion w.r.t.\ the bookkeeping variables $\eta_i \in \mathbb C$, one recovers the combinations of $f^{(w)}(z_{i,j},\tau)$ in the integrands of string amplitudes. In particular, correlators in different string theories reside at different orders in $\eta_j$, e.g.\ integrands for bosonic string amplitudes are typically found at four subleading orders in $\eta_j$ as compared to their supersymmetric counterparts. 

Under $\tau$-derivatives, the vector of Koba-Nielsen integrals over chains (\ref{intr.3}) obeys a first-order differential equation of KZB type. The homogeneity of this equation not only substantiates the claim that the $(n{-}1)!$-vectors of chains yield a Koba-Nielsen integral basis but also leads to powerful techniques to extract the $\ap$-expansions of configuration-space integrals at genus one: Based on the solution of the respective KZB-type equations via Picard iteration, the elliptic multiple zeta values in the expansion of open-string integrals \cite{Enriquez:Emzv, Broedel:2014vla} are determined in terms of iterated integrals over holomorphic Eisenstein series \cite{Mafra:2019ddf, Mafra:2019xms}.\footnote{See \cite{Broedel:2019gba} for an alternative method to obtain the $\ap$-expansion of open-string integrals at genus one from elliptic associators and \cite{Broedel:2020tmd} for its reformulation in terms of generating series.} In the closed-string case, similar first-order equations \cite{Gerken:2019cxz} and their perturbative solution \cite{Gerken:2020yii} clarified the relation between modular graph forms \cite{DHoker:2015wxz, DHoker:2016mwo} and iterated Eisenstein integrals. This in turn paved the way for identifying modular graph forms \cite{Dorigoni:2022npe} with Brown's equivariant iterated Eisenstein integrals \cite{Brown:2017qwo, Brown:2017qwo2}. In view of these mathematical developments, it is a burning question if the closure of the chains (\ref{intr.3}) under $\partial_{\tau}$ is merely a coincidence or really identifies an F-IBP basis at genus one.

Instead of attempting a mathematically rigorous proof, we gather further evidence for (\ref{intr.3}) to form a basis by decomposing cyclic products and more general arrangements of Kronecker-Eisenstein series into the chain form. In this way, we not only add further credence to the conjectural basis but also arrive at practical tools to

\vspace{-0.1cm}
\begin{itemize}
\item simplify the genus-one correlators in heterotic- and bosonic-string amplitudes (say converting their cycles of $f^{(w)}(z_{i,j},\tau)$ into chains) and set the stage to investigate physical interpretations of the basis coefficients,

\vspace{-0.1cm}
\item reduce the $\ap$-expansions of genus-one integrals with Kronecker-Eisenstein integrands beyond the chain form to the all-order results of \cite{Mafra:2019ddf, Mafra:2019xms, Gerken:2020yii} and thereby facilitate the computation of low-energy effective couplings in heterotic and bosonic string theories.
\end{itemize}

\vspace{-0.1cm}
\noindent
The explicit basis decompositions in this work via F-IBP relations generalize the tree-level computations in \cite{Huang:2016tag, Schlotterer:2016cxa, He:2018pol, He:2019drm} and should be useful in placing the $(n{-}1)!$-bases on rigorous footing. It would be exciting if the expansion coefficients in our basis decompositions can be interpreted as suitable generalizations of intersection numbers which need to accommodate differential operators in the bookkeeping variables $\eta_j$ \cite{Mafra:2019ddf, Mafra:2019xms, Gerken:2020yii}.

While the original motivation for this work arises from conventional string theories with infinite spectra, our results can be exported to both ambitwistor strings \cite{Mason:2013sva, Berkovits:2013xba} and chiral strings \cite{Hohm:2013jaa, Huang:2016bdd}: As for instance discussed in \cite{Gomez:2013wza, He:2017spx, He:2018pol, He:2019drm, Kalyanapuram:2021xow}, the integration-by-parts manipulations of moduli-space integrands can be smoothly translated between these types of string theories (possibly involving a formal $\ap \rightarrow \infty$ limit). It is also conceivable that our results shed further light on massive loop amplitudes in conventional and chiral string theories as done at tree level in \cite{Guillen:2021mwp} based on Parke-Taylor decompositions.

The computation of closed-string loop amplitudes greatly simplifies in the
framework of chiral splitting \cite{DHoker:1988pdl, DHoker:1989cxq}: The introduction of loop momenta reduces closed-string correlators to double copies of meromorphic
open-string building blocks known as chiral amplitudes. However, F-IBP simplifications of chiral amplitudes are more subtle than those of the manifestly doubly-periodic $f^{(w)}(z_{i,j},\tau)$-integrands that arise after loop integration. Certain total derivatives in the punctures of chiral amplitudes may integrate to non-vanishing boundary terms in a closed-string context. We will discuss the role of these boundary terms in the quest for a chiral-splitting analogue of the $(n{-1})!$ genus-one bases of $f^{(w)}(z_{i,j},\tau)$-integrands.

A {\tt Mathematica} implementation of our results as well as chain decompositions of 
more general classes of genus-one string integrands can be found in a companion paper \cite{companion}.

\subsection*{Outline}

The present work is organized as follows: The conjectural chain bases of genus-one integrals and their building blocks are reviewed in section~\ref{section:two}. We then reduce a single cycle of Kronecker-Eisenstein series to combinations of chains, with a detailed discussion of the two-point case in section~\ref{section:three} and the $n$-point generalization in section~\ref{section:four}. 
In section~\ref{section:five}, the results and techniques are reformulated in the framework of chiral splitting, with a discussion of boundary terms beyond an $(n{-}1)!$ basis. 
Section~\ref{section:six} is dedicated to basis decompositions for products of two or three cycles of Kronecker-Eisenstein series. We present our conclusions and provide an outlook in section \ref{sec:conclusion}. Additional examples relevant for applications to specific
one-loop heterotic-string amplitudes can be found in appendix \ref{singlecycle456},
and intermediate steps for the reduction of double cycles are given in appendix \ref{sec:2cyc.4}.
Moreover, the reader is referred to section~\ref{subsec:outline} for a more detailed 
outline of this work.
 
\section{Review, notation and conventions}
\label{section:two}

\subsection{Basic definitions}
\label{sec:2.2}

One-loop string amplitudes are computed from moduli-space integrals over
correlation functions of certain worldsheet fields that carry the external-state information.
All the dependence of these genus-one correlators on the punctures $z \in \mathbb C$ and modular parameter $\tau \in \mathbb C$ with $\Im \tau>0$ can
be deduced from the Kronecker-Eisenstein series \cite{Kronecker}
\begin{equation}
F(z,\eta,\tau) \coloneqq \frac{ \theta_1'(0,\tau) \theta_1(z{+}\eta,\tau) }{\theta_1(z,\tau)  \theta_1(\eta,\tau)}
\label{1.2}\,,
\end{equation}
where the standard odd Jacobi theta function with $q \coloneqq \exp (2 \pi i \tau)$ is given by
\ba
\theta_{1}(z, \tau) \coloneqq 2 q^{1 / 8} \sin (\pi z) \prod_{n=1}^{\infty}\left(1-q^{n}\right)\left(1-q^{n} e^{2 \pi i z}\right)\left(1-q^{n} e^{-2 \pi i z}\right)\,.
\ea
Based on a non-holomorphic admixture in the exponent of \cite{BrownLev}
\begin{equation}
\Omega(z,\eta,\tau) \coloneqq \exp\Big( 
2\pi i \eta \, \frac{ \Im  z}{\Im  \tau} \Big)
\, F(z,\eta,\tau) \,,
\label{1.1}
\end{equation}
we attain a doubly-periodic completion of the meromorphic Kronecker-Eisenstein
series (\ref{1.2}) subject to
$ \Omega(z,\eta,\tau) =\Omega(z{+}1,\eta,\tau) =\Omega(z{+}\tau,\eta,\tau) $.
Laurent expansions in the bookkeeping variables $\eta \in \mathbb C$ define
Kronecker-Eisenstein coefficients $g^{(w)},f^{(w)}$ with $w \in \mathbb N_0$,
\begin{equation}
F(z,\eta,\tau) =: \sum_{w=0}^{\infty} \eta^{w-1} g^{(w)}(z,\tau)\,
\qquad {\rm and} \qquad
\Omega(z,\eta,\tau) =: \sum_{w=0}^{\infty} \eta^{w-1} f^{(w)}(z,\tau)\, ,
\label{1.1b}
\end{equation}
for instance $g^{(0)}(z,\tau)=f^{(0)}(z,\tau)=1$ as well as
$g^{(1)}(z,\tau) = \partial_z \log \theta_1(z,\tau)$
and $ f^{(1)}(z,\tau) = g^{(1)}(z,\tau)+ 2\pi i \frac{ \Im z}{\Im  \tau}$.  
While the meromorphic $g^{(w)}$ feature $B$-cycle monodromies
generated by $ F(z{+}\tau,\eta,\tau)=e^{-2\pi i \eta}F(z,\eta,\tau)$, the
doubly-periodic $f^{(w)}$ have non-vanishing antiholomorphic derivatives\footnote{Delta-function contributions to antiholomorphic derivatives
(\ref{partialbar}) and $\frac{\partial}{\partial \bar z} {g}^{(w)}(z, \tau)=\frac{\partial}{\partial \bar z} F(z, \eta,\tau)=0$ are not tracked in this work.}
\beq
\label{partialbar}
\frac{\partial}{\partial \bar z} \Omega(z,\eta,\tau) = - \frac{\pi \eta}{ {\rm Im} \,\tau} \Omega(z,\eta,\tau) \, 
\ \ \Rightarrow  \ \
\frac{\partial}{\partial \bar z} {f}^{(w)}(z, \tau)=-\frac{\pi}{ {\rm Im} \,\tau} {f}^{(w-1)}(z, \tau) \,, \ \ w\geq 1  \,.
\eeq
Laurent expansion of (\ref{1.1}) relates the two types of Kronecker-Eisenstein coefficients via 
 \ba\label{gtof}
g^{(w)}(z, \tau) = \sum_{k=0}^{w} \frac{(-\nu)^{k}}{k !} f^{(w-k)}(z, \tau) \, , \ \ \ \ 
f^{(w)}(z, \tau) = \sum_{k=0}^{w} \frac{\nu^{k}}{k !} g^{(w-k)}(z, \tau)
\,,
\ea
with  $\nu \coloneqq 2 \pi i \frac{\operatorname{Im} z}{{\rm Im} \,\tau}$.
Iterated integrals of the $g^{(w)}$ and $f^{(w)}$ give rise to different
descriptions of elliptic polylogarithms \cite{BrownLev, Broedel:2014vla, Broedel:2017kkb}
which had dramatic impact on recent developments 
in both string perturbation theory \cite{Berkovits:2022ivl} and particle physics \cite{Bourjaily:2022bwx}.

\subsubsection{Holomorphic Eisenstein series}

Evaluating the $f^{(w)}(z,\tau)$ at the origin produces holomorphic Eisenstein series 
\begin{equation}
{\rm G}_w(\tau) \coloneqq  \sum_{(m,n) \neq (0,0)} \frac{1}{(m\tau + n)^{w}} = - f^{(w)}(0,\tau)\, ,\qquad w\geq 4   
\label{1.6}
\end{equation}
of modular weight $(w,0)$,
represented via absolutely convergent double sums over integers $m,n$ if $w\geq 4$.
While the analogous $z\rightarrow 0 $ limit of $f^{(2)}(z,\tau)$ is ill-defined, we will later encounter
a non-holomorphic but modular variant of the weight-two Eisenstein series
 \begin{equation}
\hat {\rm G}_2(\tau) \coloneqq  \lim_{s\rightarrow 0}\sum_{(m,n) \neq (0,0)} \frac{1}{(m\tau + n)^{2} \,|m\tau + n|^s} =  {\rm G}_2(\tau)  - \frac{\pi}{\Im  \tau}\, .
\end{equation}
The factor of $|m\tau {+} n|^{-s}$ is necessary for absolute convergence of the double sum, and the 
meromorphic counterpart ${\rm G}_2(\tau)$ of $\hat {\rm G}_2(\tau)$ is defined through the Eisenstein summation prescription
\beq
{\rm G}_2(\tau)  \coloneqq \lim_{M\rightarrow  \infty}  \lim_{N\rightarrow \infty} 
\sum_{m=-M}^M \sum_{n=-N}^N \frac{1}{(m\tau+n)^2}
\, .
\eeq

\subsubsection{Properties of the Kronecker-Eisenstein series}

The Kronecker-Eisenstein series as well as its doubly-periodic completion
satisfy the antisymmetry property
\beq
F(-z,-\eta,\tau) = - F(z,\eta,\tau) \, , \ \ \ \ \ \ 
\Omega(-z,-\eta,\tau) = - \Omega(z,\eta,\tau)
\label{antisyprop}
\eeq
as well as Fay identities
\beq \label{fayid}
F(z_1,\eta_1,\tau)F(z_2,\eta_2,\tau) =
F(z_1,\eta_1{+}\eta_2,\tau) F(z_2{-}z_1,\eta_2,\tau)
+F(z_2,\eta_1{+}\eta_2,\tau) F(z_1{-}z_2,\eta_1,\tau)\,,
\eeq
which hold in identical form for $F(z,\eta,\tau)\rightarrow \Omega(z,\eta,\tau)$ and can be thought of as quasi- and doubly-periodic generalizations of the partial-fraction relation $\frac{1}{z_iz_j}=\frac{1}{z_i-z_j}(\frac{1}{z_j}-\frac{1}{z_i})$.

By carefully taking the limit $z_1\to z$ and $z_2\to -z$ in (\ref{fayid}),  one can derive  the following identities 
\begin{align}
F(z,\eta_1,\tau) F({-}z,\eta_2,\tau) &= F(z,\eta_1{-}\eta_2,\tau)
 \big(  g^{(1)}(\eta_2,\tau) -  g^{(1)}(\eta_1,\tau) \big) + \partial_z F(z,\eta_1{-}\eta_2) \,,
\label{variantF}
\\
\Omega(z,\eta_1,\tau) \Omega({-}z,\eta_2,\tau) &= \Omega(z,\eta_1{-}\eta_2,\tau)
 \big( \hat g^{(1)}(\eta_2,\tau) - \hat g^{(1)}(\eta_1,\tau) \big) + \partial_z \Omega(z,\eta_1{-}\eta_2) \,, \label{variant}
\end{align}
where the main difference between the meromorphic and the doubly-periodic case concerns the Eisenstein series ${\rm G}_2(\tau)$ or $\hat {\rm G}_2(\tau)$ in the expansions
\begin{align}\label{gexpand}
g^{(1)}(\eta,\tau) =&\; \partial_\eta \log \theta_1(\eta,\tau)   =  \frac{1}{\eta} - \eta {\rm G}_2(\tau)
-\sum_{n=4}^{\infty} \eta^{n-1} {\rm G}_n(\tau) \,,
\\
\label{gexpandF} 
\hat g^{(1)}(\eta,\tau) \coloneqq&\; \partial_\eta \log \theta_1(\eta,\tau)  + \frac{ \pi \eta}{\Im \tau} =  \frac{1}{\eta} - \eta \hat {\rm G}_2(\tau)
-\sum_{n=4}^{\infty} \eta^{n-1} {\rm G}_n(\tau) \,.
\end{align}
One can employ the $\eta_j$-expansions of (\ref{variantF}) and (\ref{variant})
to rewrite derivatives $\partial_z {g}^{(w)}(z, \tau)$ and $\partial_z {f}^{(w)}(z, \tau)$
in terms of bilinears in undifferentiated Kronecker-Eisenstein coefficients
and Eisenstein series. Similar rewritings of $\partial_z {g}^{(w)}(z, \tau)$
and $\partial_z {f}^{(w)}(z, \tau)$ can be attained from the expansion of
\begin{align}
\partial_{z}F(z,\eta,\tau)-\partial_{\eta}F(z,\eta,\tau)&=\big( g^{(1)}(\eta,\tau)-g^{(1)}(z,\tau) \big)
  F(z,\eta,\tau)\,,
  \label{derivativeF}
  \\ 
  \partial_{z}\Omega(z,\eta,\tau)-\partial_{\eta}\Omega(z,\eta,\tau)&= \big(\hat g^{(1)}(\eta,\tau)-f^{(1)}(z,\tau) \big)
  \Omega(z,\eta,\tau)\, ,
  \label{derivative}  
\end{align}
which straightforwardly follow from the definition (\ref{1.2}).
Upon isolating fixed orders in $\eta$, we obtain identities such as
\beq
\partial_z f^{(1)}(z,\tau) = 2 f^{(2)}(z,\tau) - \big( f^{(1)}(z,\tau) \big)^2 - {\rm \hat G}_2(\tau) 
\eeq
and more generally
\begin{align}
\partial_z f^{(w)}(z,\tau) &= (w{+}1) f^{(w+1)}(z,\tau) -  f^{(w)}(z,\tau) f^{(1)}(z,\tau) 
\label{atfixedeta} \\
&\quad
- {\rm \hat G}_2(\tau) f^{(w-1)}(z,\tau) - \sum_{n=4}^{w+1} {\rm G}_n f^{(w+1-n)}(z,\tau)
\, . \notag
\end{align}

\subsubsection{Shorthand notation}

Since the main results of this work concern configuration-space integrals over several punctures $z_1,z_2,\ldots$, it will be convenient to use the shorthand notation
\beq
\partial_j \coloneqq \frac{ \partial}{\partial z_j} \,, \ \ \ \ \ \
g_{i j}^{(w)} \coloneqq g^{(w)}(z_{i}{-}z_j,\tau)
\,, \ \ \ \ \ \
f_{i j}^{(w)} \coloneqq f^{(w)}(z_{i}{-}z_j,\tau)\,,
\eeq
and similarly
\beq
F_{i j}(\eta) \coloneqq F(z_{i}{-}z_j,\eta,\tau)
\,, \ \ \ \ \ \
\Omega_{i j}(\eta) \coloneqq \Omega(z_{i}{-}z_j,\eta,\tau) \, .
\eeq
In this notation, (\ref{derivativeF}) and
(\ref{derivative}) for instance take the alternative form
\begin{align}
\partial_{i}F_{ij}(\eta)&=
-g^{(1)}_{ij} F_{ij}(\eta)
+\big(g_1(\eta)+\partial_{\eta}\big)F_{ij}(\eta)\,,
\label{eqUsefulF}
\\
\partial_{i}\Omega_{ij}(\eta)&=-
f^{(1)}_{ij} \Omega_{ij}(\eta)+\big(\hat{g}_1(\eta)+\partial_{\eta}\big)\Omega_{ij}(\eta)\,,
\label{eqUseful}
\end{align}
where fixed orders in $\eta$ allow to eliminate
 $\partial_i g^{(w)}_{ij}$ and $\partial_i f^{(w)}_{ij}$
 in favor of undifferentiated Kronecker-Eisenstein
 coefficients, cf.\ (\ref{atfixedeta}). Similarly, the expansion of the Fay
identity (\ref{fayid}) in $\eta_1,\eta_2$ yields quadratic relations 
involving three points $z_1,z_2,z_3$~\cite{Broedel:2014vla}
\begin{align}
f^{(n)}_{12} f^{(m)}_{23} &= - f_{13}^{(m+n)}
+ \sum_{j=0}^{n}(-1)^j {m-1+j \choose j}
f_{13}^{(n-j)} f_{23}^{(m+j)} \label{faycomp} \\
& \ \ \ \ \   \ \ \ \ \   \ \ \ \ \  \,
+ \sum_{j=0}^{m}(-1)^j {n-1+j \choose j} f_{13}^{(m-j)} f_{12}^{(n+j)} 
\notag
\end{align}
which hold in identical form for $f^{(w)}_{ij}  \rightarrow g^{(w)}_{ij} $.

\subsection{The structure of one-loop string amplitudes}

Open-string $n$-point amplitudes at one loop descend from worldsheets of cylinder- and Moebius-strip topologies with modular parameter $\tau$ and punctures $z_i$ on the boundary, 
\beq\label{openamp}
\mathcal{A}_{n}=\sum_{\text {top }} C_{\text {top }} \int \limits_{D^{\tau}_{\text {top }}} %
\frac{d \tau }{(\operatorname{Im} \tau)^{\frac{D}{2}}}
 \int \limits_{D^{z}_{\text {top }}} 
\underbrace{d z_{2}   \ldots d z_{n}}_{d \mu^{\rm op}_{n}} 
\, \mathcal{I}_{n}^{\text {op}}(z_i,\tau,k_i) \,
K_{n}^{\text {op}}
(f^{(w)}, \tau, k_i, \epsilon_i,\cdots)\,,
\eeq
see \cite{green1988superstring} for the color (or Chan-Paton) factors $C_{\text {top}}$
as well as the associated integration domains $D^{\tau}_{\text {top}}$ and 
$D^{z}_{\text {top}}$ for $\tau$  and $z_i$. 
The exponent of $\Im \tau$ depends on the number $D$ of spacetime dimensions, and
the contributions ${\cal I}_{n}^{\text {op}}$ and $K_{n}^{\text {op}}$ to the 
$n$-point integrands depend on the moduli $\tau,z_i$ as well as
the momenta $k_i$ and polarizations $\epsilon_i$ of the external legs. 

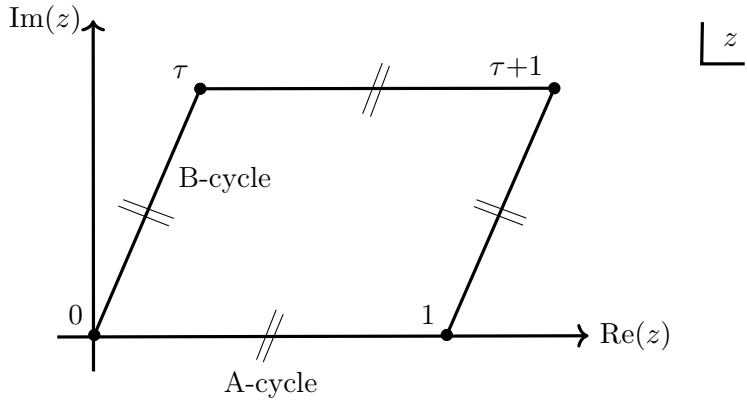
\begin{figure}
  \centering
  \begin {tikzpicture}[scale=3]
   \draw [very thick,->] (0.704353, 0.374557) coordinate(A1) node [left]{}-- (3.03464, 0.379902) coordinate (B1) node [right]{${\rm Re}(z)$};
      \draw [very thick,->] (0.86337, 0.222221) coordinate(A2) node [below]{}-- (0.861918, 1.77486) coordinate (B2) node [left]{${\rm Im}(z)$};
            \draw [very thick] (0.866759, 0.383114) coordinate(C1) node [above left]{0}
            -- (1.33171, 1.47161) coordinate (C2) node [above left]{$\tau$}
            -- (2.88462, 1.47386) coordinate (C3) node [above left]{$\tau{+}1$}
            -- (2.41283, 0.383478) coordinate (C4) node [above left]{1}
            ;
            
            \filldraw (C1) circle(.7pt) ;
             \filldraw (C2) circle(.7pt) ;
              \filldraw (C3) circle(.7pt) ;
               \filldraw (C4) circle(.7pt) ;
               
       \node at     ($ (C1)! 1/2 ! (C4) + (0,-.22)$) {A-cycle};
             \node at     ($ (C1)! 1/2 ! (C2) + (.34,.14)$) {B-cycle};
             
             \coordinate   (D0) at ( 3.65742, 1.68769)  ;
             \node at   ( 3.65742, 1.68769)  {$z$};       
             
              \draw [very thick]
               ($(3.47214, 1.80545)!1/3!(D0)$) coordinate(A1) node [left]{}-- ($(3.46738, 1.5285) !1/3!(D0)$)
               coordinate (B1) node [right]{}-- ($ (3.75044, 1.52865)!1/3!(D0)$) coordinate (B1) node [above]{} ;
               
   \draw (1.57857,0.274158)--(1.66436,0.494671);
\draw (1.60756,0.263328)--(1.69516,0.474998);
\draw (2.04386,1.36077)--(2.13005,1.58511);
\draw (2.07809,1.3479)--(2.16144,1.5727);
\draw (1.19425,0.865392)--(0.971937,0.953592);
\draw (1.21584,0.898475)--(0.993368,0.98144);
\draw (2.53208,0.95108)--(2.74731,0.86862);
\draw (2.54332,0.981167)--(2.76156,0.900012);

    \end {tikzpicture}
  \caption{The parametrization of the torus worldsheet $\mathfrak{T}_\tau$ through a parallelogram where the non-contractible A- and B-cycles are taken to be the path
  from the origin to $1$ and the modular parameter $\tau$, respectively.}
  \label{parallelogram}
\end{figure}

Closed-string one-loop amplitudes in turn are given by
\beq\label{closedamp}
\mathcal{M}_{n}= \int \limits_{\mathfrak{F}} 
 \frac{d^{2} \tau }{(2 {\rm Im} \,\tau)^{\frac{D}{2}}}
\int \limits_{\mathfrak{T}_\tau^{n-1}}
\underbrace{
d^{2} z_{2}   \ldots d^{2} z_{n}}_{d \mu_n^{\rm cl}} \,
\mathcal{I}_{n}^{\rm cl}(z_i,\tau,k_i) \,
K_{n}^{\text {cl}}
(f^{(w)},{\bar f}^{(w)}, \tau,k_i, \epsilon_i, {\bar \e}_i,\cdots)\,,
\eeq
where the modular parameter $\tau$ of the torus worldsheet
is integrated over the fundamental domain $\mathfrak{F}$ of ${\rm SL}_2(\mathbb Z)$. 
The punctures $z_2,\ldots,z_n$ are
integrated over the torus $\mathfrak{T}_\tau$, parametrized by the standard parallelogram 
with corners $0,1,\tau{+}1,\tau$ in the complex $z_i$-plane depicted in figure \ref{parallelogram}. The contribution 
$K_{n}^{\text {cl}}$ to closed-string integrands
depends on two species of polarizations $\epsilon_i$ and ${\bar \e}_i$ associated with
left- and right-moving worldsheet degrees of freedom.

Although we fixed $z_1=0$ using translation invariance on open- and closed-string
worldsheets at genus one, we shall keep $z_1$ generic throughout this work.

\subsubsection{Koba-Nielsen factors $\mathcal{I}_{n}$}

In the integrands of both open- and closed-string amplitudes (\ref{openamp})
and (\ref{closedamp}), we separate universal Koba-Nielsen factors
\ba
\mathcal{I}_{n}^{\text {op}}  = \mathcal{I}_{n}^{\rm op}(z_i,\tau,k_i) \coloneqq\,\,& \exp \bigg({-}\sum_{i<j}^{n} s_{i j}\left[\log \left|\theta_{1}(z_{i j}, \tau)\right|-\frac{\pi}{\operatorname{Im} \tau}\left(\operatorname{Im} z_{i j}\right)^{2}\right]\bigg)\, ,
\nonumber
\\
\mathcal{I}_{n}^{\rm cl}  = \mathcal{I}_{n}^{\rm cl}(z_i,\tau,k_i)\coloneqq\,\,& \exp \bigg({-}\sum_{i<j}^{n} s_{i j}\left[\log \left|\theta_{1}(z_{i j}, \tau)\right|^{2}-\frac{2 \pi}{{\rm Im} \,\tau} \left(\operatorname{Im} z_{i j}\right)^{2}\right]\bigg)\,,
\label{defkn}
\ea
from theory-dependent factors $K_{n}^{\text{op}}$ and $K_{n}^{\text {cl}}$ to be described below. 
Our conventions for the dimensionless Mandelstam invariants are fixed by,
\begin{equation}
s_{i j}=-  k_i\cdot k_j\, , \quad s_{i_1 i_2\ldots i_r}=-  \sum_{1\leq p<q \leq r} k_{i_p}\cdot k_{i_q} \, ,
\quad \alpha' = \left\{ \begin{array}{cl} 1/2 &:\ {\rm open} \ {\rm strings}\, , \\
2 &:\ {\rm closed} \ {\rm strings} \, .
\end{array} \right.
\end{equation} 
We will study the Koba-Nielsen factors (\ref{defkn}) at independent values of
all the $s_{ji}=s_{ij}$ with $1\leq i<j\leq n$ even though the amplitudes
(\ref{openamp}) and (\ref{closedamp}) are eventually evaluated on
the support of on-shell conditions for $k_j^2$ and momentum-conserving 
delta functions.\footnote{For massless two- and three-point amplitudes,
momentum conservation together with the on-shell conditions $k_j^2=0$
would enforce all the $s_{ij}$ to vanish. Starting from \cite{Minahan:1987ha},
it proved convenient to relax momentum conservation and 
on-shell conditions in intermediate steps of studying string amplitudes.} In this way,
Koba-Nielsen factors at $n=2$ and $n=3$ points are taken to depend on 
one variable $s_{12}$ and three variables $\{s_{12},s_{13},s_{23}\}$, respectively.

\subsubsection{Theory dependent integrands $K_{n}$}

The theory-dependent open- and closed-string integrands 
$K_{n}^{\text{op}}$ and $K_{n}^{\text {cl}}$ carry the entire polarization
dependence of the amplitudes (\ref{openamp}) and (\ref{closedamp}),
and the main goal of this work is to provide tools for their systematic
simplification. As a key result of \cite{Dolan:2007eh, Broedel:2014vla, Gerken:2018jrq},
their dependence on the punctures in massless amplitudes of type I, type II and
heterotic theories\footnote{The analogous decompositions of $n$-point one-loop integrands in
orbifold compactifications with reduced supersymmetry to combinations of
$f^{(w)}$ are discussed in \cite{Berg:2016wux} building upon earlier results 
in \cite{Bianchi:2006nf, Bianchi:2015vsa}. For bosonic strings, the
$K_{n}^{\text{op}}$ and $K_{n}^{\text {cl}}$ at genus one straightforwardly boil 
down to derivatives
$ f_{ij}^{(1)}$ and $\partial_i f_{ij}^{(1)}$ of the bosonic Green function.} 
is entirely expressible in terms of the Kronecker-Eisenstein
coefficients $f^{(w)}$ reviewed in section \ref{sec:2.2}. Closed-string integrands 
$K_{n}^{\text {cl}}$ may additionally feature powers of $\frac{\pi}{\Im \tau}$
due to Wick contractions between left- and right-movers. Hence, the
theory-dependent integrands take the schematic form
\begin{align}
K_{n}^{\text {op}} &=Z^{\rm op}(\tau,D,\ldots) \sum N^{\rm op}(\e_i, k_i)
\Big( f_{i_{1} j_{1}}^{\left(k_{1}\right)} f_{i_{2} j_{2}}^{\left(k_{2}\right)} \cdots \Big) \,,
\label{stringint} \\
K_{n}^{\text {cl}} &= Z^{\rm cl}(\tau,D,\ldots) \sum_{w} \sum N^{\rm cl}_w (\e_i, {\bar \e}_i, k_i) \bigg( \frac{\pi}{\Im \tau} \bigg)^w
\Big( f_{i_{1} j_{1}}^{\left(k_{1}\right)} f_{i_{2} j_{2}}^{\left(k_{2}\right)} \cdots \Big) \Big( \bar f_{p_{1} q_{1}}^{\left(r_{1}\right)} \bar f_{p_{2} q_{2}}^{\left(r_{2}\right)} \cdots \Big)\,,
\notag
\end{align}
where the {\it numerators} $N^{\rm op}(\e_i, k_i)$ and $N^{\rm cl} (\e_i, {\bar \e}_i, k_i)$ 
separate the polarization dependence from the moduli $z_i,\tau$. In heterotic or bosonic string amplitudes as well as (toroidal or orbifold) compactifications, one additionally encounters
$z_i$-independent partition functions $Z^{\rm op}(\tau,D,\ldots)$ and $Z^{\rm cl}(\tau,D,\ldots)$. The latter depend on the number $D$ of spacetime dimensions and compactification details
if $D<10$ such as radii of circular dimensions or twist parameters of orbifolds, see for instance \cite{Green:1982sw}.

The structure (\ref{stringint}) of genus-one integrands $K_{n}^{\text{op}}$ and $K_{n}^{\text {cl}}$
is established for massless external states, but expected to also capture one-loop amplitudes of massive modes.

\subsection{One-loop string integrals and their F-IBP relations}

In order to simplify the Koba-Nielsen integrals over the expressions (\ref{stringint}) for
$K_{n}^{\text {op}}$ and $K_{n}^{\text {cl}}$, it is essential to minimize the
number of independent products $ f_{i_{1} j_{1}}^{\left(k_{1}\right)} f_{i_{2} j_{2}}^{\left(k_{2}\right)} \ldots$ of Kronecker-Eisenstein kernels. To some extent, this
can be achieved via algebraic identities at the level of the integrand such
as Fay relations (\ref{faycomp}) or the removal of $z_i$-derivatives of
$f^{(w)}_{ij}$ via (\ref{atfixedeta}). However, the main driving force for
reductions of open- and closed-string integrands (\ref{stringint}) to a basis
is integration by parts to be reviewed in this section. We reiterate that the entirety
of integration-by-parts relations and Fay identities will be referred to as F-IBP.

\subsubsection{Koba-Nielsen derivatives and IBP relations}

The integration domains $D_{\rm top}$ in open-string amplitudes (\ref{openamp})
only leave one real degree of freedom for each modulus $z_j$ and $\tau$. Accordingly,
there is no separate reference to their complex conjugates $\bar z_j$ and $\bar \tau$
as in the closed-string setting (\ref{closedamp}) where the integration domains
$\mathfrak{F}$ and $\mathfrak{T}_\tau$ have two real dimensions. We can
therefore write the derivatives w.r.t.\ punctures $z_i$ of both Koba-Nielsen factors (\ref{defkn}) 
in the unified form
\ba\label{baseafter}
\partial_i \mathcal{I}^{\bullet}_{n}=-\bigg(\sum_{j \neq i}^{n} x_{i,j} \bigg) \mathcal{I}^{\bullet}_{n}\,,\qquad x_{i,j} \coloneqq s_{ij} f^{(1)}_{ij}\, ,
\ea
with derivatives $\partial_i$ in the sense of real analysis for
 $\bullet \rightarrow {\rm op}$ and holomorphic derivatives
 $\partial_i$ (as opposed to $\bar\partial_{ i} = \frac{ \partial }{ \partial \bar z_i}$) 
 in case of $\bullet \rightarrow {\rm cl}$.

In the context of open- and closed-string amplitudes
(\ref{openamp}) and (\ref{closedamp}), total derivatives
$\partial_i$ w.r.t.\ the punctures typically act on the $z_i$-dependence
of $K_{n}^{\text {op}} $ and $K_{n}^{\text {cl}}$ besides the respective
Koba-Nielsen factor. These two types of contributions are gathered in the notation
\ba
\label{defnabla}
\nabla_i \varphi \coloneqq
  \partial_i\varphi -\bigg(\sum_{j \neq i}^{n} x_{i,j} \bigg) \varphi  
  =\frac{1}{\mathcal{I}^{\bullet}_{n}}\partial_i (\varphi\,  \mathcal{I}^{\bullet}_{n} )\,, 
\ea 
for arbitrary contributions $\varphi = \varphi (z_i,\tau)$ to open- or
closed-string integrands (\ref{stringint}). The images of these operators
$\nabla_i$ integrate to zero within both open- and closed-string settings,
\be 
\label{totalderivative}
\int_{ D^{z}_{\text {top}} } d \mu^{\rm op}_{n}\,  \mathcal{I}^{\rm op}_{n} \, \nabla_i \varphi 
= 0 
= \int_{ \mathfrak{T}_\tau^{n-1} } d \mu^{\rm cl}_{n}\,  \mathcal{I}^{\rm cl}_{n} \, \nabla_i \varphi \,,
\ee
which relies on the following two salient points:
\begin{itemize}
\item [(i)] Both ${\cal I}_{n}^{\bullet}$ and the $\varphi$ of interest 
descend from correlation functions 
on the torus and are therefore doubly-periodic 
under $z_i \rightarrow z_i{+}1$ and $z_i \rightarrow z_i{+}\tau$. As
will be illustrated in section \ref{sec:revCS}, this is 
particularly important in a closed-string context.
\item [(ii)] The Koba-Nielsen factors ${\cal I}_{n}^{\rm op}$ and ${\cal I}_{n}^{\rm cl}$
exhibit local behaviour $|z_{i,j}|^{-s_{ij}}$ and $|z_{i,j}|^{-2s_{ij}}$ when pairs of 
punctures $z_i,z_j$ collide. Upon analytic continuation to the kinematic region
where ${\rm Re}(s_{ij})<0$, boundary terms $z_i \rightarrow z_j$
are suppressed (and similarly $z_i \rightarrow z_j+m\tau+n$ for $m,n\in \mathbb Z$
by double-periodicity).
\end{itemize}
Since the images of these total derivative do not contribute to open- and closed-string
amplitudes \eqref{openamp} and \eqref{closedamp}, we identify
 \ba
\label{IBPomega}
\nabla_i \varphi \coloneqq
  \partial_i \varphi -\bigg(\sum_{j \neq i}^{n} x_{i,j} \bigg) \varphi   \cong 0\,,\quad \forall\, {\text{doubly periodic}}~ \varphi(z_j, \tau)
\ea
in the simplification of $K_{n}^{\text {op }} $ and $K_{n}^{\text {cl }}$.
Note that the operator $\nabla_{i}$ does not obey the Leibniz rule
known from $\partial_i$ but instead acts on products via
\be
\label{nablarule}
\nabla_{i}( \varphi_1 \varphi_2 )=(\nabla_{i} \varphi_1 ) \varphi_2+\varphi_1 \partial_{i}(\varphi_2)= (\partial_i \varphi_1 ) \varphi_2+\varphi_1 \nabla_{i}(\varphi_2) \,.
\ee
On the other hand, two operators $\nabla_{i}$ and $\nabla_{j}$ are easily checked to still commute with each other,
\ba
\label{commute}
(\nabla_{i}\nabla_{j}-\nabla_{j}\nabla_{i})\varphi=
 \varphi (\partial_i{+}\partial_j)x_{i,j} =0\,,~ \quad {\rm for }~ i\neq j\,.
\ea

\subsubsection{Conjectural bases of genus-one string integrals}

It is conjectured in \cite{Mafra:2019ddf, Mafra:2019xms} that any $n$-point open-string one-loop integrand
relevant to $K_n^{\rm op}$ in (\ref{stringint}) can be lined up with an $(n{-}1)!$-basis
of generating functions. The latter are composed of 
products \eqref{intr.3} of doubly-periodic Kronecker-Eisenstein series
(\ref{1.1}), 
\ba
\label{omegabasis}
{\bm \Omega}_{12\cdots n} \coloneqq 
{\Omega}_{12}({\eta}_{23 \ldots n}){\Omega}_{23}(  {\eta}_{3 \ldots n}) \ldots {\Omega}_{n-1, n}( {\eta}_{n})\,, 
\ea
and the $(n{-}1)!$-counting arises from permutations 
in the labels $i=2,3,\ldots,n$ of $z_i$ and $\eta_i$. In contrast to the differences $z_{i,j} = z_{i}{-}z_j$
of the punctures, the multi-index notation for the $n{-}1$ bookkeeping variables $\eta_2,\eta_3,\ldots,\eta_n$ in (\ref{omegabasis}) refers to the sums
\beq
\eta_{ij\cdots k}=\eta_{i}+\eta_{j}+\ldots+\eta_{k}\, .
\label{exactsum}
\eeq
As illustrated in figure \ref{fig:graph} we visualize each factor of
${\Omega}_{i j}({\eta}) = {\Omega}({z}_{i}{-}{z}_j,{\eta},{\tau})$ through
an edge connecting vertices $z_i$ and $z_j$ for the punctures.
The first arguments $z_{12},z_{23},\ldots,z_{n-2,n-1}$ and $z_{n-1,n}$ of the
products in (\ref{omegabasis}) then lead to a chain structure akin to
the Parke-Taylor factors (\ref{intr.1}) in an ${\rm SL}_2$-frame, where
one of the punctures on a genus-zero surface is mapped to $\infty$.
In contrast to the genus-zero kernels $z_{ij}^{-1}$, the
${\Omega}_{i j}({\eta})$ depend on additional bookkeeping
variables $\eta$ that can be attributed to the edges in figure
\ref{fig:graph}. For a given tree-level graph, the combinations
(\ref{exactsum}) of $\eta_i$ for each edge can be determined
from the rules in section~4.1 of \cite{Broedel:2020tmd}.

 \begin{figure}[h]
  \centering
  \begin {tikzpicture}[line width = 0.3mm, scale=0.7]
\draw(-1.55,0)node{$\Omega_{ij}(\eta) \cong$};
\draw (0,0) -- (2,0);  
\draw (0,0)node{$\bullet$}node[below]{$z_i$};
\draw (2,0)node{$\bullet$}node[below]{$z_j$};
\draw(1,0.4)node{$\eta$};
\draw(3.95,0)node{$\Longrightarrow$};
\draw(6.5,0)node{${\bm \Omega}_{12\cdots n} \cong$};
\draw(8,0) -- (12.4,0);
\draw(17,0) -- (14.6,0);
\draw[dashed](14.6,0) -- (12.4,0);
\draw(8,0)node{$\bullet$} node[below]{$z_1$};
\draw(10,0)node{$\bullet$} node[below]{$z_2$};
\draw(12,0)node{$\bullet$} node[below]{$z_3$};
\draw(15,0)node{$\bullet$} node[below]{$z_{n-1}$};
\draw(17,0)node{$\bullet$} node[below]{$z_n$};
\draw(9,0.4)node{$\eta_{23\ldots n}$};
\draw(11,0.4)node{$\eta_{3\ldots n}$};
\draw(16,0.4)node{$\eta_{n}$};
    \end {tikzpicture}
  \caption{Graphical representation of Kronecker-Eisenstein series $\Omega_{ij}(\eta) = {\Omega}({z}_{i}{-}{z}_j,{\eta},{\tau})$ and their products ${\bm \Omega}_{12\cdots n}$ in the integrand (\ref{omegabasis}) of the conjectural chain basis (\ref{zdef}) of open-string genus-one integrals}
  \label{fig:graph}
\end{figure}

The $(n{-}1)!$-element set of configuration-space integrals \cite{Mafra:2019ddf, Mafra:2019xms}
 \ba 
 Z^{\tau}_{\vec{\eta}}( {\rm top} \, |\, \a_2,\cdots,\a_n) \coloneqq \int_{D^{z}_{\text {top}}} d \mu_{n}^{\rm op} \, \mathcal{I}_{n}^{\rm op} \,{\bm\Omega}_{1\a_2\,\cdots\,\a_n}
 \label{zdef}
 \ea
with permutations $\alpha_2,\alpha_3,\ldots,\alpha_n$ of $2,3,\ldots,n$ is
claimed to generate the $\tau$-integrands of open-string amplitudes
in \eqref{openamp} upon expansion in $\eta_2,\eta_3,\ldots,\eta_n$.
For open-superstring amplitudes that preserve 16 or 8 supercharges,
the configuration-space integrals \eqref{openamp} typically occur at
homogeneity degrees $\eta_j^{-4}$ and $\eta_j^{-2}$ relative to those
of open bosonic strings \cite{Tsuchiya:1988va, Stieberger:2002wk, Bianchi:2006nf, Broedel:2014vla, Bianchi:2015vsa, Berg:2016wux, Gerken:2018jrq}. Nevertheless, the 
study of $\alpha'$-expansions
for any number of supercharges is greatly facilitated when studying 
complete generating functions (\ref{zdef}) and their differential equations
in $\tau$ instead of the component integrals at fixed orders in $\eta_j$ \cite{Mafra:2019ddf, Mafra:2019xms}.

Similarly, any closed-string one-loop integrand $K_n^{\rm cl}$ is 
conjectured to be generated by linear combination of products
${\bm \Omega}_{1\a_2\a_3\cdots \a_n} \bar{\bm \Omega}_{1\b_2\b_3\cdots \b_n} $
\cite{Gerken:2019cxz}, with complex conjugates
\begin{equation}
\bar{\bm \Omega}_{12\cdots n} \coloneqq 
\overline{\Omega_{12}(  \eta_{23 \ldots n}) \Omega_{23}(\eta_{3 \ldots n}) \ldots  \Omega_{n-1, n}(  \eta_{n})}\, .
\end{equation}
More precisely, the $(n{-}1)! \times (n{-}1)!$ matrix of generating functions
 \ba 
 Y^{\tau}_{\vec{\eta}}(\a \,|\, \beta) \coloneqq \int_{ \mathfrak{T}_\tau^{n-1}} d \mu_{n}^{\rm cl} \, \mathcal{I}_{n}^{\rm cl} \,{\bm \Omega}_{1\a_2\a_3\cdots \a_n} \bar{\bm \Omega}_{1\b_2\b_3\cdots \b_n} 
 \label{ydef}
 \ea
indexed by two independent permutations $\alpha = \alpha_2,\ldots,\alpha_n$ 
and $\beta = \beta_2,\ldots,\beta_n$
of $2,3,\ldots,n$ is claimed to contain any closed-string configuration-space integral in (\ref{closedamp}) upon expansion in the $2n{-}2$ bookkeeping variables $\eta_j$ and $\bar \eta_j$.
Closed-string integrands in type II, heterotic and bosonic theories are again encountered at different orders in $\eta_j$ and $\bar \eta_j$ since the simplifications from spacetime supersymmetry typically cancel the contributions from higher orders in the $\eta,\bar \eta$-expansions.

The main goal of this paper is to validate the above conjectural bases (\ref{zdef}) 
and (\ref{ydef}) by reducing infinite families of non-trivial examples to linear combinations of
$ Z^{\tau}_{\vec{\eta}}( {\rm top} \, |\, \alpha)$ and 
$ Y^{\tau}_{\vec{\eta}}(\a \,|\, \beta)$. More specifically, we will spell out the explicit
form of the $(n{-}1)!$ reductions for generating functions of genus-one integrals
where the graphical representation of the Kronecker-Eisenstein arguments $z_{i,j}$
in figure \ref{fig:graph} features cycles instead of the chains in~(\ref{omegabasis}).

\subsubsection{Shuffle relations from Fay identities}
\label{sec:shuf}

The permutations of ${\bm \Omega}_{12\cdots n}$ in (\ref{zdef}) single out the first puncture
$z_1$ to reside at the end of the chain structure of the product in \eqref{omegabasis}. Similarly,
the bookkeeping variables $\eta_2,\eta_3,\ldots,\eta_n$ related to the punctures $z_2,z_3,\ldots,z_n$
via $\partial_{\bar z_j}{\bm \Omega}_{12\cdots n}= \frac{ \pi \eta_j }{\Im \tau} {\bm \Omega}_{12\cdots n}$ exclude $\eta_1$.
One can refer to all pairs $(z_j,\eta_j)$ with $j=1,2,\ldots,m$ and all the $m!$ permutations 
of the ordered set $\alpha= \alpha _1 \a_2\cdots \a_m$
on an equal footing by introducing $\eta_1$ through the delta function in the following 
more general definition of ${\bm\Omega}_{\alpha }$, 
\begin{equation}
\label{deflongomega2}
{\bm\Omega}_{\alpha _1 \a_2\cdots \a_m } \coloneqq \delta\bigg( \sum_{i=1}^m \eta_{\a_i} \bigg) \prod_{i=1}^{m-1}  
\Omega_{\alpha_i\, \alpha_{i+1}} ( \eta_{\alpha_{i+1}\, \cdots\, \alpha_{m}  })    \,.
\end{equation}
Here $\{\alpha _1, \a_2,\cdots, \a_m\} $ could be any subset of $\{1,2,\cdots, n\}$ with two or more elements.  
When $\a_1=1$ and $m=n$, the above generalized definition reduces to the original one \eqref{omegabasis}. As a particular virtue of the definition (\ref{deflongomega2}),
the antisymmetry property (\ref{antisyprop}) of the individual 
Kronecker-Eisenstein factors translates into the reflection identity
\beq
{\bm\Omega}_{\alpha^{\rm T}} = (-1)^{|\alpha|-1} {\bm\Omega}_{\alpha}
\label{omrefl}
\eeq
of the chain product for any ordered set $\alpha=\alpha_1\alpha_2\ldots \alpha_{|\alpha|}$ with $|\alpha|$ elements
and reversal $\alpha^{\rm T} = \alpha_{|\alpha|}\ldots \alpha_2 \alpha_1$. Moreover, iterations of
the Fay identities \eqref{fayid} among pairs of Kronecker-Eisenstein factors
imply the following chain identities
\ba
\label{fayprac}
{\bm\Omega}_{\alpha,i,\beta}= (-1)^{|\alpha|} {\bm\Omega}_{i,\alpha^{\rm T}}  {\bm\Omega}_{i,\beta}= (-1)^{|\alpha|} \sum_{\rho\in \alpha^{\rm T}\shuffle \beta} {\bm\Omega}_{i,\rho} \,, 
\ea 
which resemble Kleiss-Kuijf relations of gauge-theory tree amplitudes 
\cite{Kleiss:1988ne} and only leave $(m{-}1)!$ out of the
$m!$ permutations of ${\bm\Omega}_{\alpha _1 \a_2\cdots \a_m }$
independent \cite{Mafra:2019xms, Gerken:2019cxz, Broedel:2020tmd}. 
The shuffle $\alpha\shuffle \beta$ of ordered sets $\alpha,\beta$ in the summation range of (\ref{fayprac})
gathers all permutations of the composed ordered set $\alpha \beta$ that preserve
the order among the elements of $\alpha$ and $\beta$. The simplest two- and three-point
examples of (\ref{omrefl}) and (\ref{fayprac}) are
\ba
{\bm \Omega}_{12} &= - {\bm \Omega}_{21} \, , \ \ \ \ \ \
{\bm \Omega}_{123} =  {\bm \Omega}_{321}
 \,,
\\
{\bm\Omega}_{213}&=
 -{\Omega}_{12}(\eta_2) {\Omega}_{13}(\eta_3) 
=  -{\Omega}_{12}(\eta_{23}) {\Omega}_{23}(\eta_{3})
- {\Omega}_{13}(\eta_{23}) {\Omega}_{32}(\eta_{2})
 = -{\bm\Omega}_{123}-{\bm\Omega}_{132} \,.
  \non
\ea
Note that the chain identities (\ref{fayprac})
following from iterated Fay identities can be equivalently written as
\ba
\sum_{\rho\in \alpha \shuffle \beta} {\bm\Omega}_{\rho} =0\, ,\qquad \forall \, \alpha, \beta\neq \emptyset\,.
\label{fayshuffle}
\ea 
Based on (\ref{fayprac}) or (\ref{fayshuffle}), any product of $n{-}1$ 
Kronecker-Eisenstein series whose first arguments $z_{i,j}$ form a
tree graph can be reduced to the conjectural $(n{-}1)!$ chain basis (\ref{zdef}),
see e.g.\ appendix E of \cite{Gerken:2019cxz} or section 4.1 of \cite{Broedel:2020tmd}.
Upon Laurent expansion in the $\eta_{i\ldots j}$ variables, this
implies algebraic rearrangements of the Kronecker-Eisenstein
coefficients $f^{(w)}_{ij}$ in the open- and closed-string integrands 
(\ref{stringint}). These simplifications terminate in the chain basis
of $f^{(k_1)}_{1\alpha_2}f^{(k_2)}_{\alpha_2\alpha_3} \ldots
f^{(k_{n-1})}_{\alpha_{n-1}\alpha_n}$ with $z_1$ at one end without any need for
integration by parts (\ref{IBPomega}) provided that the starting
point already had the structure of a tree graph.

\subsubsection{Kronecker-Eisenstein cycles}

The main target of this work for F-IBP reduction into the conjectural chain
bases are Koba-Nielsen integrals over cyclic products of Kronecker-Eisenstein 
series. The explicit basis
reductions of the associated cycles of coefficients $f^{(w)}_{ij}$ can then
be obtained as corollaries upon $\eta$-expansion.

In order to set the stage for the products of cycles in section \ref{section:six}, we define cycles
\begin{equation}\label{defc}
\CC_{(12\cdots m)}(\xi) \coloneqq \delta \bigg( \sum_{i=1}^m \eta_{i} \bigg) \Omega_{12}(\eta_{23\cdots m}{+}\xi)\Omega_{23}(\eta_{3\cdots m}{+}\xi) 
\cdots 
\Omega_{m-1,m}(\eta_{ m}{+}\xi)
\Omega_{m,1}(\xi) 
\end{equation}
for general multiplicities $2\leq m\leq n$. Already for a single cycle at $m=n$, the reduction 
to products (\ref{deflongomega2}) of chain topology is beyond the reach of algebraic
manipulations due to \eqref{antisyprop} or \eqref{fayid} and calls for
integration by parts.

The $m$ factors of $\Omega_{ij}$ in (\ref{defc}) are associated with $m$ 
independent bookkeeping variables which can be taken as $\eta_2,\ldots,\eta_m,\xi$ 
on the support of the delta function. The arguments of the individual factors 
$\Omega_{j-1,j}(\eta_{j \ldots m}{+}\xi)$ in (\ref{defc}) are chosen such as to 
\begin{itemize}
\item attain simple transformation properties under cyclic shifts and cycle reversal
\begin{align}
\CC_{(12\cdots m)}(\xi{+}\eta_1) &= \CC_{(23\cdots m1)}(\xi)\,,
\label{dihedcyc}\\
\CC_{(12\cdots m)}(-\xi) &= (-1)^m \CC_{(m\cdots 21)}(\xi)\,,
\notag
\end{align} 
where the relabelling $\CC_{(12\cdots m)} \rightarrow \CC_{(23\cdots m1)}$
affects both the $z_i$ and $\eta_i$
\item preserve the antiholomorphic differential equations of the chains (\ref{deflongomega2}),
\beq
\partial_{\bar z_j} \CC_{(12\cdots m)}(\xi)= \frac{ \pi \eta_j }{\Im \tau} \CC_{(12\cdots m)}(\xi)
\ \ \leftrightarrow \ \
\partial_{\bar z_j}{\bm \Omega}_{12\cdots m}= \frac{ \pi \eta_j }{\Im \tau} {\bm \Omega}_{12\cdots m}\, ,
\quad \  \forall\, 1\leq j\leq m\,.
\label{antiholocyc}
\eeq
\end{itemize}
At two points, for instance, we obtain $\CC_{(12)}(\xi)= \delta(\eta_{12})\Omega_{12}(\eta_2{+}\xi)\Omega_{21}(\xi)$ such that $\eta_2 = - \eta_1$ brings the cyclic image into the form 
$\CC_{(21)}(\xi) = \CC_{(12)}(\xi{+}\eta_1)$. The reflection properties in turn specialize to
$\CC_{(21)}(-\xi)= \delta(\eta_{12}) \Omega_{21}(\eta_1{-}\xi)\Omega_{12}(-\xi) =\CC_{(12)}(\xi) $. 

As will be detailed in the next section, the main results of this paper are explicit
F-IBP decomposition formulae for single cycles $\CC_{(12\cdots n)}(\xi)$ or products
$\CC_{(12\cdots i)}(\xi_1) \CC_{(i+1\cdots j)}(\xi_2)\ldots$ into the conjectural
chain basis of (\ref{zdef}). The reductions will be performed in an open-string context,
i.e.\ in absence of complex-conjugate ${\bar\Omega}_{ij}$, but can be easily adapted
to closed strings: First, the reduction formulae in later sections will track the
$\nabla_j$-derivatives (\ref{IBPomega}) discarded in this process. Second,
the additional IBP contributions in presence of chains or cycles of ${\bar\Omega}_{ij}$ 
can be straightforwardly inferred from the complex conjugates of the
differential equations (\ref{antiholocyc}).

Note that the breaking of an $n$-point cycle $\CC_{(12\cdots n)}(\xi)$ via F-IBP
will resemble the genus-zero IBP reduction \cite{Schlotterer:2016cxa} of products of 
two Parke-Taylor factors (\ref{intr.1}): The ${\rm SL}_2$-fixing of one puncture at genus zero to
$z_j \rightarrow \infty$ breaks one of the Parke-Taylor cycles. In a graphical representation
of $z_{ij}^{-1}$ by edges as in figure \ref{fig:graph}, the $z_j \rightarrow \infty$ fixing at genus 
zero always reduces the loop order by one. Similarly, the basis
reductions of $k$ genus-one cycles $\CC_{(\ldots)}(\xi_i)$ necessitate F-IBP
identities of the same combinatorial complexity as those for genus-zero reductions of
$(k{+}1)$ Parke-Taylor factors \cite{He:2018pol, He:2019drm}.

\subsubsection{A more detailed outline}
\label{subsec:outline}

In the subsequent sections of this paper, we start with the F-IBP reduction of
the two-point cycle $\CC_{(12)}(\xi_1)$ in section~\ref{section:three} to illustrate the
key ideas in breaking cycles of Kronecker-Eisenstein series. Section~\ref{section:four}
culminates in an elegant closed formula (\ref{rhorho})
for the breaking of a single length-$m$ cycle,
with numerous further details and intermediate steps up to six points.
A parallel derivation of a formula to break any meromorphic Kronecker-Eisenstein cycle
in the chiral-splitting framework is given in section~\ref{section:five}. 

More complicated cases of doubly-periodic integrands including products
of cycles $\CC_{(\ldots)}(\xi_i)$ can be reduced to the conjectural chain basis
by repeated applications of the single-cycle formula (\ref{rhorho}).
In section~\ref{section:six}, we show a first non-trivial application of
the iterative procedure to break a product of two cycles of arbitrary multiplicities $m_1,m_2$
and spell out all examples up to and including $m_1{+}m_2=6$.  
For the iterative breaking of three or more cycles, it is convenient
to introduce additional combinatorial tools including labelled forests 
to organize the results in a compact way. Simple examples of
triple cycles at low multiplicities can be found at the end of section~\ref{section:six} 
while we refer the readers to a companion paper \cite{companion} for F-IBP
reductions of arbitrary numbers and multiplicities of cycles.

\section{Warm-up at two points}
\label{section:three}

In this section, we discuss the simplest case of F-IBP reductions and demonstrate the breaking of a length-two
cycle $\CC_{(12)}(\xi)= \Omega_{12}(\eta_{ 2}{+}\xi) \Omega_{21}(\xi) $ in (\ref{defc}) as 
a warm-up example. According to the coincident limit \eqref{variant} of Fay identities, we have
\ba\label{trade2}
\CC_{(12)}(\xi) =   \Omega_{12}(\eta_{ 2})  \left(\hat{g}^{(1)}(\xi, \tau)-\hat{g}^{(1)}(\eta_{ 2}{+}\xi, \tau)\right)-\partial_{2} \Omega_{12}( \eta_{ 2}) 
\ea
with $\hat{g}^{(1)}$ defined by (\ref{gexpandF}).
In presence of $n$-point Koba-Nielsen factors (\ref{defkn}),
the $z_2$-derivative in the second term of (\ref{trade2})
can be rewritten as 
\begin{align}
\big( \partial_{2} \Omega_{12}(\eta_{ 2})     \big) \mathcal{I}^\bullet_{n}& =  \partial_{2}\big( \Omega_{12}(\eta_{ 2})      \mathcal{I}^\bullet_{n}  \big) 
-\Omega_{12}(\eta_{ 2})   
\big(  \partial_{2} \mathcal{I}_{n}^\bullet \big) \label{ibp2pt} \\
&= 
\bigg({-}s_{12} f_{12}^{(1)} +\sum_{i=3}^{n}  s_{2i} f_{2i}^{(1)} + \partial_{2} \bigg) \big(
\Omega_{12}(\eta_{ 2})     \mathcal{I}^\bullet_{n} \big)\,,
\notag
\end{align}
where $\bullet$ may refer to either open or closed strings.
Together with the algebraic identity \eqref{eqUseful} to 
rewrite $f_{12}^{(1)} \Omega_{12}(\eta_{ 2})
= \partial_2 \Omega_{12}(\eta_2)+ \big(\hat{g}^{(1)}(\eta_2)+\partial_{\eta_2}\big)\Omega_{12}(\eta_2)$, this can be solved for the derivative $\partial_{2} \Omega_{12}(\eta_{ 2})  $,
\beq
(1+s_{12})\big( \partial_{2} \Omega_{12}(\eta_{ 2})     \big) \mathcal{I}^\bullet_{n}
= \bigg( \sum_{i=3}^{n}  s_{2i} f_{2i}^{(1)} 
- s_{12} \big(\hat{g}^{(1)}(\eta_2)+\partial_{\eta_2}\big)
+ \partial_{2} \bigg) \big(
\Omega_{12}(\eta_{ 2})     \mathcal{I}^\bullet_{n} \big)
\eeq
and insertion into \eqref{trade2} yields the desired reduction identity
 \begin{align}
&\CC_{(12)} (\xi) \mathcal{I}^\bullet_{n}   = \frac{1}{1+s_{12}}  \label{2ptfinalpre}\\
&\ \ \times
\bigg(s_{12} \partial_{\eta _2}-\hat g^{(1)}(\eta _2)+\left(1{+}s_{12}\right) v_1(\eta _2,\xi)-
\sum_{i=3}^{n}  s_{2i} f_{2i}^{(1)}-\partial_{2}
\bigg) \big( \Omega _{12} (\eta_2)  \mathcal{I}^\bullet_{n}  
\big)\,. \notag
\end{align}
We will use the shorthand $ v_1(\eta, \xi ) $ for the elliptic function
of two variables
\begin{align}
 v_1(\eta, \xi )&: =\hat g^{(1)}(\eta)+\hat g^{(1)}(\xi)-\hat g^{(1)}(\eta{+}\xi)=g^{(1)}(\eta)+g^{(1)}(\xi)- g^{(1)}(\eta{+}\xi)
\notag \\
&\phantom{:}=
\frac{1}{\eta} + \frac{1}{\xi} - \frac{1}{\eta{+}\xi} + \sum_{k=4}^{\infty} {\rm G}_k
 \sum_{\ell=1}^{k-2} {k{-}1\choose \ell} \eta^\ell \xi^{k-1-\ell}  \,,
 \label{defv1}
 \end{align}
which is ubiquitous to the generalizations in later sections.
We shall furthermore introduce
 \ba
 \label{mm12}
\MM_{12}(\xi) \coloneqq \left( s_{12} \partial_{\eta _2}-\hat g^{(1)}(\eta _2)+(1{+}s_{12}) v_1(\eta _2,\xi) \right)\Omega _{12} (\eta_2)
\ea
for the contribution from the two-point chain $\Omega _{12} (\eta_2)$
and employ the extended derivative $\nabla_i$ in (\ref{defnabla}) to
rewrite \eqref{2ptfinalpre} without explicit reference to the Koba-Nielsen factor,
  \ba\label{2ptfinal}
\CC_{(12)} (\xi)  =\,\,&\frac{1}{1+s_{12}}
\bigg(\MM_{12}(\xi)- \Omega _{12} (\eta_2) 
\sum_{i=3}^{n}  s_{2i} f_{2i}^{(1)}
-  \nabla_{2}   \Omega _{12} (\eta_2)   
\bigg)\,.
\ea
In applications to closed-string integrands (such as (\ref{coupling2}) below), the cycle
$\CC_{(12)}(\xi) $ may be multiplied by additional $z_2$-dependent factors.
That is why the $\nabla_2$-derivative in \eqref{2ptfinal} is tracked, and one may
still drop it in simplified situations without further $z_2$ dependence. 
The first two terms $\sim \MM_{12}(\xi)$ and $\sim \Omega _{12} (\eta_2) $ 
in (\ref{2ptfinal}) are considered as accomplishing
the reduction to a chain basis since
\begin{itemize}
\item each contribution to $\MM_{12}(\xi)$ in (\ref{mm12}) is a $z_i$-independent
linear operator (multiplication or $\eta_2$-derivative) acting on the chain-basis 
element $ \Omega _{12} (\eta_2)$ -- said operators
can be pulled out of the integral over the punctures $z_i$
in one-loop string amplitudes,
\item the products of $ \Omega _{12} (\eta_2)$ with the extra terms 
$s_{2i} f_{2i}^{(1)}$ in case of $(n\geq 3)$-point Koba-Nielsen factors
do not feature any loops in the sense of figure \ref{fig:graph} -- 
there merely realize specific components in the $\eta$-expansion
of higher-point chains ${\bm \Omega} _{12i}$.
\end{itemize}

\subsection{Connection with the two-point chain basis}
\label{sec:conbas}

We shall now employ (\ref{2ptfinal}) to
illustrate the role of the open- and closed-string integrals $Z^{\tau}_{\vec{\eta}}$ and $Y^{\tau}_{\vec{\eta}}$ in (\ref{zdef}) and (\ref{ydef}) as a two-point basis. This specializes
the Koba-Nielsen factor to $n=2$ and removes the sum over $s_{2i} f_{2i}^{(1)}$ from (\ref{2ptfinal}).
Its corollary in the open-string setting of (\ref{openamp}) is
\begin{align}
\int_{D_{\rm top}^z} dz_2 \,\mathcal{I}_{2}^{\rm op}\, \CC_{(12)}(\xi)
&=\frac{1}{1+s_{12}} \int_{D_{\rm top}^z} dz_2 \,\mathcal{I}_{2}^{\rm op}\, \MM_{12}(\xi) \label{opex2} \\
&=\left( \frac{s_{12} \partial_{\eta _2}-\hat g^{(1)}(\eta _2) }{1+s_{12}}+ v_1(\eta _2,\xi)\right) Z^{\tau}_{\eta_2}({\rm top} \, | \, 2)\,,
\notag 
\end{align} 
where the coefficient of the basis integral $Z^{\tau}_{\eta_2}$ is a differential
operator in the bookkeeping variable $\eta_2$. Similar operator-valued coefficients
of the conjectural $(n{-}1)!$-basis integrals (\ref{zdef}) were encountered in the 
$\tau$-derivatives of $Z^{\tau}_{\vec{\eta}}$ and $Y^{\tau}_{\vec{\eta}}$
determined in \cite{Mafra:2019xms} and \cite{Gerken:2019cxz}, respectively.

In an application of the cycle reduction (\ref{2ptfinal}) to the closed-string setting
(\ref{closedamp}), the $\nabla_2$-term turns out to be essential: In presence of
a complex conjugate integrand $\overline{\Omega_{12}( \eta_2)}$, we obtain
an additional contribution from the IBP relation
\beq
\overline{\Omega_{12}( \eta_2)}  \nabla_{2} \big( \Omega _{12} (\eta_2)   \big)  =\nabla_{2} \big( \overline{\Omega_{12}( \eta_2)} \Omega _{12} (\eta_2)   \big)-\Omega _{12} (\eta_2) 
\partial_2 \big( \overline{\Omega_{12}( \eta_2)}  \big) 
\label{2ptibp}
\eeq
due to (\ref{nablarule}) and find
\ba 
\label{coupling2}
\int_{\mathfrak{T}_\tau} \! \! d^2 z_2 \,\mathcal{I}_{2}^{\rm cl}\, \CC_{(12)}(\xi) \overline{\Omega_{12}( \eta_2)} 
=\,&\frac{1}{1+s_{12}} \int_{\mathfrak{T}_\tau} \! \! d^2 z_2 
\,\mathcal{I}_{2}^{\rm cl}\,
\Big( \MM_{12} (\xi) \,\overline{\Omega_{12}(\eta_2)} 
+  \Omega _{12} (\eta_2) \partial_2 \big(  \overline{\Omega_{12}(\eta_2)}  \big)  \Big)\,,
\ea 
since only the first term $\nabla_{2} \big( \overline{\Omega_{12}(\eta_2)} \Omega _{12} (\eta_2)   \big)$ in (\ref{2ptibp}) integrates to zero.
With the derivative $\partial_2 \big(\overline{\Omega_{12}(\eta_2)}  \big)= \frac{\pi {\bar\eta}_2}{ {\rm Im} \,\tau}   \overline{\Omega_{12}(\eta_2)} $ in \eqref{partialbar} intertwining left-
and right-movers, this gives rise to the last term $\sim \frac{\bar \eta_2}{\Im \tau}$ in
\ba 
\int_{\mathfrak{T}_\tau} d^2 z_2
\,\mathcal{I}_{2}^{\rm cl}\,
\CC_{(12)}(\xi) \overline{\Omega_{12}(\eta_2)} 
=\left( \frac{s_{12} \partial_{\eta _2}-\hat g^{(1)}(\eta _2) }{1+s_{12}}+ v_1(\eta _2,\xi) + \frac{\pi {\bar\eta_2}}{ {\rm Im} \,\tau (1+s_{12})}  \right) Y^{\tau}_{\eta_2}(2\,|\, 2)\,.
\label{clex2}
\ea 
The $\eta_2$-expansions of (\ref{opex2}) and (\ref{clex2}) 
are readily obtained from those of $\hat g^{(1)}(\eta _2)$ in (\ref{gexpandF})
and its doubly periodic combination $ v_1(\eta, \xi )$ in (\ref{defv1}). Upon insertion into (\ref{opex2}) and (\ref{clex2}), these expansions
generate F-IBP reductions of
Koba-Nielsen integrals over products of coefficients $f_{12}^{(k_1)} f_{12}^{(k_2)}$ that may
appear in open- and closed-string integrands (\ref{stringint}). In both cases, we
encounter denominators $1{+}s_{12}$ that signal tachyon propagation in applications
to bosonic string amplitudes and occur in the analogous integral reductions at genus 
zero \cite{Huang:2016tag, Schlotterer:2016cxa, He:2018pol, He:2019drm}.
In applications to heterotic or supersymmetric theories, such factors of $(1{+}s_{12})^{-1}$ do not translate into poles of the amplitude, either by the zeros of the accompanying torus integrals or by factors of $1{+}s_{12}$ in the associated numerators $N^{\rm op},N^{\rm cl}_w$ in (\ref{stringint}).

\section{Breaking of single cycles at arbitrary length}
\label{section:four}

In  this section, we develop a general method to break a single 
cycle $\CC_{(12\cdots m)}(\xi)$ in (\ref{defc}) of arbitrary length $m$ 
and spell out the coefficients
in its F-IBP decomposition into $(m{-}1)!$ chain integrands 
${\bm \Omega}_{1\alpha_2\alpha_3\cdots \alpha_m}$ in (\ref{omegabasis}).  
The accompanying Koba-Nielsen factors ${\cal I}_n^{\rm op},{\cal I}_n^{\rm cl}$
in (\ref{defkn}) are kept at independent multiplicity $n \geq m$ to make the results of this 
section applicable to the breaking of multiple cycles in section \ref{section:six} and \cite{companion}.

\subsection{Length-three cycles}
\label{section:four.1}

We start by adapting the key ideas in the two-point example of section \ref{section:three}
to length-three cycles $\CC_{(123)}(\xi)= \Omega_{12}(\eta_{ 23}{+}\xi) 
\Omega_{23}(\eta_{ 3}{+}\xi)\Omega_{31}(\xi)$. Apart from chain
integrands such as ${\bm \Omega} _{123}=\Omega _{12}(\eta _{23}) \Omega _{23}(\eta _3)$
in the final formula (\ref{3ptfinal}) below, we will encounter the following $z_i$-derivatives 
in intermediate steps, 
\ba\label{4zderi}
\partial_{3} {\bm \Omega} _{123}\, ,\quad \partial_{2}  {\bm \Omega} _{132}\,,\quad \partial_{2}   {\bm \Omega} _{123}\,,\quad \partial_{3} {\bm \Omega} _{132}
\ea 
which fall into two pairs under relabellings $(z_2,\eta_2) \leftrightarrow (z_3,\eta_3)$ such as $\partial_{2}{\bm \Omega}_{132} =\partial_{3}{\bm \Omega}_{123} \big| _{2\leftrightarrow 3}$. Hence, the main task is to derive a system of
F-IBP identities to solve for all of $\CC_{(123)}(\xi)$ and the four derivatives in (\ref{4zderi})
in terms of ${\bm \Omega} _{123}$ and ${\bm \Omega} _{132}$ without any $z_i$-derivatives.

\subsubsection{The F-IBP equation system for chain derivatives}

\begin{figure}[h]
  \centering
  \begin {tikzpicture}[line width = 0.3mm, scale=0.7]
\draw(-1.85,0)node{$f^{(1)}_{12} \Omega_{12}(\eta) \cong$};
\draw (0,0) -- (2,0);  
\draw (0,0)node{$\bullet$}node[below]{$z_1$};
\draw (2,0)node{$\bullet$}node[below]{$z_2$};
\draw[dotted](0,0) .. controls (0.7,0.9) and (1.3,0.9) .. (2,0);
\draw(1,-0.4)node{$\eta$};
\scope[yshift=1.1cm]
\draw(6.2,0)node{$f^{(1)}_{23}{\bm \Omega}_{123} \cong$};
\draw(8,0) -- (12,0);
\draw(8,0)node{$\bullet$} node[below]{$z_1$};
\draw(10,0)node{$\bullet$} node[below]{$z_2$};
\draw(12,0)node{$\bullet$} node[below]{$z_3$};  
\draw(9,-0.4)node{$\eta_{23}$};
\draw(11,-0.4)node{$\eta_3$};
\draw[dotted](10,0) .. controls (10.7,0.9) and (11.3,0.9) .. (12,0);
\endscope
\scope[yshift=-1.1cm]
\draw(6.2,0)node{$f^{(1)}_{13}{\bm \Omega}_{123} \cong$};
\draw(8,0) -- (12,0);
\draw(8,0)node{$\bullet$} node[below]{$z_1$};
\draw(10,0)node{$\bullet$} node[below]{$z_2$};
\draw(12,0)node{$\bullet$} node[below]{$z_3$};  
\draw(9,-0.4)node{$\eta_{23}$};
\draw(11,-0.4)node{$\eta_3$};
\draw[dotted](8,0) .. controls (9.1,0.9) and (10.9,0.9) .. (12,0);
\endscope
 \end {tikzpicture}
  \caption{Examples of $f$-$\Omega$ cycles at length two and length three.
  Factors of $\Omega_{ij}(\eta)$ and $f^{(1)}_{ij}$ are represented
  by solid and dotted lines between vertices $z_i$ and $z_j$, respectively.}
  \label{fig:fom}
\end{figure}
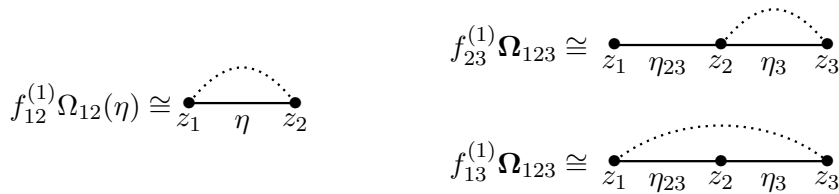

The first source of identities is the three-point analogue of the IBP relation
\eqref{ibp2pt}, 
\ba
 \partial_{3} {\bm \Omega}_{123}   = 
\Big({-}s_{13} f_{13}^{(1)}-s_{23} f_{23}^{(1)} +\sum_{i=4}^{n}  s_{3i} f_{3i}^{(1)}+\nabla_3   \Big)
{\bm \Omega}_{123}  
\,,
\label{ibp3pt}
\ea
where we again employ the operator $\nabla_i$ in (\ref{defnabla})
to capture the Koba-Nielsen factor in
$( \partial_{3} {\bm \Omega}_{123} ) \mathcal{I}^\bullet_{n} =
  \partial_{3}( {\bm \Omega}_{123}\mathcal{I}^\bullet_{n} ) 
-{\bm \Omega}_{123}
(  \partial_{3} \mathcal{I}^\bullet_{n})$. The right-hand side of (\ref{ibp3pt}) features a
chain of Kronecker-Eisenstein series as visualized in figure \ref{fig:graph}.
However, the factors of $f_{13}^{(1)}$ and $f_{23}^{(1)}$ introduce a second species of edges
 which are visualized by dotted lines in figure \ref{fig:fom} and
lead to cycles similar to  $\CC_{(12)}(\xi)$ and $\CC_{(123)}(\xi)$. 
Cycles with a mix of $f^{(w)}_{ij}$ and $\Omega_{ij}(\eta)$ factors will be referred to as
$f$-$\Omega$ cycle, and the contributions $ f_{23}^{(1)} { \Omega}_{23} (\eta_3)$ and 
$ f_{13}^{(1)} {\bm \Omega}_{123}$ to (\ref{ibp3pt}) furnish simple examples at
length two and three, respectively.

The length-two example of $f$-$\Omega$ cycles, i.e.\ the pair $f^{(1)}_{ij} \Omega_{ij}(\eta)$, 
can be traded for a derivative of  $\Omega_{ij}(\eta)$ via \eqref{eqUseful}.  Hence, for  $ f_{23}^{(1)} {\bm \Omega}_{123}$, we have 
\ba\label{fomega3f1}
 f_{23}^{(1)} {\bm \Omega}_{123}  =\,\,& \Omega_{12}(\eta_{23})  \Big(\partial_{3}\Omega_{23}(\eta_{3})+\big(\hat{g}^{(1)}(\eta_{3})+\partial_{\eta_{3}}\big)\Omega_{23}(\eta_{3})\Big)
\nl
=\,\,&
\partial_{3}{\bm \Omega}_{123}+\big(\hat{g}^{(1)}(\eta_{3})-\partial_{\eta_{2}}+\partial_{\eta_{3}}\big){\bm \Omega}_{123}\,.
\ea
In passing to the second line, we have rearranged $z_i$- and $\eta_i$-derivatives
of the individual Kronecker-Eisenstein factors to act on the entire chains. 
Throughout this work, we will extensively use
 \beq
\partial_{i}\Omega_{ij}(\eta) = -\partial_{j}\Omega_{ij}(\eta)
\, , \ \ \ \ \partial_{\eta_i}\Omega(z,\eta_{ij \ldots k}) = \partial_{\eta_j}\Omega(z,\eta_{ij\ldots k}) 
\label{movegraph}
\eeq
to iteratively reorganize expressions $\Omega \partial \Omega$ 
as total derivatives $ \partial  (\Omega \Omega )$ as long as the products $(\Omega \Omega )$ 
do not form a cycle. Simple applications of (\ref{movegraph}) include
$\Omega_{12}(\eta_{23}) \partial_{\eta_{3}}\Omega_{23}(\eta_{3})
= (\partial_{\eta_{3}} {-} \partial_{\eta_{2}}) {\bm \Omega}_{123}$ and
$( \partial_{3}\Omega_{13}(\eta_{23})  ) \Omega_{32}(\eta_{2})
= (\partial_{3}{+}\partial_{2}){\bm \Omega}_{132}$.

The $f$-$\Omega$ cycle $ f_{13}^{(1)} {\bm \Omega}_{123}$
of length three in (\ref{ibp3pt}) reduces to length two by virtue of the Fay identity
\ba
 f_{13}^{(1)} {\bm \Omega}_{123}  =\,\,& f_{13}^{(1)} \Omega_{13}(\eta_3)\Omega_{12}(\eta_2) - f_{13}^{(1)} \Omega_{13}(\eta_{23}) \Omega_{32}(\eta_2)\,.
 \label{simfay}
\ea
Applying  \eqref{eqUseful} to
$ f_{13}^{(1)} \Omega_{13}(\eta_3)$ and
$ f_{13}^{(1)} \Omega_{13}(\eta_{23})$ then results in
\begin{align}
 f_{13}^{(1)} {\bm \Omega}_{123}  &=  \Big(\partial_{3}\Omega_{13}(\eta_{3})+\big(\hat{g}^{(1)}(\eta_{3})+\partial_{\eta_{3}}\big)\Omega_{13}(\eta_{3})
 \Big)
  \Omega_{12}(\eta_2)
\label{fomega3p1} \\
&\quad
  -\Big(\partial_{3}\Omega_{13}(\eta_{23})+\big(\hat{g}^{(1)}(\eta_{23})+\partial_{\eta_{3}}\big)\Omega_{13}(\eta_{23})
 \Big)\Omega_{32}(\eta_2)\,. 
 \notag
\end{align}
By rewriting the $z_i$- and $\eta_i$ derivatives via (\ref{movegraph}) to act on the complete
chains as in (\ref{fomega3f1}), we arrive at the simplified form
\ba\label{fomega3f2}
 f_{13}^{(1)} {\bm \Omega}_{123} =\,\,& \partial_{3}{\bm \Omega}_{123}-\partial_{2}{\bm \Omega}_{132}+\hat{g}^{(1)}(\eta_{3}) ({\bm \Omega}_{123}+{\bm \Omega}_{132})
+\partial_{\eta_{3}}{\bm \Omega}_{123} 
  -\hat{g}^{(1)}(\eta_{23}){\bm \Omega}_{132}
\,.
\ea
Inserting \eqref{fomega3f1} and \eqref{fomega3f2} into \eqref{ibp3pt} results in one 
equation for the four derivatives in \eqref{4zderi},
\ba&
  \left(1{+}s_{13}{+}s_{23}\right)  \partial_{3}{\bm \Omega}_{123}  -s_{13}  \partial_{2}{\bm \Omega}_{132} = {\bm h}_{3| 23} \, ,
  \label{ibp3ptff1}
  \ea
where the right-hand side is formally free of $z_i$-derivatives $\partial_i {\bm \Omega}$ 
and denoted by ${\bm h}_{3| 23}$ for future reference
\ba
{\bm h}_{3| 23}
\coloneqq\,\,& \Big({-}(s_{13}{+} s_{23}) \big( \hat{g}^{(1)}(\eta_{3})+ \partial_{\eta_{3}} \big) +s_{23}\,\partial_{\eta_{2}} +\sum_{i=4}^{n}  s_{3i} f_{3i}^{(1)}  +\nabla_3 \Big)
{\bm \Omega}_{123}  
\nl&
 + s_{13}\big( \hat{g}^{(1)}(\eta_{23}) -\hat{g}^{(1)}(\eta_{3}) \big)
{\bm \Omega}_{132}  \,.
\label{b323}
\ea
The above steps to rewrite the derivative $\partial_3{\bm \Omega}_{123}$
w.r.t.\ the endpoint $z_3$ of the chain in (\ref{ibp3pt}) can be straightforwardly adapted to
$( \partial_{2} {\bm \Omega}_{123}  ) \mathcal{I}^\bullet_{n} =  
\partial_{2} ( {\bm \Omega}_{123}\mathcal{I}^\bullet_{n}  ) -{\bm \Omega}_{123}(  \partial_{2} \mathcal{I}^\bullet_{n})$, where the derivative is now taken w.r.t.\ the puncture $z_2$
in the middle of the chain ${\bm \Omega}_{123}$. After manipulations similar to those
in (\ref{fomega3f1}) to (\ref{fomega3f2}), we arrive at a second equation for the $\partial_i {\bm \Omega}$
\ba
\label{ibp3ptff2}
&
\left(1{+}s_{12}\right)  \partial_{2}{\bm \Omega}_{123} +\left(s_{12}{-}s_{23}\right) \partial_{3}{\bm \Omega}_{123} 
= {\bm h}_{2| 23}\,,
  \ea
where the building block ${\bm h}_{2| 23}$ on the right-hand side is again free of 
$z_i$-derivatives $\partial_i {\bm \Omega}$ but not a relabelling of
${\bm h}_{3| 23}$ in (\ref{b323}):
\ba
{\bm h}_{2| 23}
\coloneqq\,\,& \Big(  s_{23} \big( \hat{g}^{(1)}(\eta_{3})+ \partial_{\eta_{3}} \big)
-s_{12}\, \hat{g}^{(1)}(\eta_{23})
 -(s_{12}{+}s_{23})\,\partial_{\eta_{2}} +\sum_{i=4}^{n}  s_{2i} f_{2i}^{(1)}  +\nabla_2 \Big)
{\bm \Omega}_{123}    \,.
\label{otherb3}
\ea
Relabelling $2\leftrightarrow 3$ in \eqref{ibp3ptff1} and \eqref{ibp3ptff2} by
trading $(z_2,\eta_2,s_{12}) \leftrightarrow (z_3,\eta_3,s_{13})$ 
yields another pair of equations,
\ba
  \left(1{+}s_{12}{+}s_{23}\right)  \partial_{2}{\bm \Omega}_{132}  -s_{12} \partial_{3}{\bm \Omega}_{123} =\,\,&  {\bm h}_{2| 32}\,,
  \nl 
  \left(1{+}s_{13}\right)  \partial_{3}{\bm \Omega}_{132} +\left(s_{13}{-}s_{23}\right)  \partial_{2}{\bm \Omega}_{132}
=\,\,& {\bm h}_{3| 32} \,,
\label{ibp3ptff3}
\ea
where ${\bm h}_{2| 32}={\bm h}_{3| 23}\big|_{2\leftrightarrow3}$
and ${\bm h}_{3| 32}  = {\bm h}_{2| 23}  \big|_{2\leftrightarrow3}$
are obtained by relabelling (\ref{b323}) and (\ref{otherb3}).

Based on the four equations in (\ref{ibp3ptff1}), (\ref{ibp3ptff2}) and (\ref{ibp3ptff3}), we
can now expand the four $z_i$-derivatives \eqref{4zderi} in the chain basis,
\ba\label{sol123}
\partial_{3}{\bm \Omega}_{123} &= \frac{ {\bm h}_{3|23} (1{+}s_{12}{+}s_{23})+{\bm h}_{2|32} s_{13} }{(1{+}s_{23}) (1{+}s_{123}) }\, ,
\\
\partial_{2}{\bm \Omega}_{123} &= 
 \frac{{\bm h}_{2|23} (1{+}s_{23}) (1{+}s_{123})-  {\bm h}_{2|32}s_{13} (s_{12}{-}s_{23})-{\bm h}_{3|23} (s_{12}{-}s_{23}) (1{+}s_{12}{+}s_{23}) }{(1{+}s_{12}) (1{+}s_{23}) (1{+}s_{123}) }
\,.
\notag
\ea
The remaining two derivatives $\partial_{2}{\bm \Omega}_{132}$ and $\partial_{3}{\bm \Omega}_{132}$ are obtained by relabelling $2\leftrightarrow 3$. By the expressions (\ref{b323})
and (\ref{otherb3}) for 
${\bm h}_{i|ji}$ and ${\bm h}_{i|i j}$, the right-hand sides of (\ref{sol123}) 
are in a chain basis: Similar to the comments below (\ref{2ptfinal}), each contribution to
${\bm h}_{i|ji},{\bm h}_{i|i j}$ is one of
\begin{itemize}
\item a linear, $z_i$-independent operation (multiplication or $\eta_i$-derivative) of the 
chains ${\bm \Omega}_{123}$ or ${\bm \Omega}_{132}$ which can be pulled out of
the integrals over the punctures,
\item a Koba-Nielsen derivative $\nabla_i$ which will contribute 
factors of $\frac{ \pi \bar \eta_i }{\Im \tau}$ in closed-string applications
with complex conjugate chains and cycles,
see (\ref{antiholocyc}) and (\ref{clex2}),
\item products of ${\bm \Omega}_{123}$ or ${\bm \Omega}_{132}$ with $f_{3i}^{(1)}$ or $f_{2i}^{(1)}$ with $i\geq 4$ which fall into four-point chain bases
(possibly after rearranging ${\bm \Omega}_{123} f_{2i}^{(1)}$
or ${\bm \Omega}_{132} f_{3i}^{(1)}$ via Fay identities).
\end{itemize}

\subsubsection{Breaking the length-three cycle}

The chain-basis reduction \eqref{sol123} of derivatives $\partial_i {\bm \Omega}_{123}$
is the key to break the cycle $\CC_{(123)}(\xi)$. Using Fay identities, the
length-three cycle can be reduced to length two,
\ba\label{cc123p1}
\CC_{(123)}(\xi)=
\Omega _{12}(\eta _{23}) \Omega _{23}(\eta _{3}{+}\xi) \Omega _{32}(\xi)-\Omega _{13}(\eta _{23}) \Omega _{23}(\eta _{3}+\xi) \Omega _{32}(\eta _{23}{+}\xi)\,,
\ea  
and we can now follow \eqref{trade2} to convert the bilinears in $\Omega_{23}$ 
to $z$-derivatives. We again apply (\ref{movegraph}) to reorganize derivatives,
\begin{align}
\Omega _{12}(\eta _{23})  \Omega _{23} (\eta _{3}{+}\xi) \Omega _{32} (\xi ) &=  \Omega _{12}(\eta _{23}) \Omega_{23}(\eta_{ 3})  \big(\hat{g}^{(1)}(\xi)-\hat{g}^{(1)}(\eta_{ 3}{+}\xi)\big)
+\Omega _{12}(\eta _{23}) \partial_{2} \Omega_{23}( \eta_{ 3} ) \notag \\
&=  \big(\hat{g}^{(1)}(\xi)-\hat{g}^{(1)}(\eta_{ 3}{+}\xi)\big) {\bm \Omega} _{123}
- \partial_{3}  {\bm \Omega} _{123}\,, 
\end{align}
and simplify the second term on the right-hand side of \eqref{cc123p1} 
with the same methods,
\begin{align}
\CC_{(123)}(\xi)&=-\partial_{3}  {\bm \Omega} _{123}+\partial_{2} {\bm \Omega} _{132}
+\big(\hat{g}^{(1)}(\xi)-\hat{g}^{(1)}(\eta _{3}{+}\xi)\big) {\bm \Omega} _{123}  \notag \\
&\quad +\big(\hat{g}^{(1)}(\eta _{23}{+}\xi)-\hat{g}^{(1)}(\eta _{3}{+}\xi)\big) {\bm \Omega} _{132}\,.
\label{cc123tozd}
\end{align}
 Thus, the chain-basis expansion of the length-three cycle $\CC_{(123)}(\xi)$
reduces to that of the $z_i$-derivatives (\ref{4zderi}) of the chains 
${\bm \Omega}_{123}$ and ${\bm \Omega}_{132}$.  
Substituting the solutions \eqref{sol123} into \eqref{cc123tozd}, we get
\ba
\CC_{(123)}(\xi)=\frac{{\bm h}_{2|32}-{\bm h}_{3|23}}{1+s_{123}} 
+\big(\hat{g}^{(1)}(\xi)-\hat{g}^{(1)} (\eta _{3}{+}\xi )\big) {\bm \Omega} _{123} +\big(\hat{g}^{(1)}(\eta _{23}{+}\xi )-\hat{g}^{(1)}(\eta _{3}{+}\xi )\big) {\bm \Omega} _{132}\,.
\label{closetofinal}
\ea
Note in particular that only the first tachyon pole from the
expressions (\ref{sol123}) for $\partial_3{\bm \Omega}_{123} \sim (1{+}s_{123})^{-1} (1{+}s_{23})^{-1}$ is left. Moreover, only $ {\bm h}_{3|23}$ and $ {\bm h}_{2|32}$ given by  (\ref{b323}) appear in  (\ref{sol123}) whereas their relabelling-inequivalent counterparts
${\bm h}_{2|23},{\bm h}_{3|32}$ determined by (\ref{otherb3})
dropped out from (\ref{closetofinal}).
 After inserting \eqref{b323}, our final result is
\begin{align}\label{3ptfinal}
\CC_{(123)}(\xi)
=\,\,&\frac{\MM_{123}(\xi)}{1+s_{123}}- \frac{1}{1+s_{123}}
\Bigg[\Bigg(   \sum_{i=4}^{n}  x_{3,i}  
+\nabla_3
   \Bigg){\bm \Omega}_{123} 
-
\Bigg(   \sum_{i=4}^{n}  x_{2,i}  
+\nabla_2
   \Bigg){\bm \Omega}_{132} 
 \Bigg]
 \,,
\end{align}
with $x_{3,i} = s_{3i} f^{(1)}_{3i}$ and
\begin{align}
\MM_{123}(\xi)
\coloneqq\,\,&
\Big[  \big( (s_{13} {+}s_{23} ) \partial_{\eta _3}-s_{23} \partial_{\eta _2} -\hat g^{(1)} \left(\eta _3\right)-s_{12} v_1(\eta _3,\eta _2)  \big){\bm \Omega} _{123} 
-(2\leftrightarrow 3) \Big]
\nl&
+(1{+}s_{123})\big( v_1(\eta _3,\xi) {\bm \Omega} _{123} - v_1(\eta _2,\eta _{3}{+}\xi ){\bm \Omega} _{132} \big)
 \,.
 \label{mm123}
\end{align}
The elliptic function $v_1(\eta _3,\xi)$ is defined in (\ref{defv1}),
and each term on the right-hand side of (\ref{3ptfinal}) is in a chain
basis by the discussion below (\ref{sol123}).

\subsection{Length-four cycles}
\label{section:four.2}

Also for the length-four cycle $\CC_{(1234)}(\xi)= \Omega_{12}(\eta_{ 234}{+}\xi) \Omega_{23}(\eta_{ 34}{+}\xi)\Omega_{34}(\eta_{ 4}{+}\xi)\Omega_{41}(\xi)$, we follow the strategy of section \ref{section:four.1} and start by reducing
the 18 permutations of the $z_i$-derivatives $\partial_{2}{\bm  \Omega }_{1234 } ,
\partial_{3}{\bm  \Omega }_{1234 } , \partial_{4}{\bm  \Omega }_{1234 } $ to a chain basis,
\ba 
\label{18zderi}
\partial_{i}{\bm  \Omega }_{1, \rho(2,3,4) } \qquad ~{\rm with}~ 2\leq i\leq 4\, , \  \rho\in S_3\,.
\ea 

\subsubsection{The F-IBP equation system for chain derivatives}

As a convenient starting point for the F-IBP manipulations to expand 
the total of 18 derivatives (\ref{18zderi}) in a chain basis, we rewrite
\ba
 \partial_{4} {\bm \Omega} _{1234}   = 
\Big({-}s_{14} f_{14}^{(1)}-s_{24} f_{24}^{(1)}-s_{34} f_{34}^{(1)} +\sum_{i=5}^{n}  s_{4i} f_{4i}^{(1)}+\nabla_4  \Big)
{\bm \Omega} _{1234}  
\label{ibp4pt}
\ea
using $( \partial_{4} {\bm \Omega} _{1234} ) \mathcal{I}_{n}^\bullet =  \partial_{4} ( {\bm \Omega} _{1234}\mathcal{I}_{n}^\bullet  ) 
-{\bm \Omega} _{1234}
 (  \partial_{4} \mathcal{I}_{n}^\bullet )
$. The products of $ {\bm \Omega}_{1234}$ with $f_{34}^{(1)}, f_{24}^{(1)} $ and $f_{14}^{(1)}$ on the right-hand side of (\ref{ibp4pt}) form
 $f$-$\Omega$ cycles in the sense of figure \ref{fig:fom} of length two, three and four, respectively. 
 Similar to (\ref{simfay}), 
 repeated Fay identities allow to reduce all of these $f$-$\Omega$ cycles to length two, i.e.\ to products of $f_{i4}^{(1)}  \Omega_{i4}$ with chains. These products in turn can be readily 
 converted to derivatives acting on a single $\Omega_{ij}$, see (\ref{eqUseful}).  
Since none of these derivatives act on $\Omega_{ij}$ within cycles,
all of the $\partial_i$ can be moved to act on full-fledged
chains ${\bm \Omega} _{1,\rho(2,3,4)}$ by iterative
use of (\ref{movegraph}). After applying these steps to
all the products $ f_{j4}^{(1)} {\bm \Omega} _{1234}  $ in (\ref{ibp4pt}) with $j=1,2,3$, we arrive at
\ba\label{ibp4ptff1}
\! \! \!\left(1{+}s_{14}{+}s_{24}{+}s_{34}\right)  \partial_{4}   {\bm \Omega} _{1234}-   \left(s_{14}{+}s_{24}\right) \partial_{3} {\bm \Omega} _{1243}-  s_{14} \partial_{3}  {\bm \Omega} _{1423}+  s_{14}\partial_{2}   {\bm \Omega} _{1432}= {\bm h}_{4|234}\,,
\ea
where the following combination is free of cycles or $z_i$-derivatives,
\ba
 {\bm h}_{4|234}\coloneqq\,\,&
\bigg({-}(s_{14}{+}s_{24}{+}s_{34})  \big( \hat{g}^{(1)}(\eta_{4})  +\partial_{\eta_{4}}   \big) + s_{24} \partial_{\eta_{2}}   +  s_{34} \partial_{\eta_{3}}  +\sum_{i=5}^{n}  s_{4i} f_{4i}^{(1)} +\nabla_4  \bigg)
{\bm \Omega} _{1234}  
\nl&
+\big(\hat{g}^{(1)}(\eta_{34}) - \hat{g}^{(1)}(\eta_{4})\big) \big ( (s_{14}{+}s_{24}) {\bm \Omega} _{1243}+ s_{14} {\bm \Omega} _{1423} \big) \nl
&+\big(\hat{g}^{(1)}(\eta_{34})-\hat{g}^{(1)}(\eta_{234}) \big) s_{14} {\bm \Omega} _{1432}
\label{b4234}\,.
\ea
The analogous IBPs $( \partial_{i} {\bm \Omega} _{1234}   ) \mathcal{I}^\bullet_{n} =  \partial_{i} ( {\bm \Omega} _{1234}\mathcal{I}^\bullet_{n} ) 
-{\bm \Omega} _{1234} (  \partial_{i} \mathcal{I}^\bullet_{n} )
$ with $i=2,3$, yield two other equations
\ba
 \left(1{+}s_{12}\right)  \partial_{2}{\bm \Omega} _{1234}+\left(s_{12}{-}s_{23}\right)  \partial_{3} {\bm \Omega} _{1234}+\left(s_{12}{-}s_{23}{-}s_{24}\right) \partial_{4}   {\bm \Omega} _{1234}+s_{24} \partial_{3}   {\bm \Omega} _{1243}&= {\bm h}_{2|234}\,,
\nl
 \left(1{+}s_{13}{+}s_{23}\right) \partial_{3}  {\bm \Omega} _{1234}+\left(s_{13}{+}s_{23}{-}s_{34}\right) \partial_{4}  {\bm \Omega} _{1234}-s_{13} \partial_{2}   {\bm \Omega} _{1324}-s_{13} \partial_{2}   {\bm \Omega} _{1342}&= {\bm h}_{3|234}\,,
 \label{otheribp4}
\ea
where $ {\bm h}_{2|234},  {\bm h}_{3|234}$ are free of $z$-derivatives, take a form
similar to (\ref{b4234}) but cannot be obtained from its relabellings.
Permutations of \eqref{ibp4ptff1} and \eqref{otheribp4} in the labels $2,3,4$ of 
$z_i, \eta_i, s_{ij}$ yield a system of 18 equations, 
which can be solved for the 18 $z_i$-derivatives in \eqref{18zderi}. In particular, we arrive 
at an expression which is completely determined by permutations of $ {\bm h}_{4|234}$ in (\ref{b4234}),
\begin{align}
\label{sol1234}
&\left(1{+}s_{34}\right) \left(1{+}s_{234}\right) \left(1{+}s_{1234}\right)\partial_{4}{\bm \Omega} _{1234} 
=  -{\bm h}_{2|432} s_{14} \left(1{+}s_{23}{+}s_{34}\right)
-{\bm h}_{2|342} s_{13} s_{24} \\
&\quad \quad
+{\bm h}_{3|243} \big(s_{14}  (1{+}s_{234} )+s_{24}  (1{+}s_{12}{+}s_{234} )\big)
+{\bm h}_{3|423}s_{14} \left(1{+}s_{23}{+}s_{34}\right)
  \notag \\
&\quad \quad
+ {\bm h}_{4|234} \big(s_{12} (1{+}s_{23}{+}s_{34} )+ (1{+}s_{13}{+}s_{23}{+}s_{34} )  (1{+}s_{234})\big) +{\bm h}_{4|324} s_{13} s_{24} \,, \notag
\end{align}
i.e.\ where all permutations of $ {\bm h}_{2|234}$ and ${\bm h}_{3|234}$ in $2,3,4$ dropped out.

\subsubsection{Breaking the length-four cycle}
\label{brkC1234}

The next step is to apply the chain expansion \eqref{sol1234} of
$\partial_{4}{\bm \Omega} _{1234}$ and its 
relabellings to break the length-four cycle. 
Similar to the length-three strategy in (\ref{cc123p1}), we
reduce $\CC_{(1234)}(\xi)$ to a combination of length-two 
cycles (possibly multiplied by chains) by virtue of Fay identities,
\begin{align}
\label{cc1234p1}
\CC_{(1234)}(\xi)&=-\Omega _{14}(\eta _{234})
\big[  \Omega _{23}(\eta _{34}{+}\xi ) \Omega _{32}(\eta _{4}{+}\xi) \big] 
\Omega _{42}(\eta _{23})  \\
&\quad +\Omega _{12}(\eta _{234}) \Omega _{23}(\eta _{34}) \big[ \Omega _{34} (\eta _{4}{+}\xi) \Omega _{43}(\xi) \big] \notag \\
&\quad +\Omega _{14}(\eta _{234}) 
\big[ \Omega _{23}(\eta _{34}{+}\xi ) \Omega _{32}(\eta _{234}{+}\xi ) \big] 
 \Omega _{43}(\eta _{23}) \notag \\
 &\quad -\Omega _{12}(\eta _{234}) \Omega _{24}(\eta _{34}) 
 \big[ \Omega _{34}(\eta _{4}{+}\xi ) \Omega _{43}(\eta _{34}{+}\xi) \big]
\nonumber
\,.
\end{align}
The length-two cycles on the right-hand side are once more converted
to derivatives $\partial \Omega$ using \eqref{variant}, for instance,
 \ba
  \Omega _{23}(\eta _{34}{+}\xi ) \Omega _{32}(\eta _{4}{+}\xi)= \Omega_{23}(\eta_{ 3})  \big(\hat{g}^{(1)}(\eta_{ 4}{+}\xi)-\hat{g}^{(1)}(\eta_{ 34}{+}\xi )\big)
+\partial_{2} \Omega_{23}( \eta_{ 3} )\,.
  \ea
As before, the rearrangement of derivatives via (\ref{movegraph}) leads to
$\partial_i$ acting on full-fledged chains, for example
$ \Omega _{14}(\eta _{234})\Omega _{42}(\eta _{23})
 \partial_{2} \Omega_{23}( \eta_{ 3} ) =-   \partial_{3}  {\bm \Omega} _{1423}$.
Moreover, Fay identities
(e.g.\ their shuffle representation in section \ref{sec:shuf}) allow to express
all the products of three $\Omega_{ij}$ in \eqref{cc1234p1} in terms of
the six chains ${\bm  \Omega }_{1, \rho(2,3,4) }$.
Based on these manipulations, the expression (\ref{cc1234p1}) for the 
cycle $\CC_{(1234)}(\xi)$ reduces to four-point chains and their $z$-derivatives,
\ba\label{cc1234tozd}
\CC_{(1234)}(\xi)=\,\,& \partial_{3} {\bm \Omega} _{1243}+\partial_{3} {\bm \Omega} _{1423}-\partial_{4} {\bm \Omega} _{1234}-\partial_{2}{\bm \Omega} _{1432}
\nl &
+{\bm \Omega} _{1234} \big(\hat g^{(1)}(\xi )-\hat g^{(1)} (\eta _{4}{+}\xi)\big)
+{\bm \Omega} _{1432}\big(\hat g^{(1)} (\eta _{34}{+}\xi)-\hat g^{(1)} (\eta _{234}{+}\xi)\big)
\nl\,\,&+({\bm \Omega} _{1243} +{\bm \Omega} _{1423}) \big(\hat g^{(1)} (\eta _{34}{+}\xi)-\hat g^{(1)} (\eta _{4}{+}\xi)\big)\,.
\ea
Substituting the chain-expansion of $\partial_{4} {\bm \Omega} _{1234}$ in 
\eqref{sol1234} and its relabellings into (\ref{cc1234tozd}), we~get
\ba
\CC_{(1234)}(\xi)=\,\,&\frac{{\bm h}_{3|243}+{\bm h}_{3|423}
-{\bm h}_{4|234}-{\bm h}_{2|432}}{1+s_{1234}} +
{\bm \Omega} _{1234} \big(\hat g^{(1)}(\xi)-\hat g^{(1)} (\eta _{4}{+}\xi)\big)
\\ &
+({\bm \Omega} _{1243} +{\bm \Omega} _{1423}) \big(\hat g^{(1)} (\eta _{34}{+}\xi)-\hat g^{(1)} (\eta _{4}{+}\xi)\big)
\nl\,\,&
+{\bm \Omega} _{1432} \big(\hat g^{(1)} (\eta _{34}{+}\xi)-\hat g^{(1)} (\eta _{234}{+}\xi)\big)
\,.
\non
\ea 
In spite of the multiple tachyon poles $(1{+}s_{jk})^{-1}(1{+}s_{ijk})^{-1}(1{+}s_{1ijk})^{-1}$ in the expression (\ref{sol1234}) for the individual $\partial_k {\bm \Omega} _{1ijk}$,
only the four-particle pole in $(1{+}s_{1234})$ is left. Plugging in the expression \eqref{b4234} for ${\bm h}_{4|234}$ and its relabellings leads to the final result
\ba\label{4ptfinal}
\CC_{(1234)} (\xi) 
=\,\,& \frac{\MM_{1234}(\xi) }{1+s_{1234}}
- 
\frac{1}{1+s_{1234}}
\Bigg[ \Big(
\sum_{i=5}^{n}  s_{4i} f_{4i}^{(1)} 
 +\nabla_4 \Big)
{\bm \Omega} _{1234} +
 \Big(
\sum_{i=5}^{n}  s_{2i} f_{2i}^{(1)} 
 +\nabla_2 \Big)
{\bm \Omega} _{1432} 
\nl&
\qquad \qquad \qquad \qquad \qquad \qquad
-\Big(  \sum_{i=5}^{n}  s_{3i} f_{3i}^{(1)} +\nabla_3 \Big)
\big( {\bm \Omega} _{1423}  + {\bm \Omega} _{1243}  \big)  \Bigg]
\,,
\ea
with 
\ba
\notag
&\!\!\! \MM_{1234} (\xi) \coloneqq (1{+}s_{1234}) \big[  {\bm \Omega} _{1234} v_1(\eta _4, \xi)
{+}{\bm \Omega} _{1432} v_1(\eta _2,\eta _{34}{+}\xi)
{-} \big( {\bm \Omega} _{1423} {+} {\bm \Omega} _{1243} )  v_1(\eta _3,\eta _{4}{+}\xi) \big]
\\
&\!\!\!+
\Big[ 
\big((s_{14}{+}s_{24}{+}s_{34}) \partial_{\eta _4}
{-}s_{24} \partial_{\eta _2}
{-}s_{34} \partial_{\eta _3}
{-}\hat g^{(1)}(\eta _4)
{-}(s_{13}{+}s_{23}) v_1(\eta _4,\eta _3)
{-}s_{12} v_1(\eta _4,\eta _{23})
\big){\bm \Omega} _{1234} \nl
&\!\!\!
\ \ \,  -\big( (s_{13} {+}s_{23} {+}s_{34} )\partial_{\eta _3}
{-}s_{23} \partial_{\eta _2}
{-}s_{34} \partial_{\eta _4}
{-}\hat g^{(1)}(\eta _3)
{-}(s_{12}{+}s_{24}) v_1(\eta _3,\eta _2)
{-}s_{14} v_1(\eta _3,\eta _4) \big){\bm \Omega}_{1423} 
\nl
&\!\!\!
 \ \ \,  +\big(2\leftrightarrow4\big)\Big]
-s_{13}
 ({\bm \Omega} _{1324}+{\bm \Omega} _{1342}) 
\big(
 \hat g^{(1)}(\eta_{3} )  -\hat g^{(1)}(\eta_{23} )  -\hat g^{(1)}(\eta_{34} )  
+\hat g^{(1)} (\eta_{234} )  
\big)
\,, \label{mm1234}
\ea
where the relabelling $2\leftrightarrow 4$ 
applies to the second and third lines and concerns the labels of all of $\eta_i,s_{ij}$ and 
${\bm \Omega} _{1ijk}$.
Note that the combination in the last line can be written
as a combination of the elliptic function in (\ref{defv1}),
\begin{align}
v_1(\eta_3,-\eta_{23})+ v_1(\eta_{234},-\eta_{34}) &=   \hat g^{(1)}(\eta_{3} )  -\hat g^{(1)}(\eta_{23} )  -\hat g^{(1)}(\eta_{34} )  
+\hat g^{(1)} (\eta_{234} )  \label{vfourg} \\
&= g^{(1)}(\eta_{3} )  -g^{(1)}(\eta_{23} )  -g^{(1)}(\eta_{34} )  
+g^{(1)} (\eta_{234} )  \,. \notag
\end{align}

\subsection{Generalization to cycles of arbitrary length}
\label{section:four.3}

With the experience from the previous sections, we shall now
perform the chain reduction of cycles $\CC_{(12\cdots m)}(\xi)$ of arbitrary length
$m\geq 2$ in presence of Koba-Nielsen factors $ \mathcal{I}^\bullet_{n}$ at $n\geq m$ points. 
Following (\ref{cc123p1}) and (\ref{cc1234p1}), a key step is to apply Fay identities 
to rewrite $\CC_{(12\cdots m)}(\xi)$ as a combination of
chains of lower length multiplying length-two cycles $\Omega_{ij} \Omega_{ji}$. 
The latter can be immediately converted to
derivatives $\partial_i \Omega_{ij}$ using \eqref{variant} and ultimately to
derivatives of length-$m$ chains via (\ref{movegraph}), resulting in the all-multiplicity formula
\begin{align}\label{cc123mtozd}
\CC_{(12\cdots m)}(\xi)
= 
 \sum_{
b  =2}^{m} \sum_{\substack{\rho \in \{2,3,\cdots, b-1\}
\\ 
\,\, \shuffle \{m,m-1,\cdots, b+1 \}}} \!\!\!\!\!\!\!\!\!\!
 (-1)^{m-b} \Big(  
\hat g^{(1)} \left(\eta_{b+1,\cdots,m}{+}\xi \right)-\hat g^{(1)} \left(\eta_{b,\cdots,m}{+}\xi 
 \right)-\partial_{b}
\Big){\bm \Omega}_{1,\rho,b} \,,
 \end{align}
see (\ref{cc123tozd}) and (\ref{cc1234tozd}) for the examples at $m=3$ and $m=4$.
The leftover task is to set up and solve an F-IBP equation system that reduces the
$(m{-}1)(m{-}1)! $ derivatives of the shuffle-independent chains
 \ba 
\label{genezderi}
\partial_{i}{\bm  \Omega} _{1, \rho(2,3,\cdots, m) } \qquad ~{\rm with}~ 2\leq i\leq m
\, , \  \rho\in S_{m-1}
\ea 
to chains without any $z_i$-derivatives.

\subsubsection{The F-IBP equation system for chain derivatives}

As a generalization of (\ref{ibp3pt}) and (\ref{ibp4pt}), our starting point to generate
F-IBP relations is
\bea
\partial_m {\bm \Omega}_{12\cdots m}  = \bigg({-} \sum_{\ell=1}^{m-1} s_{\ell,m} f^{(1)}_{\ell,m}
+ \sum_{i=m+1}^n s_{m,i} f^{(1)}_{m,i} + \nabla_m \bigg) {\bm \Omega}_{12\cdots m} \, ,
\label{genFIBP}
\eea
based on $( \partial_{m} {\bm \Omega}_{12\cdots m} ) \mathcal{I}^\bullet_{n} 
=  \partial_{m}( {\bm \Omega}_{12\cdots m}\mathcal{I}^\bullet_{n} ) 
-{\bm \Omega}_{12\cdots m}
(  \partial_{m} \mathcal{I}^\bullet_{n})$. The products of $f^{(1)}_{\ell,m}$ and
${\bm \Omega}_{12\cdots m}$ form $f$-$\Omega$ cycles in sense of figure \ref{fig:fom}
whose lengths range between two and $m$. Fay identities of the chains then
reduce any $f$-$\Omega$ cycle to length two, which can be immediately converted to 
derivatives acting on a single $\Omega$. By the absence of cycles at this stage, these
derivatives can again be rearranged via (\ref{movegraph}) to act on full-fledged chains, and 
we obtain the following $m$-point generalization of (\ref{ibp3ptff1}) and (\ref{ibp4ptff1})
\begin{align}
\partial_{m} {\bm \Omega}_{12\cdots m} +
 \sum_{
b  =2}^{m} \sum_{\rho \in \{2,3,\cdots, b-1\}\shuffle \{m,m-1,\cdots, b+1 \}} \!\!\!\!\!\!\!\!\!\!
 (-1)^{m-b} S_{m, \rho}
 \partial_{b} 
{\bm \Omega}_{1,\rho,b} = {\bm h}_{m|23\cdots m}\,,
\label{mderatmpt}
 \end{align}
where the following generalization of \eqref{b323} and \eqref{b4234} is free of cycles
or $z_i$-derivatives
  \begin{align}
 &{\bm h}_{m|23\cdots m} \coloneqq \Bigg(
 \sum_{i=2}^{m-1}s_{i,m}\partial_{\eta_i}-  \sum_{i=1}^{m-1} s_{i,m} \big(\partial_{\eta_m}{+} \hat{g}^{(1)}(\eta_m) \big)+ \! \!\sum_{i=m+1}^{n}  s_{mi} f_{mi}^{(1)}+ \nabla_m  \Bigg){\bm \Omega} _{12\cdots m} 
\nl
  &\quad \quad +
   \sum_{
b  =2}^{m} \sum_{\rho \in \{2,\cdots, b-1\}\shuffle \{m,\cdots, b+1 \}} \!\!\!\!\!\!\!\!\!\!
 (-1)^{m-b} S_{m, \rho} 
    \left(\hat{g}^{(1)}(\eta_{b+1,\cdots,m})-\hat{g}^{(1)}(\eta_{b,\cdots,m})\right) 
\,{\bm \Omega} _{1,\rho,b}\,.
  \label{bm23m}
 \end{align}
 Moreover, both of (\ref{mderatmpt}) and (\ref{bm23m}) feature the following
 shorthand $S_{j, \rho}$ for sums of Mandelstam variables
\begin{align}
S_{j, \rho} \coloneqq
\left\{
\begin{array}{ll}
s_{1j }+
\sum\limits_{  i \in \rho} s_{  i j} &: \ {\rm if}
~ j\notin \rho\,,
\\
s_{ 1 j}+ \! \! \! \! \! \sum\limits_{ i \in \rho \atop i \text { precedes } j \text { in } \rho}
\! \! \! \! \! s_{  i j} \ \ \ \ 
&: \ {\rm if}~ j\in \rho\,,
\end{array}
\right.
\label{defsjrho}
 \end{align}
e.g.\ $S_{4, 23}= S_{4, 234}=s_{14}{+}s_{24}{+}s_{34}$ as well as
$S_{4, 423}=s_{14}$ and $S_{5, 42536}=s_{15}{+}s_{45}{+}s_{25}$.
 
The same methods can be used to obtain representations of $\partial_i {\bm \Omega} _{1,\rho(2,\cdots,m)} $ similar to (\ref{mderatmpt}) for any $i=2,3,\ldots,m{-}1$ and $\rho \in S_{m-1}$.
This leads to a total of $(m{-}1)(m{-}1)!$ F-IBP relations which suffice
to solve for the chain derivatives in \eqref{genezderi}.  
 
 \subsubsection{General formula for $m$-cycles}

The only chain derivatives in the expression (\ref{cc123mtozd}) for the length-$m$ cycle 
are w.r.t.\ the endpoint, i.e.\ $\partial_{m} {\bm \Omega}_{12\cdots m}$ 
and its $(m{-}1)!$ permutations in $2,3,\ldots,m$.
Hence, the solution of the F-IBP relations (\ref{mderatmpt}) implements its reduction to undifferentiated chains
\begin{align}
\CC_{(12\cdots m)}(\xi)
= 
 \sum_{
b  =2}^{m} \sum_{\substack{\rho \in \{2,\cdots, b-1\}
\\ 
\shuffle \{m,\cdots, b+1 \}}} \!\!\!\!\!\!\!\!\!
 (-1)^{m-b} \bigg(
\frac{-{\bm h}_{b | \rho,b}}{1{+}s_{12\cdots m}}
{+} \hat g^{(1)}(\eta_{b+1,\cdots,m}{+}\xi){-}\hat g^{(1)}(\eta_{b,\cdots,m}{+}\xi)
\bigg){\bm \Omega}_{1,\rho,b} \,.
 \end{align}
Inserting the expression \eqref{bm23m} 
for ${\bm h}_{b | \rho,b}$ leads to the following closed formula for arbitrary
cycle length $m$ and Koba-Nielsen multiplicity $n\geq m$, 
 \begin{align}\label{rhorho}
 \CC_{(12\cdots m)}(\xi)
=\,\,& \frac{\MM_{12\cdots m}(\xi)}{ 1+s_{12\cdots m}}-
 \frac{1}{ 1+s_{12\cdots m}}
 \sum_{
b  =2}^{m}
 \!\!\!\!\!\!\!\!\!
 \sum_{\substack{\rho \in \{2,3,\cdots, b-1\}
\\
\qquad \shuffle \{m,m-1,\cdots, b+1 \}} }
\!\!\!\!\!\!\!\!\!\!\!\!
 (-1)^{m-b} \Bigg(
 \sum_{i=m+1}^{n}  x_{b,i}  +
\nabla_b 
\Bigg)
{\bm \Omega}_{1,\rho,b} 
 \,,
 \end{align}
see \eqref{2ptfinal}, \eqref{3ptfinal} and \eqref{4ptfinal} for examples at $m=2,3$ and $4$.
While the contributions from $ x_{b,i} $ yield $(m{+}1)$-point chains upon multiplication
by ${\bm \Omega}_{1,\rho,b} $, the numerator $\MM_{12\cdots m}(\xi)$ generalizing 
(\ref{mm12}), (\ref{mm123}) and (\ref{mm1234}) is expressed in terms of the conjectural 
$(m{-}1)!$ basis of $m$-point chains ${\bm \Omega}_{1,\alpha(2,3,\ldots,m)}$ 
with $\alpha \in S_{m-1}$,
  \begin{align}
&\!\!\!\!\MM_{12\cdots m}(\xi) \coloneqq
 \sum_{b  =2}^{m} 
\!\!\!\!\!\!
\sum_{\substack{\rho \in \{2,3,\cdots, b-1\}
\\
\quad \shuffle \{m,m-1,\cdots, b+1 \}
}} 
\!\!\!\!\!\!\!\!\!\!\!\!\!\!\!\!
 (-1)^{m-b} \Bigg(
\sum_{i=1}^m \! s_{i b} \,\partial_{\eta_b}{-}
    \sum_{i=2}^m  \! s_{i b}\, \partial_{\eta_i}
  {+}  (1{+}s_{12\cdots m}) v_1(\eta_{b}, \,\eta_{b+1,\cdots,m}{+}\xi )     
  \nl
  &\quad  \quad  \quad \quad  \quad  
   -  \hat g^{(1)}(\eta_b) 
  -\sum_{i=2}^{b-1}  { S}_{i,\rho} 
v_1(\eta_{b },\,\eta_{i,i+1,\cdots, b-1 })
-\sum_{i=b+1}^m { S}_{i,\rho} 
v_1(\eta_{b },\,\eta_{b+1,b+2,\cdots, i })   
 \Bigg){\bm \Omega}_{1,\rho,b}   
\notag \\
&\quad  + \!\!\!\!\!\!\!\!\! \!\!\!\sum_{1\leq p<u<v<w<q\leq m+1} 
 \!\!\!  \!\!\! \!\!\!\!\!\!\!\!\!
(-1)^{m+u+v+w}\, \Big(
v_1(\eta_{u+1,\cdots,w-1},-\eta_{u,\cdots,w-1})+v_1(\eta_{u,\cdots,w}-\eta_{u+1,\cdots,w}) 
 \Big) 
\label{rhorhoMM} \\
&\quad \quad  \quad  \quad \quad     \times
\Big(\sum_{i=q}^m s_{vi}+\sum_{i=1}^p s_{vi} \Big)  \! \! \! \!
 \sum_{\substack{\rho \in \{2,3,\cdots,p\}\shuffle \{m,m-1,\cdots, q\}\\
\gamma \in \{p+1,p+2,\cdots,u-1\}\shuffle \{v-1,v-2,\cdots, u+1\}\\
\pi \in \{v+1,v+2,\cdots,w-1\}\shuffle \{q-1,q-2,\cdots, w+1\}
 }} 
 \sum_{\sigma\in \{\gamma,u\}\shuffle \{ \pi, w\}} 
 \!\!\!\!\!\!\!
 \quad 
  {\bm \Omega}_{1,\rho,v,\sigma}
 \,,
 \non
 \end{align}
which has been checked up to and including $m=10$. In sections \ref{subsubsec:a} and \ref{subsubsec:b} below,
we shall give a more detailed discussion of the contributions ${\bm \Omega}_{1,\rho,b} $ in
the second line and $ {\bm \Omega}_{1,\rho,v,\sigma}$ in the fourth line. The
sums of $v_1$ functions in the third line can be rewritten as
\begin{align}
&v_1(\eta_{u+1,\cdots,w-1},-\eta_{u,\cdots,w-1})+v_1(\eta_{u,\cdots,w},-\eta_{u+1,\cdots,w})  \\
& = \hat g^{(1)} (\eta_{u+1,\cdots,w-1} ) 
 -\hat g^{(1)} (\eta_{u,\cdots,w-1} ) 
 -\hat g^{(1)} (\eta_{u+1,\cdots,w} )
 +\hat g^{(1)} (\eta_{u,\cdots,w} )  \notag \\
 & = g^{(1)} (\eta_{u+1,\cdots,w-1} ) 
 - g^{(1)} (\eta_{u,\cdots,w-1} ) 
 - g^{(1)} (\eta_{u+1,\cdots,w} )
 + g^{(1)} (\eta_{u,\cdots,w} ) \, ,
  \notag
 \end{align}
see (\ref{vfourg}) for a length-four example.
Starting from $m=6$, the expression \eqref{rhorhoMM} 
does not involve all the $(m{-}1)!$ independent chains (e.g.\ ${\bm \Omega}_{132546}$ and ${\bm \Omega} _{153246}$ do 
not occur in $\MM_{123456}$).

We emphasize that  \eqref{rhorho} is an exact formula instead of an equivalence relation under IBP, since we have tracked the Koba-Nielsen derivatives $\nabla_ b(\cdots)$. Hence,
our chain reduction of length-$m$ cycles can be readily applied in a closed-string context:
Similar to the two-point example in (\ref{2ptibp}) to (\ref{clex2}), any contribution
of $\nabla_b {\bm \Omega}_{1,\rho,b} $ to (\ref{rhorho}) in presence of
$\overline{{\bm \Omega}_{1,\sigma(2,\cdots,m)}}$ or $\overline{\CC_{(12\cdots m)}(\xi)}$ translates
into $-\frac{\pi \bar \eta_b}{\Im \tau} {\bm \Omega}_{1,\rho,b}$.

The expansion of our key result (\ref{rhorho}) in the
bookkeeping variables $\eta_{23\ldots m}{+}\xi , \, \eta_{3\ldots m}{+}\xi ,$ $\cdots,\eta_m{+}\xi, \, \xi$
yields the F-IBP reduction of cycles $f^{(k_1)}_{12} f^{(k_2)}_{23}\ldots 
f^{(k_{m-1})}_{m-1,m} f^{(k_m)}_{m1}$ to chains. In this way, the complexity of 
closed- and open-string integrands (\ref{stringint}) can be dramatically reduced,
see section \ref{sec4p4} below for examples.

 \subsubsection{Examples of contributions ${\bm \Omega}_{1,\rho,b} $ to (\ref{rhorhoMM})}
 \label{subsubsec:a}
 
The contributions to $\MM_{12\cdots m}(\xi)$ of the form ${\bm \Omega}_{1,\rho,b} $ in 
the first two lines of (\ref{rhorhoMM}) are illustrated up to four points 
by \eqref{mm12}, \eqref{mm123} and \eqref{mm1234}. We shall here add
examples of ${\bm \Omega}_{1,\rho,b} $ in the analogous reduction
of length-five chains,
 \begin{align}
\MM_{12345}(\xi)\big|_{{\bm \Omega} _{12345}}&= (s_{15}{+}s_{25}{+}s_{35}{+}s_{45}) \partial_{\eta _5} 
- s_{25} \partial_{\eta _2}-s_{35} \partial_{\eta _3}-s_{45} \partial_{\eta _4}
 -\hat g^{(1)}(\eta _5)\nl
&\quad
-s_{12} v_1(\eta _5,\eta _{234})
-(s_{13}{+}s_{23}) v_1(\eta _5,\eta _{34})-(s_{14}{+}s_{24}{+}s_{34}) v_1(\eta _5,\eta _4)
\nl
&\quad +(1{+}s_{12345}) v_1(\eta _5,\xi)
\,,
\nl
\MM_{12345}(\xi)\big|_{{\bm \Omega} _{15234}}&=  -(s_{14}{+}s_{24}{+}s_{34}{+}s_{45}) \partial_{\eta _4}
+s_{24} \partial_{\eta _2}+s_{34} \partial_{\eta _3}+s_{45} \partial_{\eta _5}+\hat g^{(1)}(\eta _4)
\nl
&\quad +s_{15} v_1(\eta _4,\eta _5)+(s_{12}{+}s_{25}) v_1(\eta _4,\eta _{23})+(s_{13}{+}s_{23}{+}s_{35}) v_1(\eta _4,\eta _3)
\nl
&\quad -(1{+}s_{12345}) v_1(\eta _4,\eta _{5}{+}\xi)\,,
\end{align}
where neither ${\bm \Omega} _{12345}$ nor ${\bm \Omega} _{15234}$ arise in 
the last two lines of (\ref{rhorhoMM}) at $m=5$.

 \subsubsection{Examples of contributions $ {\bm \Omega}_{1,\rho,v,\sigma}$ to (\ref{rhorhoMM})}  
 \label{subsubsec:b}
 
The contributions to $\MM_{12\cdots m}(\xi)$ of the form $ {\bm \Omega}_{1,\rho,v,\sigma}$ in 
the last two lines of (\ref{rhorhoMM}) firstly arise at $m=4$, see the terms $\sim s_{13}$ in the
last line of (\ref{mm1234}). While the coefficients of these chains $ {\bm \Omega}_{1,\rho,v,\sigma}$ are simple combinations of elliptic $v_1$ functions and Mandelstam variables, 
the coupled summation ranges call for further illustration:
The condition $1 \leq p<u<v<w<q \leq m{+}1$ in the third line of (\ref{rhorhoMM})
implies $p{+}4 \leq q$ and $p{+}2 \leq v \leq q{-}2$. When $q=m{+}1$, which can be identified as 1  modulo $m$, the set $\{m, m{-}1, \cdots, q\}$ is understood as the empty set $\emptyset$ and $\sum_{i=q}^{m} s_{v i}=0$.  

The way how a given choice of $(p,u,v,w,q)$ translates into
 the ordered sets $\rho,\sigma$ of ${\bm \Omega}_{1,\rho,v,\sigma}$
 in the fourth line of (\ref{rhorhoMM})
is illustrated in figure \ref{fig:illus} below.
 The outermost summation variables $p,q$ delimit ordered sets
$\{2,3,\cdots, p\}$ and $\{m,m{-}1,\cdots, q\}$ whose shuffle gives rise to $\rho$.
The remaining summation variables $u,v,w$ then separate the ordered set $\{p{+}1,p{+}2,\ldots, q{-}1\}$ into four parts which determine $\sigma$ through an iteration of shuffles and reversals
detailed in figure \ref{fig:illus}.

 \begin{figure}[h]
 \centering
\begin {tikzpicture}[scale=1.6]
\draw [thick] (0,0)--(4,0)  ;
    \filldraw (0,0) circle (1pt) node[below=0pt]{$p$} 
       (1,0) circle (1pt) node[below=0pt]{$u$} 
         (2,0) circle (1pt) node[below=0pt]{$v$} 
           (3,0) circle (1pt) node[below=0pt]{$w$}
              (4,0) circle (1pt) node[below=0pt]{$q$}  ;
\draw[->](0.33,0) -- (0.34,0) ;
\draw[->](0.66,0) -- (0.67,0) ;
\draw[<-](1.33,0) -- (1.34,0) ;
\draw[<-](1.66,0) -- (1.67,0) ;
\draw[->](2.33,0) -- (2.34,0) ;
\draw[->](2.66,0) -- (2.67,0) ;
\draw[<-](3.33,0) -- (3.34,0) ;
\draw[<-](3.66,0) -- (3.67,0) ;
\node at (0.5,0.2) {$\overbrace{~~~~~~}$} ;     
\node at (1.5,0.2) {$\overbrace{~~~~~~}$} ;     
\node at (2.5,0.2) {$\overbrace{~~~~~~}$} ;     
\node at (3.5,0.2) {$\overbrace{~~~~~~}$} ;   
\draw[->] (0.5,0.45) -- (0.8,0.9);   
\draw[->] (1.5,0.45) -- (1.2,0.9); 
\draw[->] (2.5,0.45) -- (2.8,0.9);   
\draw[->] (3.5,0.45) -- (3.2,0.9);  
\draw(1,0.65)node{$\shuffle$};  
\draw(3,0.65)node{$\shuffle$}; 
\draw(1,1.05)node{$\gamma$};  
\draw(3,1.05)node{$\pi$};     
\draw(1.5,1.05)node{$u$};  
\draw(3.5,1.05)node{$w$};    
\node at (1.25,1.3) {$\overbrace{~~~~~~}$} ;   
\node at (3.25,1.3) {$\overbrace{~~~~~~}$} ;   
\draw[->](1.25,1.55) -- (2.05,2.1) ;
\draw[->](3.25,1.55) -- (2.45,2.1) ;
\draw(2.25,1.85)node{$\shuffle$};
\draw(2.25,2.2)node{$\sigma$};
      \draw (0,0) arc (90:270:.5);
       \draw (4,0) arc (90:-90:.5);
       \draw (0,-1)--(4,-1);
       \draw[->](1.34,-1)--(1.33,-1);
        \draw[->](0.67,-1)--(0.66,-1);
       \draw[<-](3.34,-1)--(3.33,-1);
        \draw[<-](2.67,-1)--(2.66,-1);
         \filldraw (2,-1) circle (1pt) node[above=0pt]{$1$};
 \node at (1,-1.2) {$\underbrace{~~~~~~~~~~~~~}$} ;   
 \node at (3,-1.2) {$\underbrace{~~~~~~~~~~~~~}$} ; 
\draw[->](1,-1.45) -- (1.8,-2.2) ;
\draw[->](3,-1.45) -- (2.2,-2.2) ;
\node at (2,-1.95) {$\shuffle$} ; 
\node at (2,-2.3) {$\rho$} ;
\end{tikzpicture}
  \caption{Illustration of the way the ordered sets $\rho$ and $\sigma$ 
specifying the chains ${\bm \Omega}_{1,\rho,v,\sigma}$ in the last two 
lines of (\ref{rhorhoMM}) are formed for a given choice
of the summation variables $p,u,v,w,q$. Arrows indicate where ordered
  sets are reversed before the shuffle.}
  \label{fig:illus}
\end{figure}
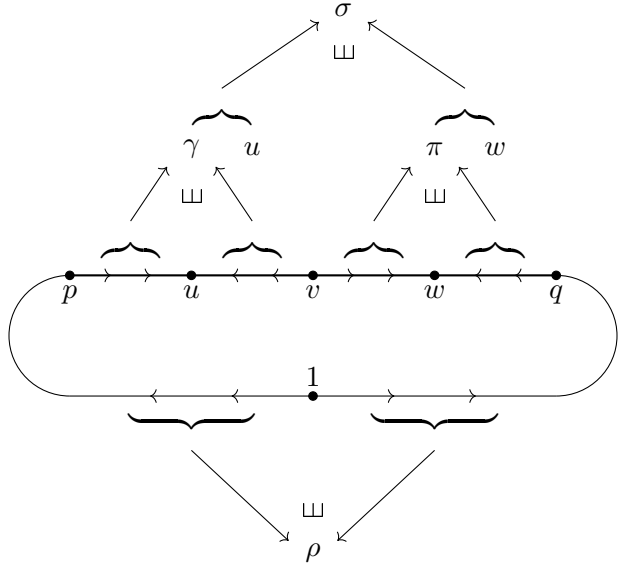    

 In the length-four example \eqref{mm1234}, the only admissible choice
 for the summation variables is $p=q=1, \, u=2, \, v=3, \, w=4$. All of 
 $\rho, \gamma,\pi$ in the fourth line of (\ref{rhorhoMM}) then reduce
 to empty sets and $\sigma= 2\shuffle 4=\{24, 42\}$, leading
 to the contributions ${\bm \Omega}_{1324}{+}{\bm \Omega}_{1342}$.
At length $m=5$, the last two lines of  (\ref{rhorhoMM}) yield
a total of 16 contributions ${\bm \Omega}_{1ijkl}$ to $\MM_{12345}$
after expanding out all shuffles,
\ba
&
\label{single5secbasis}
\big( 
v_1(\eta_3,-\eta_{23}) + v_1(\eta_{234},-\eta_{34})
\big) s_{13} \left({\bm \Omega} _{13254}+{\bm \Omega} _{13524}+{\bm \Omega} _{13542}\right)
\\
\,\,&
-\big(
v_1(\eta_{34},-\eta_{234}) + v_1(\eta_{2345},-\eta_{345})
\big) 
s_{13} \left({\bm \Omega} _{13245}+{\bm \Omega} _{13425}+{\bm \Omega} _{13452}\right)
\nl\,\,&
-\big(
v_1(\eta_4,-\eta_{34}) + v_1(\eta_{345},-\eta_{45})
\big) (s_{14}{+}s_{24}) ({\bm \Omega} _{12435}
+{\bm \Omega} _{12453})
 -
(23\leftrightarrow 54 )\,,
\nonumber
\ea
where the simultaneous relabelling $2\leftrightarrow 5$ and $3\leftrightarrow 4$
applies to all the three lines and concerns the labels of all of $\eta_i,s_{ij}$ and 
${\bm \Omega} _{1ijkl}$, for instance, $v_1(\eta_3,-\eta_{23}) s_{13} {\bm \Omega} _{13254}  \big|_{23\leftrightarrow 54}= v_1(\eta_4,-\eta_{45})  s_{14} {\bm \Omega} _{14523} $.
 
 The first example where both $\gamma$ and $\pi$ in the last line of (\ref{rhorhoMM})
 are non-empty sets appears at length $m=6$. For example, the choice 
 $p=q=1, \, u=2, \, v=4, \, w=6 $ corresponds to the following contribution to $\MM_{123456}$
 \ba
&\big(
v_1(\eta_{345},-\eta_{2345}) + v_1(\eta_{23456},-\eta_{3456})
\big) s_{14}
\\
& \times ({\bm \Omega} _{143256}
+{\bm \Omega} _{143526}
+{\bm \Omega} _{143562}
+{\bm \Omega} _{145326}
+{\bm \Omega} _{145362}
+{\bm \Omega} _{145632}
)
\,.
\non
 \ea 
After expanding out all shuffles, it becomes evident that 
the number of terms ${\bm \Omega} _{1,\rho,b}$ 
in the first two lines of (\ref{rhorhoMM}) is equal to $\sum_{b=2}^m{m-2\choose b-2}=2^{m-2}$.
Importantly, it should be noted that  each  ${\bm \Omega} _{1,\rho,v,\sigma}$ appearing in the last two lines  of  (\ref{rhorhoMM}) is distinct for different choices of 
$\{p,u,v,w,q\}$ and $\{\rho,\gamma, \pi, \sigma\}$. Consequently, the total number of  ${\bm \Omega} _{1,\rho,v,\sigma}$
 in the last two lines can be determined as follows,
\begin{align}
&\sum_{1\leq p<u<v<w<q\leq m+1}
{m{+}p{-}q\choose m{-}q{+}1}{v{-}p{-}2\choose  v{-}u{-}1}{q{-}v{-}2\choose  q{-}w{-}1}{q{-}p{-}2\choose q{-}v{-}1}  =2^{2 m - 5} - (m {-} 1) 2^{m - 3}\,,
\end{align}
where the counting at $m=2,3,4,5,6,7,8$ is $0, 0, 2, 16, 88, 416, 1824$, respectively, which coincides with the sequence on the right-hand side inferred from \cite{oeis431}. 

\subsection{Applications}
\label{sec4p4}

With the general formula \eqref{rhorho}, we now have the means to break a Kronecker-Eisenstein cycle \eqref{defc} of arbitrary length $m$. However, when dealing with string integrands in (\ref{stringint}), the focus shifts to handling cycles $f^{(k_1)}_{12} f^{(k_2)}_{23}\ldots 
f^{(k_{m-1})}_{m-1,m} f^{(k_m)}_{m1}$ of Kronecker-Eisenstein {\it coefficients} at given $k_i\neq0$ instead of their {\it generating series} $\Omega_{ij}(\eta)$. A prime example of this is the cycle of $f^{(1)}_{ij}$ found in the elliptic functions $V_m(1,2,\ldots,m)$ to be reviewed below which enter correlators of various string theories. 

In this section, we utilize \eqref{rhorho} as a generating function of component formulae for breaking the cycles of Kronecker-Eisenstein coefficients. We also introduce several helpful techniques to simplify the computation process.

\subsubsection{Elliptic functions}

An elegant construction of elliptic functions of $m$ punctures $z_1,z_2,\ldots,z_m$
is based on cyclic products of meromorphic or equivalently doubly-periodic\footnote{The non-holomorphic exponential in \eqref{1.1} drops out from the cyclic products of $\Omega$ in (\ref{1.1a}).} Kronecker-Eisenstein series at the same second argument $\eta$ \cite{Dolan:2007eh, Broedel:2014vla, Tsuchiya:2017joo},
\begin{align}
&F(z_{12},\eta,\tau) F(z_{23},\eta,\tau) \ldots F(z_{m,1},\eta,\tau)  \notag \\
&= \Omega(z_{12},\eta,\tau) \Omega(z_{23},\eta,\tau) \ldots \Omega(z_{m,1},\eta,\tau)
\notag \\
&=: \eta^{-m} \sum_{w=0}^\infty \eta^w V_w(1,2,\ldots,m|\tau)  \ . \label{1.1a}
\end{align}
The elliptic functions $V_w(1,2,\ldots,m)$ in the $\eta$-expansion are indexed
by their holomorphic modular weight $w\geq 0$. At fixed multiplicity $m$, their instances
with weight ${0\leq w \leq m}$ are sufficient to recover cases with $w>m$ via linear 
combinations with holomorphic Eisenstein series ${\rm G}_k$ as coefficients,
e.g.\ $V_4(1,2)=3 {\rm G}_4$ or $V_5(1,2,3) = 3 {\rm G}_4 V_1(1,2,3)$. 
The $\eta$-expansion of the left-hand side of (\ref{1.1a}) yields
the following representations in terms of $f^{(w)}_{ij}$ (manifesting doubly-periodicity) or equivalently $g^{(w)}_{ij}$ (manifesting meromorphicity)
\begin{align}
V_w(1,2,\ldots,m) &= \! \! \! \! \!  \sum_{k_1{+}k_2{+}\ldots{+}k_m=w}\! \! \! \! \!  
f^{(k_1)}_{12} f^{(k_2)}_{23}\ldots 
f^{(k_{m-1})}_{m-1,m} f^{(k_m)}_{m1} 
\label{altVw}  \\
&= \! \! \! \! \!    \sum_{k_1{+}k_2{+}\ldots{+}k_m=w} \! \! \! \! \!  
g^{(k_1)}_{12} g^{(k_2)}_{23}\ldots 
g^{(k_{m-1})}_{m-1,m} g^{(k_m)}_{m1}\,,
\notag
\end{align}
with cyclic identification $z_{m+1}=z_1$, for instance
\begin{align}
V_0(1,2,\ldots,m) &=1 \,, \ \ \ \ \ \  V_1(1,2,\ldots,m) = \sum_{j=1}^m g^{(1)}_{j,j+1}= \sum_{j=1}^m f^{(1)}_{j,j+1}
\,, \notag
\\
V_2(1,2,\ldots,m) &= \sum_{j=1}^m f^{(2)}_{j,j+1} 
+\sum_{i=1}^m \sum_{j=i{+}1}^m f^{(1)}_{i,i+1}f^{(1)}_{j,j+1} \ .
\label{1.1c}
\end{align}
The notation for $v_1$ in the definition \eqref{defv1} is motivated by the relation 
\beq
v_1(z_{12},z_{23}) = V_1(1,2,3)
\eeq
with the elliptic function $V_1(1,2,\ldots,m)$ at $m=3$.

\subsubsection{Breaking $V_m(1,2,\ldots,m)$ }

Based on Fay identities, any $V_w(1,2,\ldots,m)$ with $w<m$ can be 
expressed in terms of expansion coefficients in permutations of
the chain ${\bm\Omega}_{12\cdots m }$ in (\ref{deflongomega2}). 
However, this is not the case for $V_w(1,2,\ldots,m)$ at $w=m$ since it includes the 
$f$-cycle $f^{(1)}_{12}f^{(1)}_{23} \ldots f^{(1)}_{m-1,m}  f^{(1)}_{m1} $. To handle 
this specific case, we will isolate suitable components of the generating-function
identity \eqref{rhorho} due to IBP to break the $f$-cycle.
As discussed at the four-point level in \cite{Gerken:2018jrq}, the breaking of cycles of  $f^{(1)}_{ij}$ has immediate applications in the gauge sector of heterotic-string
amplitudes.

According to \eqref{1.1a}, the function $V_m(1,2,\ldots,m)$ can be produced from the Kronecker-Eisenstein cycle via
\begin{align}
\label{equation451}
V_m(1,2,\cdots ,m)&=   \Big(\Omega_{m,1}(\xi)  \prod_{i=1}^{m-1}\Omega_{i,i+1}(\xi) \Big)  \big|_{\xi^0}  = 
\Big(\lim_{\eta_m\to0} \cdots\lim_{\eta_3\to0}\lim_{\eta_2\to0}\CC_{(12\cdots m)}(\xi)\Big)\big|_{\xi^0}\,.
\end{align}
On the right-hand side of   \eqref{equation451}, one can also trade the operations of taking limits for taking coefficients,  
\begin{align}
\lim_{\eta_m\to0} \big(\cdots \big(\lim_{\eta_3\to0}\big(\lim_{\eta_2\to0}\CC_{(12\cdots m)}(\xi)\big)\big)\cdots \big)&=\big(\cdots \big(\big( \CC_{(12\cdots m)}(\xi)\big|_{\eta_2^0}  \big)\big|_{\eta_3^0}  \big)\cdots \big|_{\eta_m^0}  \big) \notag \\
&=:\CC_{(12\cdots m)}(\xi)\big|\big|_{\eta_2^0,\eta_3^0,\cdots,\eta_m^0}\,,
\label{coeff}
\end{align}
where in the last equality, we have introduced the shorthand notation $||_{\eta_i^0,\eta_j^0,\ldots}$ for the process of taking limits in a specific order. Applying this operation to the single-cycle formula \eqref{rhorho},  we break  $V_m(1,2,\cdots,m)$ into pieces free of cycles,
\begin{align}
\label{vopen}
V_m(1,2,\cdots, m) 
=\,\,& \frac{\MM_{12\cdots m}(\xi)\big|\big|_{\eta_2^0,\eta_3^0,\cdots,\eta_{m}^0,\xi^0}}{ 1+s_{12\cdots m}}-
 \frac{1}{ 1+s_{12\cdots m}}\sum_{
b  =2}^{m}   (-1)^{m-b} \\
 &\quad 
 \times \! \! \! \!  \sum_{\rho \in \{2,3,\cdots, b-1\}\shuffle \{m,m-1,\cdots, b+1 \}}  \! \! \! \bigg(
\sum_{i=m+1}^{n}  s_{bi} f_{bi}^{(1)} 
 +\nabla_b 
\bigg){\bm \Omega}_{1,\rho,b}\big|\big|_{\eta_2^0,\eta_3^0,\cdots,\eta_{m}^0}   
 \,. \notag
 \end{align}
At length $m=2$ in presence of an $n$-point Koba-Nielsen factor, this specializes to
\begin{align}
V_2(1,2) 
=\,\,&
 \frac{1}{ 1+s_{12}}   \bigg(\MM_{12}(\xi)\big|\big|_{\eta_2^0,\xi^0}-
f_{12}^{(1)}   \sum_{i=3}^{n}  s_{2i} f_{2i}^{(1)} 
-\nabla_2   f_{12}^{(1)}  
\bigg)
 \,,
 \end{align}
where the expansion of $\hat g^{(1)}$ in \eqref{gexpand} yields
\ba
\MM_{12}(\xi)\big|\big|_{\eta_2^0,\xi^0}=\,\,& s_{12} \partial_{\eta _2}\Omega _{12} (\eta_2)\big|_{\eta_2^0}-\hat g^{(1)}(\eta _2)\Omega _{12} (\eta_2)\big|_{\eta_2^0} +\left(1{+}s_{12}\right) v_1(\eta _2,\xi) \Omega _{12} (\eta_2)\big|\big|_{\eta_2^0,\xi^0}
\nl
=\,\,& 2s_{12} f_{12}^{(2)}+\hat {\rm G}_2
\ea
and results in
\beq
V_2(1,2)  =  \frac{1}{ 1+s_{12}}  \bigg(
 2s_{12} f_{12}^{(2)}+\hat {\rm G}_2
 -f_{12}^{(1)}   \sum_{i=3}^{n}  s_{2i} f_{2i}^{(1)} 
-\nabla_2   f_{12}^{(1)}  
\bigg)
\, .
\label{doubperv12}
\eeq
This example illustrates the convenience of taking coefficients of $\eta_2^0$ and $\xi^0$ in  \eqref{coeff} as compared to taking limits: The operation $\big|\big|_{\eta_2^0,\xi^0}$ can be individually employed to separate parts of \eqref{rhorho}. However, we emphasize that the operation $\big|\big|_{\eta_2^0,\eta_3^0,\ldots,\eta_m^0}$ defined in \eqref{coeff} carries the information on the ordering used to extract the coefficients. This ordering is inherited from the ordering used to take the limits. Once an ordering of ${\eta_2^0,\eta_3^0,\ldots,\eta_m^0}$ is chosen,  it must be consistently applied to every term on the right-hand side of 
 \eqref{vopen}. The non-trivial dependence on the ordering is exemplified by\footnote{The difference in (\ref{exdiff}) can be understood from the fact that $\frac{\eta_3}{\eta_2+\eta_3}\big|\big|_{\eta_2^0,\eta_3^0}=\sum_{r=0}^\infty \left(-\frac{\eta_2}{\eta_3}\right)^r\big|\big|_{\eta_2^0,\eta_3^0}=1  $ whereas $\frac{\eta_3}{\eta_2+\eta_3}\big|\big|_{\eta_3^0,\eta_2^0}=-\sum_{r=1}^\infty \left(-\frac{\eta_3}{\eta_2}\right)^r\big|\big|_{\eta_3^0,\eta_2^0}=0 $. }
\ba
{\bm \Omega}_{123} \big|\big|_{\eta_2^0,\eta_3^0} =f^{(1)}_{12}f^{(1)}_{23}+f^{(2)}_{12}+f^{(2)}_{23}\,,\qquad \! \! \!
{\bm \Omega}_{123} \big|\big|_{\eta_3^0,\eta_2^0} =f^{(1)}_{12}f^{(1)}_{23}+f^{(2)}_{12}\neq {\bm \Omega}_{123} \big|\big|_{\eta_2^0,\eta_3^0} \,.
\label{exdiff}
\ea
For any alternative ordering, such as $\big|\big|_{\eta_m^0,\cdots,\eta_3^0,\eta_2^0} $, 
 it is evident from the definition (\ref{defc}) that $\CC_{(12\cdots m)}\big|\big|_{\eta_2^0,\eta_3^0,\cdots,\eta_m^0} =\CC_{(12\cdots m)}\big|\big|_{\eta_m^0,\cdots,\eta_3^0,\eta_2^0}$ while
\ba
\MM_{12\cdots m}(\xi)\big|\big|_{\eta_2^0,\eta_3^0,\cdots,\eta_{m}^0,\xi^0}=\MM_{12\cdots m}(\xi)\big|\big|_{\eta_{m}^0,\cdots,\eta_3^0,\eta_2^0,\xi^0}
\ea
can only be established by applying Fay identities.

\subsubsection{Tools for extracting coefficients}
\label{sec:toolcoeff}

Notice that the length-$m$ chains ${\bm \Omega}_{1\rho(2\cdots m)}$ in the
expression (\ref{rhorhoMM}) for $\MM_{12\cdots m}$ do not depend on 
the auxiliary variable $\xi$. The $\xi$-dependence entirely resides in the elliptic functions
$v_1(\eta_{p}, \eta_{p+1 \cdots m}{+}\xi ) $, and we can easily extract the
$\xi^0$ coefficients by combinations~of\footnote{This can be understood by separating 
$\hat g^{(1)}(\eta_{I}{+}\xi)$ into a non-singular part $\hat g^{(1)}(\eta_{I}{+}\xi)- \frac{1}{\eta_{I}{+}\xi}$ in $\eta_{I}{+}\xi$ and a singular one $\frac{1}{\eta_{I}{+}\xi}$. The non-singular part $\hat g^{(1)}(\eta_{I}{+}\xi)- \frac{1}{\eta_{I}{+}\xi}$ admits a Taylor-expansion in $\xi$ whose zeroth-order coefficient is readily identified as $\hat g^{(1)}(\eta_{I})- \frac{1}{\eta_{I}}$. The non-singular part is considered at $| \eta_I|<|\xi|$ to obtain the geometric-series expansion $\frac{1}{\eta_{I}{+}\xi}= \frac{1}{\xi} \sum_{r=0}^\infty  \big({-}\frac{\eta_I}{\xi}
\big)^r$ and read off a vanishing coefficient of $\xi^0$.}
\begin{align}
\hat g^{(1)}(\xi){\bm \Omega}_{1\cdots} \big|\big|_{\eta_2^0,\cdots,\eta_{m}^0,\xi^0}&=0\,,
\\
 \hat g^{(1)}(\eta_{I}{+}\xi){\bm \Omega}_{1\cdots} \big|\big|_{\eta_2^0,\cdots,\eta_{m}^0,\xi^0} &= \bigg( \hat g^{(1)}(\eta_{I})- \frac{1}{\eta_I}\bigg){\bm \Omega}_{1\cdots} \big|\big|_{\eta_2^0,\cdots,\eta_{m}^0} \,, ~{\rm for} ~I\neq \emptyset\,, 
 \notag
\end{align}
such that
\ba
 v_1(\eta_{m}, \eta_{m}{+}\xi ) {\bm \Omega}_{1\cdots} \big|\big|_{\eta_2^0,\cdots,\eta_{m}^0,\xi^0}
  =\,\,&  \frac{1}{\eta_m} {\bm \Omega}_{1\cdots} \big|\big|_{\eta_2^0,\cdots,\eta_{m}^0}\,,
\label{convlem}\\
  v_1(\eta_{p}, \eta_{p+1\cdots m}{+}\xi ) {\bm \Omega}_{1\cdots} \big|\big|_{\eta_2^0,\cdots,\eta_{m}^0,\xi^0} =\,\,&    \bigg( v_1(\eta_{p}, \eta_{p+1 \cdots m} ) + \frac{1}{\eta_{p \cdots m}}-\frac{1}{\eta_{p+1\cdots m}}\bigg){\bm \Omega}_{1\cdots} \big|\big|_{\eta_2^0,\cdots,\eta_{m}^0} 
  \nonumber
\ea
for  $ p<m$. Note that the right-hand side further simplifies in view of
$v_1(\eta_{p}, \eta_{p+1,\cdots,m} ) + \frac{1}{\eta_{p,\cdots, m}}-\frac{1}{\eta_{p+1,\cdots, m}}= \frac1 {\eta_p}+ {\cal O}(\eta_i) $. Thus, (\ref{convlem}) furnishes a convenient lemma to perform the $\big|\big|_{\xi^0}$ operation in (\ref{vopen}).

Another useful identity is\footnote{This is a simple consequence of the fact that the zeroth order in $\eta_i$ of some Laurent series $\partial_{\eta_i}{\bm \Omega}_{1\cdots}(\eta_i)$ is only sensitive to the linear order of ${\bm \Omega}_{1\cdots}(\eta_i)$ in $\eta_i$ where $\partial_{\eta_i}$ acts by multiplication with $(\eta_i)^{-1}$.} 
\be
 \partial_{\eta_i} {\bm \Omega}_{1\cdots}  \big|\big|_{\eta_2^0,\cdots,\eta_{m}^0} = \frac{1}{\eta_i}    {\bm \Omega}_{1\cdots}  \big|\big|_{\eta_2^0,\cdots,\eta_{m}^0},\qquad  ~~{\rm for}~ 2\leq i \leq m\,,
 \ee
which means that all derivative operators $ \partial_{\eta_i}$ 
entering \eqref{vopen} through the expression (\ref{rhorhoMM})
for $\MM_{12\cdots m}$ can be traded for simple multiplications.

\subsubsection{Length-three  example}
\label{sec:vmat3}

At length $m=3$, the general formula \eqref{vopen} for the
chain decomposition of $V_m(1,2,\ldots,m)$ specializes to
\begin{align}
V_3(1,2,3)
=\,\,&\frac{\MM_{123}(\xi)\big|\big|_{\eta_2^0,\eta_3^0,\xi^0}}{1+s_{123}}-\frac{1}{1+s_{123}}
\Bigg[ 
\bigg( \sum_{i=4}^{n}  s_{3i} f_{3i}^{(1)}   
+\nabla_3  \bigg) 
( f^{(1)}_{12}f^{(1)}_{23}+f^{(2)}_{12}+f^{(2)}_{23})  
\nl& \quad \quad \quad \quad \quad \quad \quad \quad \quad \quad
+\bigg( \sum_{i=4}^{n}  s_{2i} f_{2i}^{(1)}   
+\nabla_2 \bigg)
( f^{(1)}_{13}f^{(1)}_{23}-f^{(2)}_{13})
 \Bigg]
 \,, \label{almostsymm}
\end{align}
see (\ref{mm123}) for $\MM_{123}(\xi)$, and the tools of section \ref{sec:toolcoeff} for taking coefficients yield
\ba
\MM_{123}(\xi)\big|\big|_{\eta_2^0,\eta_3^0,\xi^0}
=\,\,&\hat {\rm G}_2 V_1(1,2,3)+ 
{\bm \Omega} _{123} \left(\frac{s_{12}}{\eta _{23}}+\frac{2 \left(s_{13}{+}s_{23}\right)}{\eta _3}-\frac{s_{12}{+}s_{23}}{\eta _2}\right)\bigg|\bigg|_{\eta_2^0,\eta_3^0}
\nl&\qquad
+{\bm \Omega} _{132} \left(-\frac{s_{13}}{\eta _{23}}+\frac{s_{13}{+}s_{23}}{\eta _3}-\frac{2 \left(s_{12}{+}s_{23}\right)}{\eta _2}\right)\bigg|\bigg|_{\eta_2^0,\eta_3^0}
\,.
\ea
Extracting the coefficients of $\eta_2^0 ,\eta_3^0$ on the right-hand side is straightforward, for example,
\be\label{elex}
\frac{{\bm \Omega} _{123} }{\eta _{23}}\Big|\Big|_{\eta_2^0,\eta_3^0}
=\frac{{\bm \Omega} _{123} }{\eta _{3}}\Big|\Big|_{\eta_2^0,\eta_3^0}=
f^{\text{(1)}}_{23} f^{\text{(2)}}_{12}+f^{\text{(1)}}_{12} f^{\text{(2)}}_{23}+f^{\text{(3)}}_{12}+f^{\text{(3)}}_{23}\,,
\ee
such that
\ba\label{nn3}
\MM_{123}(\xi)\big|\big|_{\eta_2^0,\eta_3^0,\xi^0} =\,\,&
\hat{\text{G}}_2
 V_1(1,2,3)+\left(2
   s_{12}{+}s_{23}\right) f^{\text{(1)}}_{23} f^{\text{(2)}}_{13} +\left(2 s_{13}{+}s_{23}\right)
   f^{\text{(1)}}_{23} f^{\text{(2)}}_{12}
   \nl
 &+ f^{\text{(2)}}_{23} \left(\left(s_{12}{+}2s_{13}{+}2s_{23} \right) f^{\text{(1)}}_{12}-2 \left(s_{12}{+}s_{23}\right)
   f^{\text{(1)}}_{13}\right)
   \nl
   &+\left(2 s_{13}{-}s_{12}\right) f^{\text{(3)}}_{12}+\left(s_{13}{-}2
   s_{12}\right) f^{\text{(3)}}_{13}+\left(2 s_{12}{+}2 s_{13}{+}3 s_{23}\right)
   f^{\text{(3)}}_{23}\,.
\ea
Except for the first term $ s_{12} f^{(1)}_{12} f^{(2)}_{23}$ of the second line, the right-hand side is manifestly antisymmetric under $2 \leftrightarrow 3$. The exceptional term $ s_{12} f^{(1)}_{12} f^{(2)}_{23}$ in (\ref{nn3}) compensates for the lack of $-( \sum_{i=4}^{n}  s_{2i} f_{2i}^{(1)}   
+\nabla_2 )f^{(2)}_{23}$ on the right-hand side of (\ref{almostsymm}) such that the antisymmetry $V_3(1,2,3)=-V_3(1,3,2)$ is preserved.

Additional examples for the chain decomposition of $V_m(1,2,\ldots,m)$
up to and including $m=5$ are provided in appendix \ref{singlecycle456}. 
We have checked via Fay identities that the results obtained at $m=4$ 
points agree with those reported in the literature \cite{Mafra:2018pll,Gerken:2018jrq}.
Higher-point cases of our general results (\ref{vopen}) at $m\geq 5$
cannot be found in earlier work.

\subsection{Reformulations of the single-cycle formula}
\label{sec:alttad}

Our presentation of the F-IBP decomposition \eqref{rhorho} of Kronecker-Eisenstein cycles $\CC_{(12\cdots m)}$ singles out the first puncture $z_1$ in two respects: First, the permutations ${\bm \Omega}_{\sigma(12\ldots m)},\, \sigma \in S_m$ of the chains on the right-hand side are given in the $(m{-}1)!$-basis of ${\bm \Omega}_{1\rho(2\ldots m)}$ w.r.t.\ the Fay identities (\ref{fayprac}) where the permutations $\rho \in S_{m-1}$ do not act on $z_1$. Second, the integrations by parts reflected by the $\nabla_2,\nabla_3,\ldots,\nabla_m$ in (\ref{rhorho}) exclude derivatives $\nabla_1$ w.r.t.\ $z_1$. While a change of $(m{-}1)!$ basis of chains can be straightforwardly performed via (\ref{fayprac}), the goal of this section is to spell out reformulations of \eqref{rhorho} where a general $\nabla_{a\neq 1}$ rather than $\nabla_1$ is skipped in the IBP relations. This will result in alternative chain decompositions of $\CC_{(12\cdots m)}$ where $z_{a\neq 1}$ rather than $z_1$ enters on special footing. Such reformulations will be essential for the chain decomposition of more general arrangements of Kronecker-Eisenstein series beyond a single cycle in later sections.

Our starting point is the consequence $\sum_{b=1}^m  \partial_b{\bm \Omega}_{12\cdots m}  =0$ of translation invariance. Together with the definition \eqref{defnabla} of $\nabla_b$, we are led to the relation
\ba 
\sum_{b=1}^m\left( \nabla_b+ \sum_{i=m+1}^n x_{b,i}  \right){\bm \Omega}_{\sigma(12\ldots m)}=0
\ea 
among $\nabla_1,\nabla_2,\ldots,\nabla_m$ acting on chains with
any permutation $\sigma(12\ldots m)$ of $12\cdots m$. This can be used to eliminate
any $\nabla_a$ with $2\leq a \leq m$ in favor of $\nabla_1$ and the
remaining $\nabla_{b \neq 1,a}$ in the decomposition formula \eqref{rhorho}:
\begin{align}\label{rhorhotog}
&\sum_{b=2}^m
\Bigg(
 \sum_{i=m+1}^{n}  x_{b,i}  +
\nabla_b 
\Bigg)
 \!\!\!\!\!\!\!\!\!
 \sum_{\substack{\rho \in \{2,3,\cdots, b-1\}
\\
\qquad \shuffle \{m,m-1,\cdots, b+1 \}} }
\!\!\!\!\!\!\!\!\!\!\!\!
 (-1)^{m-b} 
{\bm \Omega}_{1,\rho,b} 
 \\
 & = \sum_{\substack{
b  =2
\\
b\neq a}}^{m} 
\Bigg(
 \sum_{i=m+1}^{n}  x_{b,i}  +
\nabla_b 
\Bigg)
\Bigg(
 \!\!\!\!\!\!\!\!\!
 \sum_{\substack{\rho \in \{2,3,\cdots, b-1\}
\\
\qquad \shuffle \{m,m-1,\cdots, b+1 \}} }
\!\!\!\!\!\!\!\!\!\!\!\!
 (-1)^{m-b} 
{\bm \Omega}_{1,\rho,b} 
- \!\!\!\!\!\!\!\!\!
 \sum_{\substack{\rho \in \{2,3,\cdots, a-1\}
\\
\qquad \shuffle \{m,m-1,\cdots, a+1 \}} }
\!\!\!\!\!\!\!\!\!\!\!\!
 (-1)^{m-a} 
{\bm \Omega}_{1,\rho,a} 
\Bigg)
\nl
\,\,& \quad\quad
-\Bigg(
 \sum_{i=m+1}^{n}  x_{1,i}  +
\nabla_1 
\Bigg)
\Bigg(
\!\!\!\!\!\!\!\!\!
 \sum_{\substack{\rho \in \{2,3,\cdots, a-1\}
\\
\qquad \shuffle \{m,m-1,\cdots, a+1 \}} }
\!\!\!\!\!\!\!\!\!\!\!\!
 (-1)^{m-a} 
{\bm \Omega}_{1,\rho,a} 
\Bigg) \, , \ \ \ \ \ \ 
a\in \{ 2,3,\ldots,m\} \,.
\non
  \end{align}
The difference of the permutation sums over ${\bm \Omega}_{1,\rho,b} $ and
${\bm \Omega}_{1,\rho,a}$ in the middle line can be simplified through the
following corollary of Fay identities:
\ba
\label{fay2identity}
\sum_{\substack{\rho \in \{2,3,\cdots, b-1\}\\
\qquad\quad \shuffle \{m,m-1,\cdots, b+1 \}} }
\!\!\!\!\!\!\!\!\!\!\!\!\!\!\!\!\!\!\!\!
 (-1)^{m-b} 
{\bm \Omega}_{1,\rho,b}- 
\!\!\!\!\!\!\!\!\!\!\!\!\!\!\!\!\!\!\!\!
\sum_{\substack{\rho \in \{2,3,\cdots, a-1\}\\
\qquad\quad \shuffle \{m,m-1,\cdots, a+1 \}} }
\!\!\!\!\!\!\!\!\!\!\!\!\!\!\!\!\!\!\!\!
 (-1)^{m-a} 
{\bm \Omega}_{1,\rho, a} =\sum_{\rho \in  A \shuffle  B^{\rm T} \atop{(a, A, b, B)=\mathbb I_m}} 
 (-1)^{|B|} 
{\bm \Omega}_{a,\rho,b} \,.
\ea
The notation $(a, A, b, B)=\mathbb I_m$ in the summation range on the right-hand side
instructs to identify the (possibly empty) ordered sets $A,B$ 
by matching $(a, A, b, B)= (1,2,\cdots, m)$ up to cyclic transformations $i \rightarrow i{+}1 \ {\rm mod} \ m$. Simple examples at $(m,a) = (3,2)$ and $(m,a) = (4,4)$ are
\ba
{\bm \Omega}_{123}+{\bm \Omega}_{132}=-{\bm \Omega}_{213}\,, \ \ \ \ \ \
-\Big( {\bm \Omega} _{1243}+{\bm \Omega} _{1423} \Big)- {\bm \Omega} _{1234}={\bm \Omega}_{4123}\,.
\ea
Based on (\ref{rhorhotog}) with the simplification of its middle line via (\ref{fay2identity}),
we can rewrite the single-cycle formula \eqref{rhorho} as follows, for any $a \in \{1,2,\cdots, m\}$,
 \begin{align}\label{rhorhog}
( 1{+}s_{12\cdots m}) \CC_{(12\cdots m)}(\xi)
=\,\,& \MM_{12\cdots m} (\xi)-
  \sum_{\substack{
b   =1 \\ b\neq a}
}^{m} \sum_{\rho \in  A \shuffle  B^{\rm T} \atop{(a, A, b, B)=\mathbb I_m}} 
\! \! (-1)^{|B|} \Bigg(
 \sum_{i=m+1}^{n}  x_{b,i} 
 +\nabla_b   
\Bigg){\bm \Omega}_{a,\rho,b} 
 \,,
 \end{align}
 where $A,B$ on the right-hand side are again determined by
 $(a, A, b, B )=(1,2, \cdots ,m)$ modulo cyclic transformations. 
This reformulation of the chain decomposition of single cycles $\CC_{(12\cdots m)}$
reduces to the original formula \eqref{rhorho} for $a=1$ and otherwise offers
the flexibility to prevent one arbitrary $\nabla_{a\neq 1}$-derivative 
from appearing on the right-hand side of (\ref{rhorhog}). By virtue
of \eqref{fayprac}, the chains ${\bm \Omega}_{a,\rho,b}$ on the right-hand side
of (\ref{rhorhog}) are readily expanded in the
basis of ${\bm \Omega}_{1\ldots}$ employed in \eqref{rhorho}.

At length $m=2$ and $m=3$, setting $a=2$ in (\ref{rhorhog}) to eliminate $\nabla_2$ leads to
\ba\label{ex21}
(1{+}s_{12}) \CC_{(12)}(\xi)=\,\,&\MM_{12}(\xi)+ \Omega_{12}(\eta_2)   \sum_{i=3}^{n}  x_{1,i} 
 +\nabla_1  \Omega_{12}(\eta_2) \,,
\\
\non
(1{+}s_{123}) \CC_{(123)}(\xi)=\,\,&\MM_{123}(\xi)- {\bm \Omega}_{231}  \sum_{i=4}^{n}  x_{1,i} + {\bm \Omega}_{213}  \sum_{i=4}^{n}  x_{3,i} 
 -\nabla_1  {\bm \Omega}_{231}    +\nabla_3  {\bm \Omega}_{213}  \, ,
\ea 
where the multiplicity $n\geq m$ of the Koba-Nielsen factor has been kept arbitrary.
 We will make frequent use of \eqref{rhorhog} when dealing with products of cycles in section \ref{section:six}.

\section{Chiral splitting}
\label{section:five}

For the integrands of closed-string genus-one amplitudes, 
manifestly doubly-periodic representations are tied to 
Wick contractions of the joint zero modes of the left- and right-moving worldsheet fields
$\partial_z X$ and $\partial_{\bar z} X$. Their Wick rules couple the 
left- and right-movers $\partial_z X$ and $\partial_{\bar z} X$ and 
lead to Lorentz contractions between the chiral halves
$\epsilon_i ,\bar \epsilon_i$ of the closed-string polarizations in \eqref{stringint}, accompanied by
factors of $\frac{\pi}{\Im \tau}$. Moreover, when applying F-IBP formulae such 
as \eqref{rhorho}, we need to be careful with holomorphic derivatives since their 
action $\partial_i \bar{f}^{(m>0)}_{ij}$ on the contributions from the opposite-chirality 
sector does not vanish. 

Both of these interactions between left- and right-movers can be sidestepped by 
virtue of chiral splitting \cite{DHoker:1988pdl, DHoker:1989cxq}: The key idea is to 
separate the joint zero mode
of the fields $\partial_z X$ and $\partial_{\bar z} X$ -- the string-theory counterpart
of the loop momentum in Feynman graphs -- from the path integral over $X$ that defines
the genus-one correlators in string amplitudes. At the level of the loop integrand of 
closed-string amplitudes, meromorphic and anti-meromorphic sectors then decouple. 
This can be viewed as the string-theory origin of the double-copy structures in
the loop integrand of (super-)gravity amplitudes which grew into a wide and
vibrant research field \cite{Bern:2019prr, Adamo:2022dcm}.

The independent chiral amplitudes from left- and right-movers in the loop integrand 
of closed-string amplitudes no longer manifest the $z_i \rightarrow z_i {+}\tau$
periodicity term by term. In particular, the doubly-periodic Kronecker-Eisenstein
coefficients $f^{(w)}_{ij}$ in the loop-integrated correlators typically translate into
their meromorphic counterparts $g^{(w)}_{ij}$ in (\ref{gtof}). Their B-cycle monodromies
under $z_i \rightarrow z_i {+}\tau$ (see (\ref{bmondr}) below) are compensated by shifts 
of the loop momentum by
the external momentum $k_i$ \cite{DHoker:1988pdl, DHoker:1989cxq}, see
\cite{Mafra:2017ioj, Mafra:2018qqe, DHoker:2020prr, Geyer:2021oox} for recent applications of this mechanism to the construction of
chiral amplitudes at different loop orders.
In this section, we will reformulate the single-cycle formula \eqref{rhorho} for generating
series of $f^{(w)}_{ij}$ in terms of the meromorphic
$g^{(w)}_{ij}$, loop momenta and a chiral Koba-Nielsen factor.

By the lack of term-by-term invariance under B-cycle shifts $z_i \rightarrow z_i {+}\tau$
in chiral amplitudes,
total derivatives w.r.t.\ $z_i$ may no longer integrate to zero. As will be detailed below, B-cycle monodromies in the primitives of IBP relations lead to boundary 
terms which we shall track in the reformulation of \eqref{rhorho} in a chiral-splitting context.\footnote{OS is grateful to Filippo Balli for discussions and collaboration 
on related topics that led to the understanding of
boundary terms as presented in section \ref{sec:bdyibp}.}
These boundary terms are no obstruction to break cycles of the meromorphic 
Kronecker-Eisenstein coefficients $g^{(w)}_{ij}$. Since the tracking of boundary 
terms can be smoothly 
incorporated into the methods of this work, our main results are compatible
with the reduction of closed-string problems to open-string ones using chiral splitting.

\subsection{Basics of chiral splitting}
\label{sec:revCS}

As shown in \cite{DHoker:1988pdl, DHoker:1989cxq}, chiral splitting allows to derive open- and closed-string amplitudes from the same chiral function ${\cal K}_n(\ell)$ of the kinematic data. Open-string $n$-point amplitudes at one loop descend from worldsheets of cylinder- and Moebius-strip topologies with punctures $z_i$ on the boundaries, 
\ba\label{openampF}
\mathcal{A}_{n}= \frac{1}  {(2 \pi i)^{D}} \sum_{\text {top }} C_{\text {top }} 
\int_{D^{\tau}_{\text {top }}} 
d \tau 
\int_{D^{z}_{\text {top }}}
d \mu^{\rm op}_{n} 
\int_{\mathbb R^D} d^{D} \ell\left|\mathcal{J}_{n}(\ell)\right|\mathcal{K}_{n}(\ell) \,.
\ea
Closed-string one-loop amplitudes in turn are given by
\ba\label{closedampF}
\mathcal{M}_{n}= \frac{1}  {(2 \pi i)^{D}} \int_{\mathfrak{F}} d^{2} \tau 
\int_{\mathfrak{T}_\tau^{n-1}} d \mu^{\rm cl}_{n} 
\int_{\mathbb R^D} d^{D} \ell\left|\mathcal{J}_{n}(\ell)\right|^{2} 
\mathcal{K}_{n}(\ell) 
\tilde{\mathcal{K}}_{n}(-\ell)
\,,
\ea
see the discussion below (\ref{openamp}) and (\ref{closedamp}) for the
integration domain of the moduli $z_i$ and~$\tau$.
As a universal part of the underlying correlation functions at fixed loop momentum, both \eqref{openampF} and \eqref{closedampF} involve the chiral Koba-Nielsen factor 
\ba
\mathcal{J}_{n}(\ell) \coloneqq \exp \bigg({-} \sum_{1 \leq i<j}^{n} s_{i j} \log \theta_{1} (z_{i j}, \tau )+ \sum_{j=1}^{n}   z_{j} \left(\ell \!\cdot\! k_{j}\right)+ \frac{\tau}{4 \pi i} \ell^{2}\bigg)\,,
\label{chiKN}
\ea 
which in contrast to the ${\cal I}_n^{\bullet}$ in (\ref{defkn}) depends 
meromorphically on both $z_i$ and $\tau$.
Even though we treat the momentum invariants $s_{ij}$ as independent variables, 
translation invariance of the chiral Koba-Nielsen factor (\ref{chiKN})
necessitates the condition $\sum_{j=1}^{n}    \left(\ell \!\cdot\! k_{j}\right) =0$,
i.e.\ momentum conservation along the direction of the loop momentum.

The leftover factors of $\mathcal{K}_{n}(\ell)$ in the loop integrands of \eqref{openampF} and \eqref{closedampF} carry the dependence on the polarizations and are referred to as chiral correlators. The $z$-dependence of $\mathcal{K}_{n}(\ell)$ is encoded in $g^{(w)}_{ij}$
with $w\leq n{-}4$ in maximally supersymmetric settings\footnote{See \cite{Mafra:2018pll} for a construction of multiparticle invariants under B-cycle monodromies $z_i \rightarrow z_i{+}\tau$ together with $\ell \rightarrow \ell - 2 \pi i k_i$ and \cite{Mafra:2018qqe} for different representations of $\mathcal{K}_{n}(\ell)$ at $n\leq 7$ points in pure-spinor superspace.} and $w \leq n$ for bosonic or heterotic strings. Given that chiral correlators $\mathcal{K}_{n}(\ell)$ are polynomials in loop 
momenta $\ell$, the loop integrals in \eqref{openampF} and \eqref{closedampF} are of straightforward Gaussian type. In the simplest case, we recover the earlier Koba-Nielsen factors ${\cal I}_n^{\bullet}$ in (\ref{defkn})
\ba
\mathcal{I}_{n}^{\text {op }} = \frac{(\operatorname{Im} \tau)^{\frac{D}{2}}}  {(2 \pi i)^{D}} 
\int_{\mathbb R^D} d^{D} \ell\left|\mathcal{J}_{n}(\ell)\right|
\,,\qquad
\mathcal{I}_{n}^{\rm cl} =\frac{(2 {\rm Im} \,\tau)^{\frac{D}{2}}}{(2 \pi i)^{D}} \int_{\mathbb R^D} d^{D} \ell\left|\mathcal{J}_{n}(\ell)\right|^{2}\,,
\ea
for open and closed strings respectively. The polynomial $\ell$-dependence of
chiral correlators gives rise to polynomials in $ \nu_{ij} \coloneqq 2\pi i \frac{\Im z_{i,j}}{ {\rm Im} \,\tau}$,
\ba
\int_{\mathbb R^D} d^{D} \ell\left|\mathcal{J}_{n}(\ell)\right|^{2} \ell^\mu&=\frac{(2 \pi i )^D}{(2 {\rm Im} \,\tau)^{\frac{D}{2}}} \mathcal{I}_{n}^{\rm cl}   \sum_{a=2}^n k_a^\mu \nu_{1a}\,,
\label{loopintegral} \\
\int_{\mathbb R^D} d^{D} \ell\left|\mathcal{J}_{n}(\ell)\right|^{2} \ell^\mu \ell^\lambda&= \frac{(2 \pi i )^D}{(2 {\rm Im} \,\tau)^{\frac{D}{2}}} \mathcal{I}_{n}^{\rm cl} \bigg[  \bigg(\sum_{a=2}^n k_a^\mu \nu_{1a}\bigg) \bigg(\sum_{b=2}^n k_b^\lambda \nu_{1b}\bigg)-  \frac \pi { {\rm Im} \,\tau} \delta^{\mu \lambda} \bigg]\,,
\notag
\ea
(with external momenta $k_a^\mu,k_b^\lambda$ and Lorentz indices $\mu ,\lambda$) which eventually conspire with the meromorphic
$g^{(w)}_{ij}$ in chiral correlators to
obtain the doubly-periodic $f^{(w)}_{ij}$ in (\ref{gtof}). Hence, the two types of terms $\sim \nu_{ij}$
and $\sim \frac{ \pi }{\Im \tau}$ on the right-hand side of (\ref{loopintegral}) 
illustrate how loop integration reproduces the manifestly doubly-periodic form 
(\ref{stringint}) of one-loop closed-string integrands.\footnote{Whenever the two loop momenta
$\ell^\mu \ell^\lambda$ in the second line of (\ref{loopintegral}) arise from the same chiral amplitude (i.e.\ both from the left- or right-movers),
the factors of $\frac{\pi}{\Im \tau}$ can be absorbed into the conversion 
of $\partial_i g^{(w)}_{ij}$ into $\partial_i f^{(w)}_{ij}$ such as
$\partial_i g^{(1)}_{ij} + \frac{ \pi }{\Im \tau} = \partial_i f^{(1)}_{ij}$. That is why 
factors of $\frac{\pi}{\Im \tau}$ are absent in the schematic form of open-string
integrands $K_n^{\rm op}$ in the first line of (\ref{stringint}).}

\subsubsection{Koba-Nielsen derivatives}

The derivatives of the Koba-Nielsen factor  $\mathcal{J}_{n}(\ell)$ in (\ref{chiKN}) 
with respect to worldsheet positions $z_i$ are given by 
\ba\label{basebefore}
\partial_i \mathcal{J}_{n}(\ell)=\bigg( \ell \!\cdot\! k_{i}  -\sum_{j\neq i}^{n}  {\tilde  x}_{i,j}  \bigg) 
\mathcal{J}_{n}(\ell)\,,\qquad {\rm with~}  {\tilde  x}_{i,j} \coloneqq s_{i j} g_{i j}^{(1)} ~ {\rm for~} j\neq i \,.
\ea
Similar to \eqref{defnabla}, we can introduce operators 
${\tilde\nabla}_i$ incorporating Koba-Nielsen derivatives
\ba
\label{defnablaJ}
{\tilde\nabla}_i {\tilde \varphi} \coloneqq \partial_i {\tilde \varphi} + \bigg ( \ell \!\cdot\! k_{i}-\sum_{j \neq i}^{n} {\tilde x}_{i, j} \bigg) {\tilde \varphi}=\frac{1}{\mathcal{J}_{n}}\partial_i (\tilde \varphi\,  \mathcal{J}_{n} )
\ea 
for arbitrary meromorphic contributions $\tilde \varphi = \tilde \varphi (z_i,\tau)$ to
chiral correlators $\mathcal{K}_{n}(\ell)$. Similar to \eqref{nablarule}, the operator 
$\tilde\nabla_{i}$ does not obey a Leibniz rule and instead acts as follows on products
\be
\label{nablaruleF}
\tilde\nabla_{i}(\tilde \varphi_1 \tilde \varphi_2)=(\tilde\nabla_{i} \tilde \varphi_1 ) \tilde \varphi_2
+\tilde \varphi_1 \partial_{i}(\tilde \varphi_2)= (\partial_i \tilde   \varphi_1 ) \tilde \varphi_2
+\tilde   \varphi_1 \tilde\nabla_{i}(\tilde  \varphi_2) \,.
\ee
On the other hand, two operators 
$\tilde\nabla_{i}$ and $\tilde\nabla_{j}$ still commute with each other as an analogue of \eqref{commute}.

For a meromorphic function $\tilde \varphi$ expressed in terms of $g_{ij}^{(w>0)}$
with B-cycle monodromies
\beq
g^{(w)}(z{+}\tau,\tau) = \sum_{k=0}^w \frac{(-2\pi i)^k}{k!} g^{(w-k)}(z,\tau) 
\, , \label{bmondr}
\eeq
integrals over total derivatives -- $\int_{D_{\rm {top}}^z} d\mu_n^{\rm op}  \mathcal{J}_n  \tilde \nabla_i \tilde \varphi$ for open strings and $\int_{\mathfrak{T}_\tau^{n-1}} d \mu^{\rm cl}_{n}   \mathcal{J}_n  \tilde \nabla_i \tilde \varphi$ for closed strings -- do not necessarily vanish: They violate doubly-periodicity which was already highlighted as salient point (i) below \eqref{totalderivative}. However, such integrals over total derivatives can still be reduced to boundary terms using Stokes' theorem. 

\subsubsection{Stokes' theorem and boundary terms}
\label{sec:bdyibp}

We shall now evaluate the integrals over total derivatives in chiral splitting
by means of Stokes' theorem,  following the perspective on boundary terms developed during the preparation of \cite{boundaryref}. Consider a typical contribution
$\Phi(z_2)=|\mathcal{J}_{n}(z_2)|^2 \tilde \varphi_i (z_2) \overline{ \tilde \varphi_j (z_2)}$
to closed-string integrands,
viewed as a function of $z_2$ (say $ \tilde \varphi_i (z_2) \rightarrow g^{(1)}_{12} g^{(1)}_{13}$ and $ \overline{ \tilde \varphi_j (z_2)} \rightarrow 1$ in the concrete example $\Phi(z_2)\rightarrow |\mathcal{J}_{n}(z_2)|^2 g^{(1)}_{12} g^{(1)}_{13}$).
Then, integrating a total $z_2$-derivative via Stokes' theorem yields

 \begin{figure}
  \centering
  \begin {tikzpicture}[scale=3]
   \draw [very thick,->] (0.704353, 0.374557) coordinate(A1) node [left]{}-- (3.03464, 0.379902) coordinate (B1) node [right]{${\rm Re}(z_2)$};
      \draw [very thick,->] (0.86337, 0.222221) coordinate(A2) node [below]{}-- (0.861918, 1.77486) coordinate (B2) node [left]{${\rm Im}(z_2)$};
            \draw [very thick] (0.866759, 0.383114) coordinate(C1) node [above left]{0}
            -- (1.33171, 1.47161) coordinate (C2) node [above left]{$\tau$}
            -- (2.88462, 1.47386) coordinate (C3) node [above right]{$\tau{+}1$}
            -- (2.41283, 0.383478) coordinate (C4) node [above left]{1}
            ;
            
            \filldraw (C1) circle(.7pt) ;
             \filldraw (C2) circle(.7pt) ;
              \filldraw (C3) circle(.7pt) ;
               \filldraw (C4) circle(.7pt) ;
               
       \node at     ($ (C1)! 1/2 ! (C4) + (0,-.22)$) {$A_2$};
             \node at     ($ (C1)! 1/2 ! (C2) + (-.10,.22)$)  {$-B_2$};
             \node at     ($ (C2)! 1/2 ! (C3) + (0,.24)$) {$-(A_2{+}\tau)$};
             \node at     ($ (C3)! 1/2 ! (C4) + (.36,.22)$) {$(B_2{+}1)$};
             \coordinate   (D0) at ( 3.65742, 1.68769)  ;
             \node at   ( 3.65742, 1.68769)  {$z_2$};       
             
              \draw [very thick]
               ($(3.47214, 1.80545)!1/3!(D0)$) coordinate(A1) node [left]{}-- ($(3.46738, 1.5285) !1/3!(D0)$)
               coordinate (B1) node [right]{}-- ($ (3.75044, 1.52865)!1/3!(D0)$) coordinate (B1) node [above]{} ;
\node at ($ (C1)! 4/8 ! (C2) $)  (C12){};
\draw [->, line width=0.7mm] (C2) -- (C12)  ;       
\node at ($ (C2)! 4/8 ! (C3) $)  (C23){};
\draw [->, line width=0.7mm] (C3) -- (C23)  ;       
\node at ($ (C3)! 4/8 ! (C4) $)  (C34){};
\draw [->, line width=0.7mm] (C4) -- (C34)  ; 
\node at ($ (C1)! 4/8 ! (C4) $)  (C14){};
\draw [->, line width=0.7mm] (C1) -- (C14)  ; 
\draw [-, line width=0.7mm] (C1) -- (C4)  ; 
\draw [-, line width=0.7mm] (C1) -- (C2)  ; 
\draw [-, line width=0.7mm] (C2) -- (C3)  ; 
\draw [-, line width=0.7mm] (C3) -- (C4)  ; 
\node at ($ (C1)! 1/2 ! (C3)$)  (C13){$\mathfrak{T}_\tau$};
\draw [->, line width=0.7mm] (C2) -- (C12)  ;
    \end {tikzpicture}
  \caption{The fundamental parallelogram $\mathfrak{T}_\tau$ at fixed loop momentum for $z_2$ and its boundary. The boundary $\partial \mathfrak{T}_\tau$ is given in terms of the integration contours $A_2$, $B_2$ for $z_2$, and their displacements $(A_2{+}\tau)$ and $(B_2{+}1)$.}
  \label{parallelogramZ2AndItsBoundary}
\end{figure}
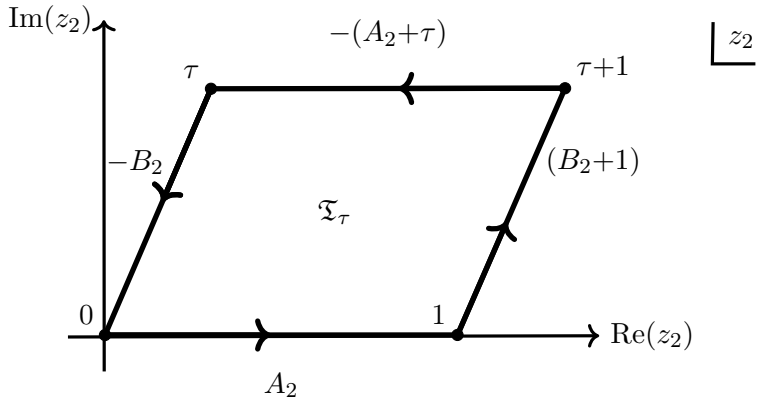

\begin{align}
\label{eq:LoopMomentumStokes}
\int_{\mathfrak{T}_\tau} d^2 z_2 \;\partial_2\big(\Phi(z_2)\big) &=
\int_{\partial \mathfrak{T}_\tau} d \bar z_2 \;\Phi(z_2) \\
 &=\int_{A_2} d \bar z_2 \; 
 \Phi(z_2) +\int_{B_2+1} d \bar z_2 \;\Phi(z_2) -\int_{A_2+\tau} d\bar z_2 \;\Phi(z_2)  -
 \int_{B_2} d\bar z_2 \;\Phi(z_2) 
 \nonumber  
 \\
 &=\int_{0}^{1} d\bar z_2 \;\Phi(z_2)  +\int_{1}^{1+\tau} d\bar z_2 \;\Phi(z_2)  -\int_{\tau}^{\tau+1} d\bar z_2 \;\Phi(z_2)  -
 \int_{0}^{\tau} d\bar z_2 \;\Phi(z_2) 
 \nonumber  \; .
\end{align}
We have decomposed the boundary $\partial \mathfrak{T}_\tau$
of the parallelogram in figure~\ref{parallelogramZ2AndItsBoundary}
 into four components, namely the homology cycles $A_2$, ${-}B_2$ and their translates $(B_2{+}1)$, ${-}(A_2{+}\tau)$ with minus signs accounting for their orientation. The four contributions to (\ref{eq:LoopMomentumStokes})
can be reorganized
into two integrals by absorbing the displacements by $1$ and $\tau$ into the~integrand,
\begin{align}
\int_{\mathfrak{T}_\tau} d^2 z_2 \;\partial_2\big(\Phi(z_2)\big) &=\int_{0}^{1} d\bar z_2 \;\big[\Phi(z_2)-\Phi(z_2{+}\tau)\big]  +\cancel{\int_{0}^{\tau} d\bar z_2 \;\big[\Phi(z_2{+}1)-\Phi(z_2)\big]} 
 \notag \\
&=\int_{0}^{1} d\bar z_2 \;\big[\Phi(z_2)-\Phi(z_2{+}\tau)\big] =: - \int_{0}^{1} d\bar z_2  \, \hat{b}_2 \Phi(z_2)\, .
\label{eq:LoopMomentumStokesSecond}
\end{align}
In passing to the second line, we have exploited the A-cycle
shift invariance $\Phi(z_2{+}1)=\Phi(z_2)$ of chiral correlators
inherited from the periodicity $g^{(w)}(z{+}1,\tau) = g^{(w)}(z,\tau)$. The last line of (\ref{eq:LoopMomentumStokesSecond}) defines the
difference operator $\hat{b}_j$ associated with a B-cycle shift of $z_j$,
\begin{align}
\hat{b}_j \Phi(z)=\Phi(z)\big{|}_{z_j\rightarrow z_j+\tau}-\Phi(z) \, .
\end{align}
Since a typical primitive $\Phi(z_2)$ in the total derivative (\ref{eq:LoopMomentumStokes})
may have B-cycle monodromies (\ref{bmondr}), the integrand $ \hat{b}_2 \Phi(z_2)$ in the last line of (\ref{eq:LoopMomentumStokesSecond}) is in general non-zero. Nevertheless, the surface integral over $\mathfrak{T}_\tau$ on the left-hand side has simplified to a boundary integral where $z_2$ is restricted to the A-cycle $(0,1)$. These boundary terms can be reconstructed
from the images of ${\tilde\nabla}_i$ in subsequent formulae for the breaking of
cycles of $g^{(w)}_{ij}$ in chiral amplitudes. Hence, by consistently retaining
${\tilde\nabla}_i$ before the loop integral, the results of this section provide the chiral-splitting
analogues of the IBP relations \eqref{IBPomega}.

\subsection{Single cycles versus chains of meromorphic $F$}

By analogy with our notation \eqref{deflongomega2} for products of doubly-periodic
Kronecker-Eisenstein series, we denote chains of their meromorphic
counterparts $F_{ij}(\eta) = F(z_i{-}z_j,\eta,\tau)$ by
\begin{equation}
\label{deflongF}
{\bm F}_{\alpha _1 \a_2\cdots \a_m } \coloneqq \delta\bigg( \sum_{i=1}^m \eta_{\a_i} \bigg) \prod_{i=1}^{m-1}  
F_{\alpha_i\, \alpha_{i+1}} ( \eta_{\alpha_{i+1}\, \cdots\, \alpha_{m}  })    \,,
\end{equation}
such that for instance ${\bm F}_{123}= \delta(\eta_{1}{+}\eta_2{+}\eta_3)
F_{12}( \eta_{23}) F_{23}( \eta_{3}) $. Since the Fay identity (\ref{fayid})
is universal to $F_{ij}(\eta)$ and $\Omega_{ij}(\eta)$, the relation
(\ref{fayprac}) for doubly-periodic chains with ordered sets $\alpha,\beta$ 
straightforwardly propagates to the $F$-chain in (\ref{deflongF}),
\ba
\label{faypracF}
{\bm F}_{\alpha,i,\beta}= (-1)^{|\alpha|} {\bm F}_{i,\alpha^{\rm T}}  {\bm F}_{i,\beta}= (-1)^{|\alpha|} \sum_{\rho\in \alpha^{\rm T}\shuffle \beta} {\bm F}_{i,\rho}\,.
\ea 
In particular, only $(m{-}1)!$ out of the
$m!$ permutations of ${\bm F}_{\alpha _1 \a_2\cdots \a_m }$
are independent under (\ref{faypracF}), and one can for instance 
take those chains with $\alpha_1=1$ as the independent representatives. 
However, it is a separate question whether the $(m{-}1)!$ independent 
chains in (\ref{faypracF}) offer a basis for cycles and more general configurations
of $F_{ij}(\eta)$. By the discussion in section \ref{sec:bdyibp}, total derivatives may lead to
non-vanishing boundary terms in chiral splitting which may be thought of
as additional basis elements for a twisted cohomology associated with
the chiral Koba-Nielsen factor (\ref{chiKN}). While the identification of
cohomology bases with a full account of boundary terms is beyond the
scope of this work, we shall spell out the F-IBP relations between cycles 
of $F_{ij}(\eta)$ and combinations of chains and boundary terms.

More specifically, as the meromorphic counterpart of the doubly-periodic
cycle in (\ref{defc}), we will be interested in the F-IBP reduction of the cycles
\begin{equation}\label{defcf}
\CCF_{(12\cdots m)}(\xi) \coloneqq \delta\bigg( \sum_{i=1}^m \eta_{i} \bigg)  F_{12}(\eta_{23\cdots m}{+}\xi)F_{23}(\eta_{3\cdots m}{+}\xi) 
\cdots 
F_{m-1,m}(\eta_{ m}{+}\xi)
F_{m 1}(\xi) 
\end{equation}
with multiplicities $2\leq m\leq n$. The arguments $\eta_{i\cdots m}{+}\xi$ are again tailored to attain the same reflection and cyclicity properties (\ref{dihedcyc}) as in the doubly-periodic case. Moreover, the antiholomorphic derivatives (\ref{antiholocyc}) of $\CC_{(12\cdots m)}(\xi)$ are converted to B-cycle monodromies
\beq
\CCF_{(12\cdots m)}(\xi) \big|_{z_j \rightarrow z_j + \tau} =e^{2\pi i \eta_j} \CCF_{(12\cdots m)}(\xi) 
\eeq
in passing to the meromorphic $\CCF_{(12\cdots m)}(\xi)$ in (\ref{defcf}).
The main result of this section will be
a  formula analogous to \eqref{rhorho} to break such a single $F$-cycle \eqref{defcf}.

\subsubsection{Breaking of length-$m$ cycles}

We recall that Fay identities together with \eqref{variant}, \eqref{derivative} and the IBP relations \eqref{IBPomega} conspire in deriving the formula \eqref{rhorho} to break a single 
$\Omega$-cycle. The building blocks $F_{ij}(\eta)$ of the meromorphic cycles (\ref{defcf})
obey almost identical identities, except for the loop-momentum dependent term $\ell {\cdot} k_i$
in the chiral Koba-Nielsen derivative \eqref{basebefore} which is absent in \eqref{baseafter}.
On these grounds, one can apply the substitution rules
\beq
{\bm \Omega}_{\alpha_1 \alpha_2\ldots } \to {\bm F}_{\alpha_1 \alpha_2\ldots }\, , \ \ \ \  x_{i,j} \to \tilde x_{i,j} \, , 
\ \ \ \  \hat g^{(1)}(\eta) \to g^{(1)}(\eta) \,,  \ \ \ \
  \nabla_b\to {\tilde\nabla}_b - \ell\!\cdot\! k_b\,,
  \label{subsrule}
  \eeq
to convert \eqref{rhorho} to a very similar identity to break the $F$-cycles (\ref{defcf}),
 \begin{align}\label{rhorhoF}
 ( 1+&s_{12\cdots m}) \CCF_{(12\cdots m)}(\xi)
=\MMF_{12\cdots m}(\xi)
 \\
 \non &\ \ \ -\sum_{b  =2}^{m} 
\!\!\!
\sum_{\substack{ \rho \in \{2,3,\cdots, b-1\}
\\ \,\, \shuffle \{m,m-1,\cdots, b+1 \}} }
\!\!\!\!\!\!\!\!\!\!\!\!\!\!\!
 (-1)^{m-b} \Bigg( {-}  \ell\cdot k_b+\!\!
 \sum_{i=m+1}^{n}  {\tilde x}_{b,i} 
 +
{\tilde\nabla}_b    
\Bigg)
{\bm F}_{1,\rho,b}  
 \,.
 \end{align}
The linear combinations of chains (\ref{deflongF}) analogous to (\ref{rhorhoMM}) are given by
   \begin{align}
&\!\!\!\!\MMF_{12\cdots m}(\xi) \coloneqq
 \sum_{b  =2}^{m} 
\!\!\!\!\!\!
\sum_{\substack{\rho \in \{2,3,\cdots, b-1\}
\\
\quad \shuffle \{m,m-1,\cdots, b+1 \}
}} 
\!\!\!\!\!\!\!\!\!\!\!\!\!\!\!\!
 (-1)^{m-b} \Bigg(
\sum_{i=1}^m \! s_{i b} \,\partial_{\eta_b}{-}
    \sum_{i=2}^m  \! s_{i b}\, \partial_{\eta_i}
  {+}  (1{+}s_{12\cdots m}) v_1(\eta_{b}, \,\eta_{b+1,\cdots,m}{+}\xi )     
  \nl
  &\quad  \quad  \quad \quad \quad  
   -  g^{(1)}(\eta_b) 
  -\sum_{i=2}^{b-1}  { S}_{i,\rho} 
v_1(\eta_{b },\,\eta_{i,i+1,\cdots, b-1 })
-\sum_{i=b+1}^m { S}_{i,\rho} 
v_1(\eta_{b },\,\eta_{b+1,b+2,\cdots, i })   
 \Bigg) {\bm F}_{1,\rho,b}   
\notag \\
&\quad  + \!\!\!\!\!\!\!\!\! \!\!\!\sum_{1\leq p<u<v<w<q\leq m+1} 
 \!\!\!  \!\!\! \!\!\!\!\!\!\!\!\!
(-1)^{m+u+v+w}\, \Big(
v_1(\eta_{u+1,\cdots,w-1},-\eta_{u,\cdots,w-1})+v_1(\eta_{u,\cdots,w},-\eta_{u+1,\cdots,w}) 
 \Big) 
\label{rhoMtild} \\
&\quad \quad  \quad  \quad \quad  \times
\Big(\sum_{i=q}^m s_{vi}+\sum_{i=1}^p s_{vi} \Big)  \! \! \! \!
 \sum_{\substack{\rho \in \{2,3,\cdots,p\}\shuffle \{m,m-1,\cdots, q\}\\
\gamma \in \{p+1,p+2,\cdots,u-1\}\shuffle \{v-1,v-2,\cdots, u+1\}\\
\pi \in \{v+1,v+2,\cdots,w-1\}\shuffle \{q-1,q-2,\cdots, w+1\}
 }} 
 \sum_{\sigma\in \{\gamma,u\}\shuffle \{ \pi, w\}} 
 \!\!\!\!\!\!\!
 \quad 
{\bm F}_{1,\rho,v,\sigma}
 \,,
 \non
 \end{align}
see figure \ref{fig:illus} for an illustration of the nested sums over ordered sets
$\rho,\gamma,\pi,\sigma$ and (\ref{defsjrho}) for the definition of $S_{i,\rho}$.
In other words, the total Koba-Nielsen derivatives $\nabla_b$ in the doubly-periodic case \eqref{rhorho}
completely determine the new class of terms $\ell\!\cdot\! k_b$
involving the loop momentum of chiral splitting. One can view the closely related
formulae (\ref{rhorho}) and (\ref{rhorhoF}) as
different manifestations of the same combinatorial 
principle of cycle-breaking. The substitution rules (\ref{subsrule}) will also be applied in sections \ref{sec:2cyc.5} and \ref{triplecyclenew} to convert F-IBP reductions of multiple $\Omega$-cycles to those of $F$-cycles.

While the total Koba-Nielsen derivatives $\nabla_b(\ldots)$ in \eqref{rhorho} can be discarded
due to double-periodicity of ${\bm \Omega}_{\alpha_1 \alpha_2\ldots }$,
the total derivatives ${\tilde\nabla}_b$ in the chiral-splitting counterpart
(\ref{rhorhoF}) generically lead to non-vanishing boundary terms (\ref{eq:LoopMomentumStokes}).
 
 \subsubsection{Length-two examples and elliptic functions}
 
The two-point example of (\ref{rhorhoF})
in presence of an $n$-point chiral Koba-Nielsen factor reads
  \ba\label{2ptfinalF}
\CCF_{(12)}(\xi)   =\,\,&\frac{1}{1+s_{12}}
\bigg(\MMF_{12}(\xi) + \ell\cdot k_2 F _{12} (\eta_2) - F _{12} (\eta_2)
\sum_{i=3}^{n}  s_{2i} g_{2i}^{(1)} - {\tilde\nabla}_2   F _{12} (\eta_2)  
\bigg)
\ea
with
 \ba
\MMF_{12}(\xi) \coloneqq \left( s_{12} \partial_{\eta _2}- g^{(1)}(\eta _2)+\left(1{+}s_{12}\right) v_1(\eta _2,\xi) \right) F_{12} (\eta_2)\,.
\ea
The generating functions (\ref{1.1a}) of the elliptic $V_w(1,2,\cdots,m)$ can
written in terms of either $\Omega$-cycles or $F$-cycles. Hence, the
manipulations of $F$-cycles in this section offer an alternative
to the breaking of $V_m(1,2,\cdots ,m)$ via (\ref{equation451}),
(\ref{coeff}) and \eqref{rhorho}. More speficially,
\beq
V_m(1,2,\cdots ,m) = \CCF_{(12\cdots m)}(\xi) \big|\big|_{\eta_2^0,\eta_3^0,\cdots,\eta_{m} ^0,\xi^0} 
\label{chiralVm}
\eeq 
together with \eqref{rhorhoF} leads to F-IBP decompositions
of cycles of $g^{(1)}_{ij}$ into $F$-chains and boundary terms
specific to chiral splitting. 

 At two points, combining (\ref{2ptfinalF}) 
with (\ref{chiralVm}) implies
\begin{align}
V_2(1,2) 
=\,\,&
 \frac{1}{ 1+s_{12}}   \bigg(\MMF_{12}(\xi)\big|\big|_{\eta_2^0,\xi^0}+ g_{12}^{(1)}     \,\ell\cdot k_2-
g_{12}^{(1)}   \sum_{i=3}^{n}  s_{2i} g_{2i}^{(1)} 
-
{\tilde\nabla}_2  g_{12}^{(1)}  
 \bigg) \label{merov12}
 \end{align}
with
\ba
\MMF_{12}(\xi)\big|\big|_{\eta_2^0,\xi^0}=\,\,& s_{12} \partial_{\eta _2} F _{12} (\eta_2)\big|_{\eta_2^0}- g^{(1)}(\eta _2) F _{12} (\eta_2)\big|_{\eta_2^0} +\left(1{+}s_{12}\right) v_1(\eta _2,\eta _3) F _{12} (\eta_2)\big|\big|_{\eta_2^0,\xi^0}
\nl
=\,\,& 2s_{12} g_{12}^{(2)}+  {\rm G}_2\,,
\ea
where the total Koba-Nielsen derivative ${\tilde\nabla}_2  g_{12}^{(1)} $
leads to non-trivial boundary terms by the B-cycle monodromy $\hat{b}_2g_{12}^{(1)}=2\pi i$,
see (\ref{eq:LoopMomentumStokesSecond}). Note that the manifestly doubly-periodic analogue of (\ref{merov12}) can be found in (\ref{doubperv12}).

\subsubsection{Reformulation of the meromorphic single-cycle formula}

The F-IBP formula (\ref{rhorho}) to break cycles of $\Omega_{ij}(\eta)$
was reformulated in \eqref{rhorhog} such as to single out an arbitrary 
$z_a$ with $a=2,3,\ldots,m$ instead of $z_1$. We can similarly
rewrite the meromorphic counterpart \eqref{rhorhoF} of the
single-cycle formula in a more flexible form,
\begin{align}\label{rhorhogf}
( 1{+}&s_{12\cdots m}) \CCF_{(12\cdots m)}(\xi)
=\,\, \MMF_{12\cdots m} (\xi)
\\
&
\nonumber
-
  \sum_{\substack{
b   =1 \\ b\neq a}
}^{m} \sum_{\substack{\rho \in  A \shuffle  B^{\rm T}
\\
(a, A, b, B )=
 \mathbb I_m
 }} 
 (-1)^{|B|} \bigg({-}\ell \!\cdot\! k_b+ 
 \sum_{i=m+1}^{n}  {\tilde x}_{b,i} 
 +{\tilde\nabla}_b   
\bigg){\bm F}_{a,\rho,b} 
 \,,
 \end{align}
 where the special leg 1 in \eqref{rhorhoF} is replaced by a general one $a \in \{1,2,\cdots, m\}$.  
 As before, the summation range defines the ordered sets $A,B$ by matching
 $(a, A, b, B )= (1,2, \cdots ,m) = (a, A, b, B )$ up to cyclic transformations 
 $i \rightarrow i{+}1 \ {\rm mod} \ m$. At $m=2$, for instance, as a comparison 
 with \eqref{2ptfinalF}, we also have
 \ba\label{2ptfinalF222}
\CCF_{(12)}(\xi)   =\,\,&\frac{1}{1+s_{12}}
\bigg(\MMF_{12}(\xi) - \ell\cdot k_1 F _{12} (\eta_2) + F _{12} (\eta_2)
\sum_{i=3}^{n}  s_{1i} g_{1i}^{(1)}
+
{\tilde\nabla}_1  F _{12} (\eta_2) 
\bigg)\,.
\ea
The reformulation (\ref{rhorhogf}) will be applied 
to break products of $F$-cycles in section~\ref{sec:2cyc.5}.

\section{Breaking of two or more cycles}
\label{section:six}

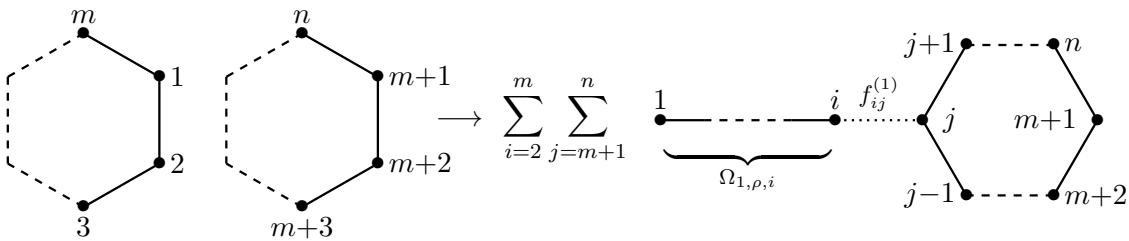
\begin{figure}[h]
  \centering
  \begin {tikzpicture}[line width = 0.3mm, scale=1.15]
\draw(0,1)node{$\bullet$}node[above]{$m$} -- (0.867,0.5);
\draw(0.867,0.5)node{$\bullet$}node[right]{$1$} -- (0.867,-0.5)node{$\bullet$}node[right]{$2$};
\draw(0.867,-0.5) -- (0,-1)node{$\bullet$}node[below]{$3$};
\draw[dashed](0,-1) -- (-0.867,-0.5);
\draw[dashed](-0.867,0.5) -- (-0.867,-0.5);
\draw[dashed](-0.867,0.5) -- (0,1);
\scope[xshift=2.5cm]
\draw(0,1)node{$\bullet$}node[above]{$n$} -- (0.867,0.5);
\draw(0.867,0.5)node{$\bullet$}node[right]{$m{+}1$} -- (0.867,-0.5)node{$\bullet$}node[right]{$m{+}2$};
\draw(0.867,-0.5) -- (0,-1)node{$\bullet$}node[below]{$m{+}3$};
\draw[dashed](0,-1) -- (-0.867,-0.5);
\draw[dashed](-0.867,0.5) -- (-0.867,-0.5);
\draw[dashed](-0.867,0.5) -- (0,1);
\endscope
\draw(4.3,0)node{$\longrightarrow$};
\draw(5.5,0)node{$\displaystyle \sum_{i=2}^m \sum_{j=m+1}^n$};
\scope[xshift=10.6cm]
\draw(-1,0) -- (-0.5,0.867);
\draw[dashed](-0.5,0.867) -- (0.5,0.867);
\draw(0.5,0.867) -- (1,0);
\draw(1,0) -- (0.5,-0.867);
\draw[dashed](0.5,-0.867)--(-0.5,-0.867);
\draw(-0.5,-0.867) -- (-1,0);
\draw(-1.5,0.3)node{\footnotesize $f^{(1)}_{ij}$};
\draw[dotted](-1,0) -- (-2,0);
\draw(-2,0)node{$\bullet$}node[above]{$i$} -- (-2.5,0);
\draw[dashed](-2.5,0) -- (-3.5,0);
\draw(-3.5,0) -- (-4,0)node{$\bullet$}node[above]{$1$};
\draw(-3,-0.5)node{$\underbrace{ \phantom{xxxxxxxxxx} }_{ \Omega_{1,\rho,i} }$};
\draw (-1,0)node{$\bullet$}node[right]{$\ j$};
\draw (-0.5,0.867) node{$\bullet$}node[left]{$j{+}1$};
\draw (0.5,0.867) node{$\bullet$}node[right]{$n$};
\draw (1,0) node{$\bullet$}node[left]{$m{+}1\ $};
\draw (0.5,-0.867) node{$\bullet$}node[right]{$m{+}2$};
\draw (-0.5,-0.867) node{$\bullet$}node[left]{$j{-}1$};
\endscope
 \end {tikzpicture}
  \caption{Tadpole graphs resulting from the breaking of the first cycle $\CC_{(12\cdots m)}(\xi_1)$ in (\ref{twocycles}) via \eqref{rhorho}. Similar to figures \ref{fig:graph} and \ref{fig:fom}, solid lines between vertices $a$ and $b$ refer to Kronecker-Eisenstein series with first argument $z_{ab}$ (with dashed lines to refer to an indefinite number of them). The dotted line represents the factors of $f^{(1)}_{ij}$ connecting the legs $i \in \{2,3,\ldots,m\}$ of the broken cycles with those of the unbroken one, $j \in \{m{+}1,\ldots,n\}$.}
  \label{fig:2cycs}
\end{figure}

This section is dedicated to applications of the single-cycle formula \eqref{rhorho} to reduce Koba-Nielsen integrals over more general arrangements of Kronecker-Eisenstein series to the conjectural chain basis. Our focus is mainly on products of two isolated $\Omega$-cycles,
\ba\label{twocycles}
\CC_{(12\cdots m)}(\xi_1)\CC_{(m+1,m+2 \cdots n)}(\xi_2)\,,
\ea
though pioneering examples of triple cycles are discussed in subsection \ref{triplecyclenew}.
The approach of this section is to break down the cycles one by one using the single-cycle formula. However, the second term of \eqref{rhorho} introduces
factors of $x_{i,j}= s_{ij} f^{(1)}_{ij}$ which connect the chains from the broken cycle 
with the unbroken cycle. The resulting $f$-$\Omega$ chains ending on a leg of the 
unbroken cycle are visualized through a tadpole graph in figure \ref{fig:2cycs} and require 
extra care in identifying the appropriate F-IBP manipulations that break the second cycle.

The $f$-$\Omega$ tadpoles in figure \ref{fig:2cycs} are still amenable to the single-cycle formula \eqref{rhorhog} which eventually leads to an expansion of the product of two cycles in (\ref{twocycles}) in an $(n{-}1)!$-element chain basis. In some cases, cycles involving all the $n$ legs of both cycles and two insertions of $f^{(1)}_{ij}$ may appear in intermediate steps, 
see the last line in the compact formula (\ref{newcyc23pre22}) below for the chain reduction 
of the two cycles in (\ref{twocycles}). Nevertheless, these cycles are readily eliminated using the
results of earlier sections and illustrate that the elimination of multiple cycles 
is most conveniently approached with a recursive strategy.

We start by illustrating the general strategy via special cases of (\ref{twocycles}),
namely two cycles of length two in section \ref{sec:2cyc.1} as well as
two cycles of length $m$ and two in section \ref{sec:2cyc.2}.
After addressing two cycles of general length in section \ref{sec:2cyc.3},
later subsections elaborate on F-IBP reductions of the meromorphic analogues 
of the cycles in (\ref{twocycles}) as well as the elimination of products of 
three cycles at six and seven points. The Koba-Nielsen derivatives discarded in this section are reinstated in 
appendix \ref{sec:2cyc.4}, and the treatment of an arbitrary number of cycles 
is presented in a companion paper \cite{companion}.

\subsection{Two cycles of length $2$ and $2$}
\label{sec:2cyc.1}

The simplest instance of the product (\ref{twocycles})
of two Kronecker-Eisenstein cycles occurs at
$n=4$ points,
 \be
\CC_{(12)}(\xi_1)\CC_{(34)}(\xi_2)=
 \Omega_{12}(\eta_{2}{+}\xi_1)\Omega_{21}(\xi_1)\,\Omega_{34}(\eta_{4}{+}\xi_2)\Omega_{43}(\xi_2)\,.
 \ee
Here and in the next two subsections, the products 
$\CC_{(12\cdots m)}(\xi_1)\CC_{(m+1,m+2 \cdots n)}(\xi_2)$ are understood to occur
in a string integrand along with the Koba-Nielsen factors in (\ref{defkn}).
Hence, we drop the total Koba-Nielsen derivative 
$\CC_{(34)}(\xi_2)\nabla_{2}  \Omega _{12} (\eta_2)  
 =\nabla_{2} \big( \CC_{(34)}(\xi_2) \Omega _{12} (\eta_2)  \big) $ 
 when breaking the first cycle $\CC_{(12)}(\xi_1)$ via \eqref{2ptfinal},
\begin{align}
\label{(12)(34)}
&(1{+}s_{12})\CC_{(12)}(\xi_1)\CC_{(34)} (\xi_2) \overset{\rm IBP}= 
 \MM_{12}(\xi_1) \CC_{(34)} (\xi_2) 
 - \Omega_{12}(\eta_2)  x_{2,34} \CC_{(34)}(\xi_2)  \, ,
    \end{align}
where $\MM_{12}(\xi_1)$ is given by (\ref{mm12}) and free of cycles. 
Here and in the rest of this work, we use the shorthand notation
\beq
x_{i,P} \coloneqq \sum_{j\in P}x_{i,j}\, ,
\eeq
where a set $P$ in the subscript of $x_{i,\cdot}$ encodes a sum over
its elements $j \in P$, e.g.\ $x_{2,34}=x_{2,3}{+}x_{2,4}$.
To address the first term on the right-hand side of \eqref{(12)(34)}, we can safely employ the relabeling of \eqref{2ptfinal} to break the second cycle $\CC_{(34)}(\xi_2)=\Omega_{34}(\eta_{4}{+}\xi_2)\Omega_{43}(\xi_2) $,
 \be\label{si1}
(1{+}s_{34})  \CC_{(34)}(\xi_2) = \MM_{34}(\xi_2)- \Omega_{34}(\eta_{4}) x_{4,12}
- \nabla_4 \Omega _{34} (\eta_4) 
\,,
 \ee
with 
 \be
\MM_{34}(\xi_2)=\bigg(
(1{+}s_{34})v_1(\eta_4,\xi_2)
+{s_{34}\partial_{\eta_4} - \hat{ g}^{(1)}(\eta_4)}
\bigg)\Omega_{34}(\eta_{4})\,.
 \ee
 This time, the Koba-Nielsen derivative in the last term of \eqref{si1} can still be dropped. 
 
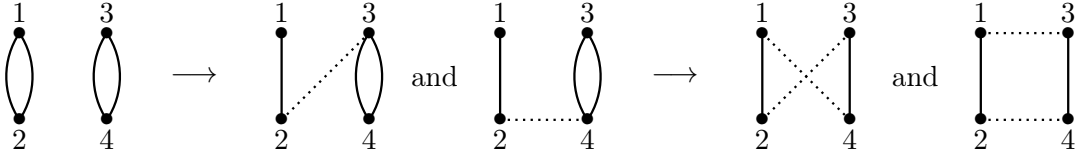
\begin{figure}[h]
  \centering
  \begin {tikzpicture}[line width = 0.3mm, scale=1.15]
\draw(-0.5,0.5) .. controls (-0.7,0.2) and (-0.7,-0.2) .. (-0.5,-0.5);
\draw(-0.5,0.5) .. controls (-0.3,0.2) and (-0.3,-0.2) .. (-0.5,-0.5);
\draw(-0.5,0.5)node{$\bullet$}node[above]{$1$}; 
\draw(-0.5,-0.5)node{$\bullet$}node[below]{$2$}; 
\draw(0.5,0.5) .. controls (0.7,0.2) and (0.7,-0.2) .. (0.5,-0.5);
\draw(0.5,0.5) .. controls (0.3,0.2) and (0.3,-0.2) .. (0.5,-0.5);
\draw(0.5,0.5)node{$\bullet$}node[above]{$3$}; 
\draw(0.5,-0.5)node{$\bullet$}node[below]{$4$}; 
\draw(1.5,0)node{$\longrightarrow$};
\scope[xshift=3cm]
\draw(0.5,0.5) .. controls (0.7,0.2) and (0.7,-0.2) .. (0.5,-0.5);
\draw(0.5,0.5) .. controls (0.3,0.2) and (0.3,-0.2) .. (0.5,-0.5);
\draw(0.5,0.5)node{$\bullet$}node[above]{$3$}; 
\draw(0.5,-0.5)node{$\bullet$}node[below]{$4$}; 
\draw(-0.5,0.5) -- (-0.5,-0.5);
\draw(-0.5,0.5)node{$\bullet$}node[above]{$1$}; 
\draw(-0.5,-0.5)node{$\bullet$}node[below]{$2$}; 
\draw[dotted](-0.5,-0.5) -- (0.5,0.5);
\endscope
\draw(4.25,0)node{and};
\scope[xshift=5.5cm]
\draw(0.5,0.5) .. controls (0.7,0.2) and (0.7,-0.2) .. (0.5,-0.5);
\draw(0.5,0.5) .. controls (0.3,0.2) and (0.3,-0.2) .. (0.5,-0.5);
\draw(0.5,0.5)node{$\bullet$}node[above]{$3$}; 
\draw(0.5,-0.5)node{$\bullet$}node[below]{$4$}; 
\draw(-0.5,0.5) -- (-0.5,-0.5);
\draw(-0.5,0.5)node{$\bullet$}node[above]{$1$}; 
\draw(-0.5,-0.5)node{$\bullet$}node[below]{$2$}; 
\draw[dotted](-0.5,-0.5) -- (0.5,-0.5);
\endscope
\draw(7,0)node{$\longrightarrow$};
\scope[xshift=8.5cm]
\draw(-0.5,0.5) -- (-0.5,-0.5);
\draw(-0.5,0.5)node{$\bullet$}node[above]{$1$}; 
\draw(-0.5,-0.5)node{$\bullet$}node[below]{$2$}; 
\draw(0.5,0.5) -- (0.5,-0.5);
\draw(0.5,0.5)node{$\bullet$}node[above]{$3$}; 
\draw(0.5,-0.5)node{$\bullet$}node[below]{$4$}; 
\draw[dotted](-0.5,0.5) -- (0.5,-0.5);
\draw[dotted](-0.5,-0.5) -- (0.5,0.5);
\endscope
\draw(9.75,0)node{and};
\scope[xshift=11cm]
\draw(-0.5,0.5) -- (-0.5,-0.5);
\draw(-0.5,0.5)node{$\bullet$}node[above]{$1$}; 
\draw(-0.5,-0.5)node{$\bullet$}node[below]{$2$}; 
\draw(0.5,0.5) -- (0.5,-0.5);
\draw(0.5,0.5)node{$\bullet$}node[above]{$3$}; 
\draw(0.5,-0.5)node{$\bullet$}node[below]{$4$}; 
\draw[dotted](-0.5,-0.5) -- (0.5,-0.5);
\draw[dotted](-0.5,0.5) -- (0.5,0.5);
\endscope
 \end {tikzpicture}
  \caption{Graphical representation of the products of the Kronecker-Eisenstein
  series (solid lines) and factors of $f^{(1)}_{ij}$ (dotted lines) in the chain decomposition
  of the product of length-two cycles $\CC_{(12)}(\xi_1)\CC_{(34)}(\xi_2)$. The arrows indicate
  applications of the two-cycle IBP relation (\ref{2ptfinal}) to break the cycles, starting with
  $\CC_{(12)}(\xi_1)$.}
  \label{fig:22cyc}
\end{figure}

The last term on the right-hand side of \eqref{(12)(34)} requires extra
caution in view of the $f$-$\Omega$ chains $\Omega_{12}(\eta_2) f^{(1)}_{2i}$ 
with $i=3,4$ that are attached to the cycle $\CC_{(34)}(\xi_2)$ through $z_3$ or $z_4$.
The first of the resulting tadpoles $\Omega_{12}(\eta_2) f^{(1)}_{23} \CC_{(34)}(\xi_2)$
can still be broken by literally following \eqref{si1} since the $\nabla_4$-derivative 
in the last term does not interfere with $f^{(1)}_{23} $:
\ba
\Omega_{12}(\eta_2) f^{(1)}_{23} \nabla_{4}   \Omega _{34} (\eta_4)  =\nabla_{4} \big(\Omega_{12}(\eta_2) f^{(1)}_{23}  \Omega _{34} (\eta_4)  \big) \overset{\rm IBP}= 0\,.
\ea   
However, the second tadpole $\Omega_{12}(\eta_2) f^{(1)}_{24} \CC_{(34)}(\xi_2)$
necessitates a reformulation of (\ref{si1}) without reference to $\nabla_{4}$ since
  \ba
   f^{(1)}_{24} \nabla_{4}  \Omega _{34} (\eta_4)    \neq  \nabla_{4} \big(f^{(1)}_{24}  \Omega _{34} (\eta_4) \big) \,.
   \ea
Following a relabelling of \eqref{ex21}, we interchange the role of $z_3$ and $z_4$ 
and break the second cycle $\CC_{(34)}(\xi)$ via
   \be\label{si2}
 (1{+}s_{34})\CC_{(34)}(\xi_2)  = \MM_{34}(\xi_2)+ \Omega_{34}(\eta_{4}) x_{3,12}
  + \nabla_3   \Omega _{34} (\eta_4)  
\,,
 \ee
leading to 
 \be\label{si2si2}
f^{(1)}_{24} (1{+}s_{34})  \CC_{(34)}(\xi_2) = f^{(1)}_{24} \big( \MM_{34}(\xi_2)+\Omega_{34}(\eta_{4})x_{3,12} 
\big)+ \nabla_3\big( f^{(1)}_{24}  \Omega _{34} (\eta_4)  \big) 
\,.
 \ee
By assembling the results of the above IBP manipulations and discarding 
total derivatives such as $\nabla_3\big( f^{(1)}_{24}  \Omega _{34} (\eta_4)  \big) $
in (\ref{si2si2}), we conclude that
  \begin{align}\label{ofof}
(1{+}s_{12})(1{+}s_{34})\CC_{(12)}(\xi_1)\CC_{(34)} (\xi_2)
&\overset{\rm IBP}= 
\MM_{12}(\xi_1)\MM_{34}(\xi_2)   \\
& \ \ \quad  
-\MM_{12} (\xi_1)  x_{4,12} \Omega_{34}(\eta_{4}) 
- x_{2,34} \Omega_{12}(\eta_{2}) \MM_{34}(\xi_2) \notag \\
& \ \ \quad 
  +\big(  x_{2,3}x_{4,1} -    x_{2,4}x_{3,1} \big) \Omega_{12}(\eta_{2})\Omega_{34}(\eta_{4})\,.
\notag
\end{align}   
We note that the right-hand side manifests the symmetry of the left-hand side
under exchange of the two cycles $\CC_{(12)}(\xi_1)$ and $\CC_{(34)} (\xi_2)$, i.e.\ under  
$(z_1,z_2,\eta_2,\xi_1) \leftrightarrow (z_3,z_4,\eta_4,\xi_2)$. Accordingly,
the outcome of \eqref{ofof} would take the same form if the
cycles had been broken in reverse order, starting with 
$\CC_{(34)} (\xi_2)$ instead of $\CC_{(12)}(\xi_1)$.
A schematic representation of the sequential breaking of the
cycles carried out in this section can be found in figure \ref{fig:22cyc}.

\subsubsection{Basis decomposition of the single-cycle terms}

The two terms in the last line of (\ref{ofof}) feature new 
$f$-$\Omega$ cycles of length four and do not yet line up 
with the desired chain form. However, they arise from the
expansion coefficients of single cycles $\CC_{(1234)}(\xi)$ and $\CC_{(1243)}(\xi)$
whose F-IBP reduction to chains has already been accomplished
in section \ref{brkC1234}. This can be exploited by rewriting
$\Omega_{12} f^{(1)}_{23}\Omega_{34} f^{(1)}_{41}$ in the last
line of \eqref{ofof} as
\be
\Omega_{12}(\eta_2) f^{(1)}_{23}\Omega_{34} (\eta_4)f^{(1)}_{41}= \Omega_{12}(\eta_2)  \Omega_{23}(\zeta_2) \Omega_{34} (\eta_4) 
\Omega_{41}(\zeta_1)\Big|_{\zeta_1^0,\zeta_2^0}\, . 
\ee
By the prescription \eqref{4ptfinal} to break length-four cycles,
this simplifies to 
\begin{align}
&(1{+}s_{1234})\Omega_{12}(\eta_2)  \Omega_{23}(\zeta_2) \Omega_{34} (\eta_4) 
\Omega_{41}(\zeta_1) 
\\
&\ \overset{\rm IBP}=
\MM_{1234}(\xi)\big|_{\eta_{234}+\xi\to\eta_2, \,\eta_{34}+\xi \to\zeta_2, \,\eta_{4}+\xi\to\eta_4, \, \xi\to\zeta_1 } \notag \\
&\ \;  =:{\hat {\bm M}}_{1234}(\eta_2,\zeta_2,\eta_4,\zeta_1)   \, ,
\non
\end{align}
where $\MM_{1234}(\xi)$ is given by (\ref{mm1234}) and free of cycles. Hence, we arrive
at the following final result for the chain decomposition of the simplest double
cycle $\CC_{(12)}(\xi_1)\CC_{(34)}(\xi_2)$:
\begin{align}
&(1{+}s_{12})(1{+}s_{34})\CC_{(12)}(\xi_1)\CC_{(34)} (\xi_2)
\label{ofof2} \\
&\ \ 
\overset{\rm IBP}= 
\MM_{12}(\xi_1)\MM_{34}(\xi_2)-\MM_{12} (\xi_1) x_{4,12} \Omega_{34}(\eta_{4}) -   x_{2,34}  \Omega_{12}(\eta_{2}) \MM_{34}(\xi_2)
   \nonumber  \\
&\ \ \quad + \frac{  s_{23}s_{41}
   {\hat{\bm M}}_{1234}(\eta_2,\zeta_2,\eta_4,\zeta_1)+     s_{24}s_{31} {\hat{\bm M}}_{1243}(\eta_2,\zeta_2,-\eta_4,\zeta_1)
   }{1+s_{1234}} \Big|_{\zeta_1^0,\zeta_2^0} \,.
   \notag
\end{align}

 \subsubsection{Application to Kronecker-Eisenstein coefficients}
 
The applications of the above results to the Kronecker-Eisenstein coefficients
occurring in actual string integrands are straightforward.  For instance, we can use \eqref{ofof} to break the two $f$-cycles in the integrand $ V_2(1,2) V_2(3,4)$ of the non-planar four-point
one-loop amplitude in the gauge sector of the heterotic string \cite{Gerken:2018jrq},
\begin{align}\label{uniform31} 
&\hspace{-0.7cm}(1{+}s_{12})(1{+}s_{34}) V_2(1,2) V_2(3,4) 
\\
\overset{\rm IBP}= \,\, 
&
\MM_{12}(\xi_1)\big|\big|_{\eta_2^0,\xi_1^0} \MM_{34}(\xi_2)\big|\big|_{\eta_4^0,\xi_2^0} -\MM_{12} (\xi_1)\big|\big|_{\eta_2^0,\xi_1^0}  x_{4,12}  \Omega_{34}(\eta_{4}) \big|_{\eta_4^0}
   \nonumber
   \\
   & -    x_{2,34}  \Omega_{12}(\eta_{2}) \big|_{\eta_2^0} \MM_{34}(\xi_2)\big|\big|_{\eta_4^0,\xi_2^0}  +\big( 
   x_{2,3}x_{4,1} -    x_{2,4}x_{3,1} 
\big) \Omega_{12}(\eta_{2})\big|_{\eta_2^0}\Omega_{34}(\eta_{4})\big|_{\eta_4^0}\,,
\nl
&\hspace{-0.64cm} = {\rm \hat G}_2^2
 + {\rm \hat G}_2 \, \big( 2s_{12}f^{(2)}_{12}+2s_{34}f^{(2)}_{34} + 
s_{14} f^{(1)}_{14} f^{(1)}_{34} - s_{23} f^{(1)}_{12} f^{(1)}_{23}-s_{24} f^{(1)}_{12} f^{(1)}_{24} +s_{24}f^{(1)}_{24} f^{(1)}_{34} 
 \big)  \nonumber  \\
&  -2 s_{34} f_{34}^{(2)} f_{12}^{(1)} (s_{23}f^{(1)}_{23}+s_{24}f^{(1)}_{24})   +2 s_{12} f_{12}^{(2)} f_{34}^{(1)} (s_{24}f^{(1)}_{24}
+s_{14}f^{(1)}_{14})  + 4 s_{12}s_{34} f_{12}^{(2)} f^{(2)}_{34} 
\nl&
+ s_{13}s_{24} f^{(1)}_{12} f^{(1)}_{24}f^{(1)}_{43} f^{(1)}_{31}
+ s_{14}s_{23} f^{(1)}_{12} f^{(1)}_{23}f^{(1)}_{34} f^{(1)}_{41} 
\ .
\non
\end{align}
The length-four cycles $s_{13}s_{24} f^{(1)}_{12} f^{(1)}_{24}f^{(1)}_{43} f^{(1)}_{31}
$ and $s_{14}s_{23} f^{(1)}_{12} f^{(1)}_{23}f^{(1)}_{34} f^{(1)}_{41} $ in the last line can
be rewritten in terms of the elliptic $V_4$-function in (\ref{altVw}), $f^{(1)}_{12} f^{(1)}_{23}f^{(1)}_{34} f^{(1)}_{41}=V_{4}(1,2,3,4)+\ldots$, where the terms in the ellipsis are free of cycles.
Then, $V_{4}(1,2,3,4)$ can be decomposed into chains using \eqref{eqb1}, and the same
applies to its relabelling $V_{4}(1,2,4,3)$ due to $f^{(1)}_{12} f^{(1)}_{24}f^{(1)}_{43} f^{(1)}_{31}$ in (\ref{uniform31}). The outcome of this procedure is consistent with
the IBP reduction of the Koba-Nielsen integral of $V_2(1,2) V_2(3,4)$ in 
the literature and related to the results in appendix~D 
of \cite{Gerken:2018jrq} via Fay identities.

 \begin{figure}[h]
  \centering
  \begin {tikzpicture}[line width = 0.3mm, scale=1.15]
\draw(1+0,1)node{$\bullet$}node[above]{$n$} -- (1-0.867,0.5);
\draw(1-0.867,0.5)node{$\bullet$}node[left]{$3$} -- (1-0.867,-0.5)node{$\bullet$}node[left]{$4$};
\draw(1-0.867,-0.5) -- (1+0,-1)node{$\bullet$}node[below]{$5$};
\draw[dashed](1+0,-1) -- (1+0.867,-0.5);
\draw[dashed](1+0.867,0.5) -- (1+0.867,-0.5);
\draw[dashed](1+0.867,0.5) -- (1+0,1);
\draw(-.867,0.5) .. controls (-.867+0.2,0.2) and (-.867+0.2,-0.2) .. (-.867,-0.5);
\draw(-.867,0.5) .. controls (-.867-0.2,0.2) and (-.867-0.2,-0.2) .. (-.867,-0.5);
\draw(-.867,0.5)node{$\bullet$}node[above]{$1$}; 
\draw(-.867,-0.5)node{$\bullet$}node[below]{$2$}; 
\draw(3,0)node{$\longrightarrow$};
\draw(4,0)node{$\displaystyle \sum_{a=3}^n$};
\scope[xshift=5.5cm]
\draw(1+0,1)node{$\bullet$}node[above]{$a{-}1$} -- (1-0.867,0.5);
\draw(1-0.867,0.5)node{$\bullet$}node[above]{$a$} -- (1-0.867,-0.5)node{$\bullet$}node[below]{$a{+}1$};
\draw(1-0.867,-0.5) -- (1+0,-1)node{$\bullet$}node[below]{$a{+}2$};
\draw[dashed](1+0,-1) -- (1+0.867,-0.5);
\draw[dashed](1+0.867,0.5) -- (1+0.867,-0.5);
\draw[dashed](1+0.867,0.5) -- (1+0,1);
\draw(-0.867,0.5)  -- (-0.867,-0.5);
\draw(-0.867,0.5)node{$\bullet$}node[above]{$1$}; 
\draw(-0.867,-0.5)node{$\bullet$}node[below]{$2$}; 
\draw[dotted](-0.867,-0.5) -- (1-0.867,0.5);
\endscope
\draw(0,-3)node{$\longrightarrow$};
\draw(1,-3)node{$\displaystyle \sum_{3\leq a<b}^n$};
\scope[xshift=2cm,yshift=-3cm]
\draw(2+0,1)node{$\bullet$}node[above]{$a$} -- (2+0,0.5) ;
\draw(2+0,-1)node{$\bullet$}node[below]{$b$} -- (2+0,-0.5) ;
\draw[dashed](2+0,-0.5) -- (2+0,0.5) ;
\draw(2+0.3,0)node[rotate = 90]{$\underbrace{ \phantom{xxxxxxxxxx} }_{  }$};
\draw(2+0.9,0)node{$\Omega_{a,\rho,b}$};
\draw(2-2,0.5) -- (2-2,-0.5);
\draw(2-2,0.5)node{$\bullet$}node[above]{$1$}; 
\draw(2-2,-0.5)node{$\bullet$}node[below]{$2$}; 
%
\draw[dotted](2+0,1) -- (2-2,0.5);
\draw[dotted](2+0,-1) -- (2-2,-0.5);
\endscope
\draw(6,-3)node{and};
\scope[xshift=7cm,yshift=-3cm]
\draw(0+2,1)node{$\bullet$}node[above]{$a$} -- (0+2,0.5) ;
\draw(0+2,-1)node{$\bullet$}node[below]{$b$} -- (0+2,-0.5) ;
\draw[dashed](0+2,-0.5) -- (0+2,0.5) ;
\draw(+0.3+2,0)node[rotate = 90]{$\underbrace{ \phantom{xxxxxxxxxx} }_{  }$};
\draw(+0.9+2,0)node{$\Omega_{a,\rho,b}$};
\draw(2-2,0.5) -- (2-2,-0.5);
\draw(2-2,0.5)node{$\bullet$}node[above]{$1$}; 
\draw(2-2,-0.5)node{$\bullet$}node[below]{$2$}; 
%
\draw[dotted](0+2,1) -- (2-2,-0.5);
\draw[dotted](0+2,-1) -- (2-2,0.5);
\endscope
 \end {tikzpicture}
  \caption{Graphical representation of the products of the Kronecker-Eisenstein
  series (solid lines) and factors of $f^{(1)}_{ij}$ (dotted lines) in the chain decomposition
  of the product $ \CC_{(12)}(\xi_1)\CC_{(3\ldots n)}(\xi_2)$. The arrows indicate
  applications of IBP relations (\ref{2ptfinal}) and (\ref{rhorho}) to break the cycles.}
  \label{fig:2mcycAlt2}
\end{figure}
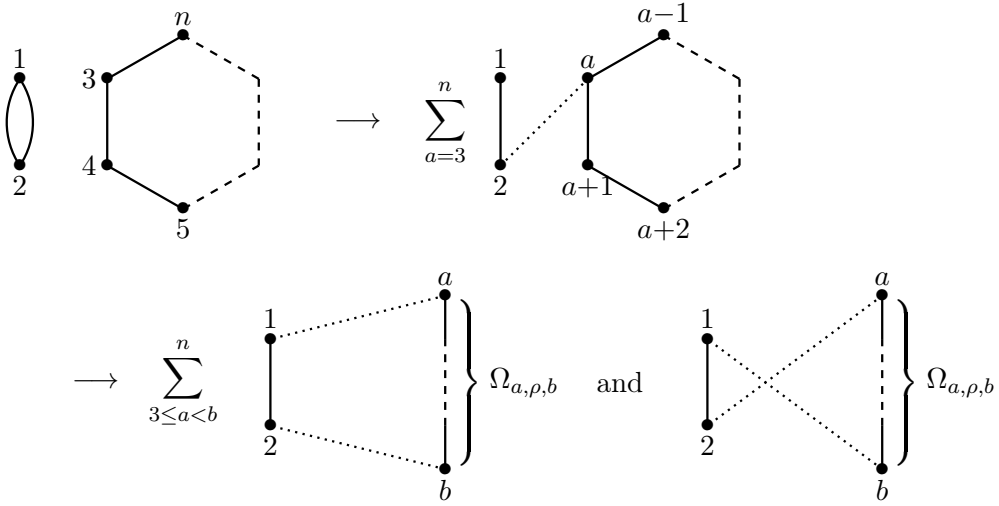

\subsection{Two cycles of length $2$ and $m$}
\label{sec:2cyc.2}

We shall now generalize the IBP reduction in the previous section
to more general products $\CC_{(12)} (\xi_1) \CC_{(34\cdots n)}(\xi_2)$, where the
length $m$ of one cycle is arbitrary, in this case $m=n{-}2\geq2$. 
 As schematically shown in figure \ref{fig:2mcycAlt2}, we first break the length-two cycle,
 \begin{align}
  &(1{+}s_{12}) \CC_{(12)}(\xi_1) \CC_{(34\cdots n)}(\xi_2)
\overset{\rm IBP}=
\MM_{12}(\xi_1)\CC_{(34\cdots n)}(\xi_2)  
  -     \Omega_{12}(\eta_2)   
\CC_{(34\cdots n)}(\xi_2) \sum_{a=3}^n x_{2,a}  
\,.  \label{newcyc23pre}  
 \end{align}
For the first term on the right-hand side, we can break the longer cycle by directly substituting \eqref{rhorho} with a relabeling $\{1{\to}3,\, 2{\to}4,\, \cdots,\, m{\to} n{=}m{+}2\}$. However, for the last term in (\ref{newcyc23pre}), we need to break the cycle $\CC_{(34\cdots n)}(\xi_2) $ in different ways based on the attachment points $a \in \{3,4,\cdots,n\}$ of $x_{2,a}  $. In this case, it is convenient to use the reorganized formula
\eqref{rhorhog} with a relabeling $\{1{\to}3,\, \cdots,\,m{\to} m{+}2\}$,
 \allowdisplaybreaks[0]
 \begin{align}
& x_{2,a}  ( 1{+}s_{34\cdots n}) \CC_{(34\cdots n)}
(\xi_2) \overset{\rm IBP}
= \,\, x_{2,a}  \MM_{34\cdots n} (\xi_2) \\
&\qquad -
  \sum_{\substack{
b  
 =3 \\ b\neq a}
}^{n} \sum_{
\substack{
 \rho \in  A \shuffle  B^{\rm T}
 \\
 (a, A, b, B )=(3,4, \ldots, n) } }
 (-1)^{|B|} 
x_{2,a} {\bm \Omega}_{a,\rho,b} x_{b,12}
 \,. \notag
 \end{align}
 The terms $x_{2,a} {\bm \Omega}_{a,\rho,b} x_{b,2} $ cancel each other when we sum over all $a,b \in \{3,4,\cdots,n\}$ with $a\neq b$ as prescribed by (\ref{newcyc23pre}), and we obtain
 \begin{align}
 \label{newcyc23pref}
  &(1{+}s_{12}) ( 1{+}s_{34\cdots n})  
  \CC_{(12)}(\xi_1)\CC_{(34\cdots n)}(\xi_2)
   \overset{\rm IBP}=  
 \MM_{12}(\xi_1) \MM_{34\cdots n}(\xi_2)
\\
&\quad
-\MM_{34\cdots n} (\xi_2) \Omega_{12}(\eta_2)   \sum_{a=3}^n x_{2,a}   
- \MM_{12} (\xi_1) \sum_{ b  =4}^{n}
\!\!\!\!
\!\!\!\!
\!\!\!\!
 \sum_{\substack{\rho \in \{4,\cdots, b-1\}\\
\qquad\quad \shuffle \{n,n-1,\cdots, b+1 \}} }
\!\!\!\!\!\!\!\!\!\!\!\!\!\!\!\!\!\!\!\!
 (-1)^{n-b} 
{\bm \Omega}_{3,\rho,b}   x_{b,12} 
\nl
&\quad  + \Omega_{12}(\eta_2)  \sum_{  3\leq a< b}^n 
\sum_{
\substack{
 \rho \in  A \shuffle  B^{\rm T}
 \\
 (a, A, b, B )=(3,4, \ldots, n) } } 
 (-1)^{|B|} \, {\bm \Omega}_{a,\rho,b}   
(x_{2,a}\, x_{b,1} -x_{2,b}\, x_{a,1} ) 
\,. \notag
 \end{align}
The first two lines are free of cycles, and the last line contains $f$-$\Omega$ cycles
$ \Omega_{12}(\eta_2) x_{2,a} {\bm \Omega}_{a,\rho,b}   x_{b,1} $ of length $n$ which 
we already know how to break. Again, reversing the order
of breaking the cycles $\CC_{(12)}(\xi_1)$ and $\CC_{(34\cdots n)}(\xi_2)$
does not change the form of the outcome of the F-IBP reduction. When specializing
the length of the second cycle to $m=2$, the ordered sets $\rho$ in the
last two lines of (\ref{newcyc23pref}) are empty, and we reproduce \eqref{ofof}. 
Additional examples at $m=3,4$ will be given below.
 
\subsubsection{Example at $n=5$ points}

For cycles of length two and $m=3$, (\ref{newcyc23pref}) reduces to
\begin{align}
&  (1{+}s_{12}) (1{+}s_{345}) \CC_{(12)}(\xi_1)\CC_{(345)} (\xi_2)
   \overset{\rm IBP}= \,\,
\MM_{12}(\xi_1)\MM_{345}(\xi_2)  \label{cyc23} \\
&\qquad -  \Omega_{12}(\eta_{2}) \MM_{345}(\xi_2) x_{2,345} 
-\MM_{12}(\xi_1) \big(
   {\bm \Omega}_{345}  x_{5,12} 
      -   
   {\bm \Omega}_{354} x_{4,12}
\big) 
\nonumber\\
&\qquad 
+\Omega_{12} (\eta_2) \big(   \left(x_{2,3}x_{5,1} -x_{2,5}x_{3,1}\right) {\bm \Omega}_{345}
  +  \left(x_{2,4}x_{3,1} -x_{2,3}x_{4,1}  \right) {\bm \Omega}_{354}    
     \nonumber
   \\
   &\qquad \quad\quad\quad\quad + 
   \left(x_{2,5}x_{4,1}-x_{2,4}x_{5,1}  \right)  {\bm \Omega}_{435}  \big)
\non
 \end{align}
which can for instance be applied to simplify the integrand
$V_2(1,2) V_3(3,4,5)$ of five-point heterotic-string amplitudes.

\subsubsection{Example at $n=6$ points}

For cycles of length two and $m=4$, (\ref{newcyc23pref}) reduces to 
\begin{align}
 &(1{+}s_{12})(1{+}s_{3456})  
 \CC_{(12)} (\xi_1)\CC_{(3456)}(\xi_2)
\overset{\rm IBP}= \,\,
\MM_{12}(\xi_1)  \MM_{3456}(\xi_2) \label{cyc24} \\
  &\quad - \MM_{3456}(\xi_2)  x_{2,3456} \Omega_{12}(\eta_{2})
-\big(
{\bm \Omega} _{3456}  x_{6,12}
      +  
{\bm \Omega} _{3654}  x_{4,12} 
-
\big({\bm \Omega} _{3465} {+}{\bm \Omega} _{3645} \big) x_{5,12} 
\big)  \MM_{12}(\xi_1)
\nonumber\\
&\quad
 + \Omega_{12}(\eta_{2})\big( 
 \left( x_{2,3}x_{6,1}-x_{2,6}x_{3,1} \right) {\bm \Omega} _{3456}
+\left(
  x_{2,3} x_{4,1}- x_{2,4} x_{3,1} \right) {\bm \Omega} _{3654}
 \nonumber
   \\
   &\quad\quad\quad
   - \left(x_{2,3}x_{5,1} -x_{2,5}x_{3,1} \right)  ({\bm \Omega} _{3465} +{\bm \Omega} _{3645})
 +  \left(x_{2,4}x_{5,1} -x_{2,5}x_{4,1} \right) {\bm \Omega} _{4365}  
   \nonumber
   \\
   &\quad\quad\quad
     + \left(x_{2,5}x_{6,1} -x_{2,6}x_{5,1} \right) {\bm \Omega _{5436}}
  - \left( x_{2,4}x_{6,1}- x_{2,6}x_{4,1}\right)  ({\bm \Omega}_{4356}+{\bm \Omega}_{4536})
   \big)
   \nonumber
 \end{align}
which can for instance be applied to simplify the integrand
$V_2(1,2) V_4(3,4,5,6)$ of six-point heterotic-string amplitudes.

 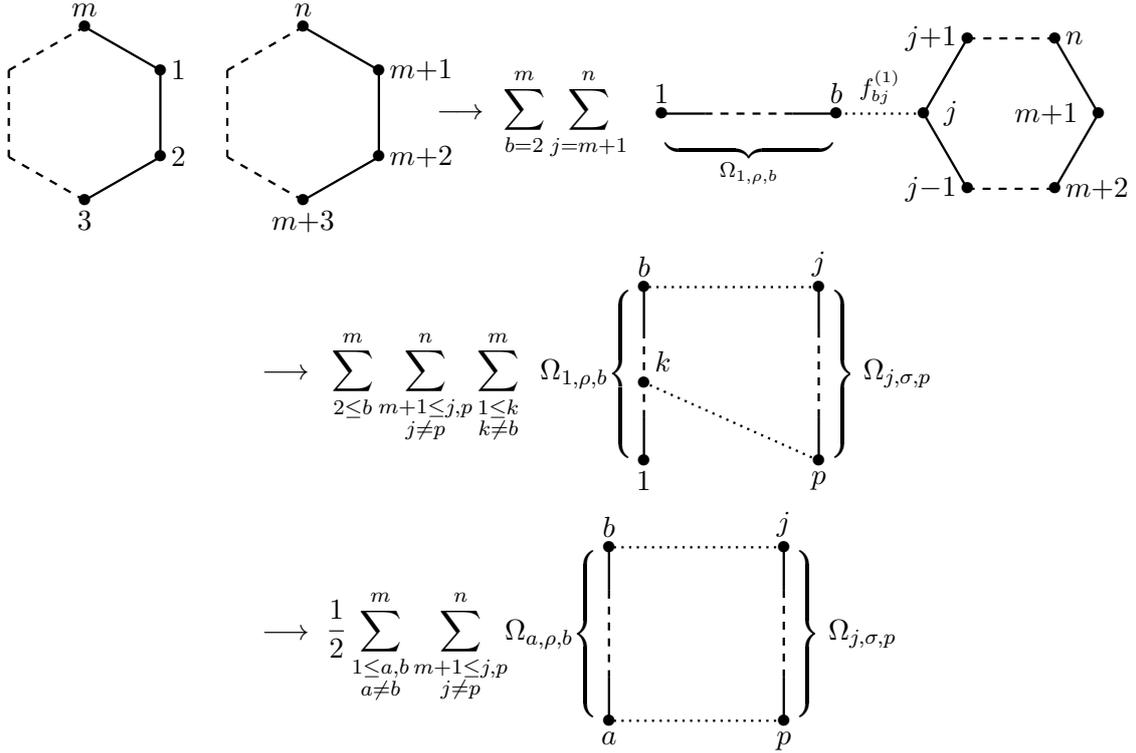
\begin{figure}[h]
  \centering
  \begin {tikzpicture}[line width = 0.3mm, scale=1.15]
\draw(0,1)node{$\bullet$}node[above]{$m$} -- (0.867,0.5);
\draw(0.867,0.5)node{$\bullet$}node[right]{$1$} -- (0.867,-0.5)node{$\bullet$}node[right]{$2$};
\draw(0.867,-0.5) -- (0,-1)node{$\bullet$}node[below]{$3$};
\draw[dashed](0,-1) -- (-0.867,-0.5);
\draw[dashed](-0.867,0.5) -- (-0.867,-0.5);
\draw[dashed](-0.867,0.5) -- (0,1);
\scope[xshift=2.5cm]
\draw(0,1)node{$\bullet$}node[above]{$n$} -- (0.867,0.5);
\draw(0.867,0.5)node{$\bullet$}node[right]{$m{+}1$} -- (0.867,-0.5)node{$\bullet$}node[right]{$m{+}2$};
\draw(0.867,-0.5) -- (0,-1)node{$\bullet$}node[below]{$m{+}3$};
\draw[dashed](0,-1) -- (-0.867,-0.5);
\draw[dashed](-0.867,0.5) -- (-0.867,-0.5);
\draw[dashed](-0.867,0.5) -- (0,1);
\endscope
\draw(4.3,0)node{$\longrightarrow$};
\draw(5.5,0)node{$\displaystyle \sum_{b=2}^m \sum_{j=m+1}^n$};
\scope[xshift=10.6cm]
\draw(-1,0) -- (-0.5,0.867);
\draw[dashed](-0.5,0.867) -- (0.5,0.867);
\draw(0.5,0.867) -- (1,0);
\draw(1,0) -- (0.5,-0.867);
\draw[dashed](0.5,-0.867)--(-0.5,-0.867);
\draw(-0.5,-0.867) -- (-1,0);
\draw(-1.5,0.3)node{\footnotesize $f^{(1)}_{bj}$};
\draw[dotted](-1,0) -- (-2,0);
\draw(-2,0)node{$\bullet$}node[above]{$b$} -- (-2.5,0);
\draw[dashed](-2.5,0) -- (-3.5,0);
\draw(-3.5,0) -- (-4,0)node{$\bullet$}node[above]{$1$};
\draw(-3,-0.5)node{$\underbrace{ \phantom{xxxxxxxxxx} }_{ \Omega_{1,\rho,b} }$};
\draw (-1,0)node{$\bullet$}node[right]{$\ j$};
\draw (-0.5,0.867) node{$\bullet$}node[left]{$j{+}1$};
\draw (0.5,0.867) node{$\bullet$}node[right]{$n$};
\draw (1,0) node{$\bullet$}node[left]{$m{+}1\ $};
\draw (0.5,-0.867) node{$\bullet$}node[right]{$m{+}2$};
\draw (-0.5,-0.867) node{$\bullet$}node[left]{$j{-}1$};
\endscope
%
%
%
\draw(0-1-.7+4,-3)node{$\longrightarrow$};
\draw(1-.5-.7+4.1,-3-0.14)node{$\displaystyle \sum_{2\leq b}^m \sum_{\substack{m+1\leq j,p\\j\neq p}}^n \sum_{\substack{1\leq k\\ k\neq b }}^m $ };
\scope[xshift=6.4cm,yshift=-3cm]
\draw(2+0,1)node{$\bullet$}node[above]{$j$} -- (2+0,0.5) ;
\draw(2+0,-1)node{$\bullet$}node[below]{$p$} -- (2+0,-0.5) ;
\draw[dashed](2+0,-0.5) -- (2+0,0.5) ;
\draw(2+0.3,0)node[rotate = 90]{$\underbrace{ \phantom{xxxxxxxxxx} }_{  }$};
\draw(2+0.9,0)node{$\Omega_{j,\sigma,p}$};
\draw[dashed](2-2,0.5) -- (2-2,-0.5);
\draw(2-2,0.5) -- (2-2,1);
\draw(2-2,-0.5) -- (2-2,-1);
\draw(2-2,1)node{$\bullet$}node[above]{$b$}; 
\draw(2-2,-1)node{$\bullet$}node[below]{$1$}; 
\draw(2-2-.6+0.3,0)node[rotate = -90]{$\underbrace{ \phantom{xxxxxxxxxx} }_{  }$};
\draw(2-2-1.7+0.9,0)node{$\Omega_{1,\rho,b}$};
%
\draw[dotted](2+0,1) -- (2-2,1);
\draw[dotted](2+0,-1) -- (2-2,-.1)node{$\bullet$}node[above right ]{$k$} ;
\endscope
\draw(0-1-.7+4,-6)node{$\longrightarrow$};
\draw(1-.5-.7+4,-6-0.14)node{$\displaystyle \frac{1}{2} \sum_{\substack{1\leq a,b
\\
a\neq b}}^m \sum_{\substack{m+1\leq j,p\\ j\neq p}}^n$};
\scope[xshift=6cm,yshift=-6cm]
\draw(2+0,1)node{$\bullet$}node[above]{$j$} -- (2+0,0.5) ;
\draw(2+0,-1)node{$\bullet$}node[below]{$p$} -- (2+0,-0.5) ;
\draw[dashed](2+0,-0.5) -- (2+0,0.5) ;
\draw(2+0.3,0)node[rotate = 90]{$\underbrace{ \phantom{xxxxxxxxxx} }_{  }$};
\draw(2+0.9,0)node{$\Omega_{j,\sigma,p}$};
\draw[dashed](2-2,0.5) -- (2-2,-0.5);
\draw(2-2,0.5) -- (2-2,1);
\draw(2-2,-0.5) -- (2-2,-1);
\draw(2-2,1)node{$\bullet$}node[above]{$b$}; 
\draw(2-2,-1)node{$\bullet$}node[below]{$a$}; 
\draw(2-2-.6+0.3,0)node[rotate = -90]{$\underbrace{ \phantom{xxxxxxxxxx} }_{  }$};
\draw(2-2-1.7+0.9,0)node{$\Omega_{a,\rho,b}$};
%
\draw[dotted](2+0,1) -- (2-2,1);
\draw[dotted](2+0,-1) -- (2-2,-1);
\endscope
 \end {tikzpicture}
  \caption{Graphical representation of the products of the Kronecker-Eisenstein
  series (solid lines) and factors of $f^{(1)}_{ij}$ (dotted lines) in the chain decomposition
  of the product $ \CC_{(12\cdots m)}(\xi_1)\CC_{(m+1,m+2\ldots n)}(\xi_2)$. The initial two arrows demonstrate the use of IBP relations, specifically those in (\ref{newcyc23predouble}) and (\ref{newcyc23pre2}), to break the cycles. The concluding arrows elucidate the simplification process that leads to (\ref{fusion2}). }
  \label{fig:mncycs}
\end{figure}

\subsection{Two cycles of general length}
\label{sec:2cyc.3}

As a further generalization of the previous F-IBP reductions, we shall now
consider the most general product $\CC_{(12\cdots m)}(\xi_1)\CC_{(m+1,m+2,\cdots, n)}(\xi_2)$ of two cycles of arbitrary length $m$ and $n{-}m\geq 2$. As schematically shown in figure \ref{fig:mncycs}, we start by breaking $\CC_{(12\cdots m)}(\xi_1)$,
  \begin{align}
  &(1{+}s_{12\cdots m}) \CC_{(12\cdots m)}(\xi_1)\CC_{(m+1,m+2,\cdots, n)}(\xi_2)
 \overset{\rm IBP}= \,\, 
\MM_{12\cdots m}(\xi_1)\CC_{(m+1,m+2,\cdots, n)}(\xi_2)
 \label{newcyc23predouble}
 \\
 &\quad\quad\quad\quad\quad\quad\quad\quad - \sum_{b  =2}^{m}
\!\!\!\!
\!\!\!\!
\!\!\!\!
 \sum_{\substack{\rho \in \{2,3,\cdots, b-1\}\\
\qquad\quad \shuffle \{m,m-1,\cdots, b+1 \}} }
\!\!\!\!\!\!\!\!\!\!\!\!\!\!\!\!\!\!\!\!
 (-1)^{m-b} 
{\bm \Omega}_{1,\rho,b}  \sum_{j=m+1}^n  x_{b,j} \CC_{(m+1,m+2,\cdots, n)}  (\xi_2)
\,.
\non
 \end{align}
The next step is to break the second cycle $\CC_{(m+1,m+2,\cdots, n)}(\xi_2)$ according the attachment point $j \in \{m{+}1,\ldots,n\}$ of the factors $x_{b,j}$ in the second line,
 \begin{align}
 x_{b,j}  ( 1{+}s_{m+1,m+2,\cdots n}) &\CC_{(m+1,m+2,\cdots, n)}(\xi_2)
 \overset{\rm IBP}= \,\,  x_{b,j}  \MM_{m+1,m+2,\cdots n}(\xi_2) \\ 
 \non
& -  \sum_{\substack{p  =m+1 \\ p\neq j}
}^{n} \sum_{
\substack{
\sigma \in  X \shuffle  Y^{\rm T}
\\
(j, X, p, Y )=(m{+}1,m{+}2, \ldots, n)
} }
\! \! \! \! \! (-1)^{|Y|} 
x_{b, j} {\bm \Omega}_{j,\sigma,p} \sum_{\substack{k=1  \\ k\neq b}}^m x_{p, k} 
 \,.
 \end{align}
 Combining the above two equations, we get
  \begin{align}
  & ( 1{+}s_{12\cdots m}) ( 1{+}s_{m+1 \cdots n})  \CC_{(12\cdots m)}(\xi_1)\CC_{(m+1, \cdots, n)} (\xi_2)
\overset{\rm IBP}= \,\,
\MM_{12\cdots m}(\xi_1)
\MM_{m+1\cdots n}(\xi_2) \notag \\
&\qquad \qquad -\MM_{12\cdots m} (\xi_1)
\sum_{
p  =m+2}^{n}
~
 \sum_{\substack{\sigma \in \{m+2,m+3,\cdots, p-1\}\\
\quad\shuffle \{n,n-1,\cdots, p+1 \}} }
\!\!\!\!\!\!\!\!\!\!\!\!\!
 (-1)^{n-m-p} 
{\bm \Omega}_{m+1,\sigma,p}  \sum_{k=1}^m  x_{p,k} 
 \label{newcyc23pre2} \\
&\qquad \qquad -\MM_{m+1,m+2,\cdots, n}(\xi_2)
\sum_{
b  =2}^{m}
\!\!\!\!
\!\!\!\!
\!\!\!\!
 \sum_{\substack{\rho \in \{2,3,\cdots, b-1\}\\
\qquad\quad \shuffle \{m,m-1,\cdots, b+1 \}} }
\!\!\!\!\!\!\!\!\!\!\!\!\!\!\!\!\!\!\!\!
 (-1)^{m-b} 
{\bm \Omega}_{1,\rho,b}  \sum_{j=m+1}^n  x_{b,j} 
\nl
&\qquad \qquad  +   
\sum_{
b  =2}^{m}
\!\!\!\!
\!\!\!\!
\!\!\!\!
 \sum_{\substack{\rho \in \{2,3,\cdots, b-1\}\\
\qquad\quad \shuffle \{m,m-1,\cdots, b+1 \}} }
\!\!\!\!\!\!\!\!\!\!\!\!\!\!\!\!\!\!\!\!
 (-1)^{m-b} 
{\bm \Omega}_{1,\rho,b} 
\!\!\!\!
\sum_{\substack{
j, p 
 =m+1 \\ j\neq p}
}^{n} 
\sum_{
\substack{
\sigma \in  X \shuffle  Y^{\rm T}
\\
(j, X, p, Y )=(m{+}1,m{+}2, \ldots, n)} }
\!\!\!\!
\!\!\!\!
\!\!\!\!
\!\!\!\!
 (-1)^{|Y|} 
x_{b,j} {\bm \Omega}_{j,\sigma,p} \sum_{\substack{k=1\\k\neq b }}^m x_{p, k} 
\,.
\non
 \end{align}
The first three lines are readily seen to be free of cycles and symmetric under exchange of
$\CC_{(12\cdots m)}(\xi_1)$ and $\CC_{(m+1,m+2,\cdots, n)}(\xi_2)$. However, this exchange
symmetry is not manifest in the last line: New $f$-$\Omega$ cycles are formed, with lengths ranging from $n{-}m{+}1$ to $n$, and they may have Kronecker-Eisenstein chains attached to them. In order to simplify the last line of (\ref{newcyc23pre2}) and expose its symmetries, we exchange the summation variables $b$ and $k$ and subsequently take their average.
By virtue of \eqref{fay2identity}, we rewrite the last line of \eqref{newcyc23pre2}  as
\begin{align}
&\sum_{\substack{a,b  =1 \\ a< b}}^{m} \,
  \sum_{\substack{
j, p 
 =m+1 \\ j\neq p}
}^{n} 
\sum_{
\substack{
\rho \in  A \shuffle  B^{\rm T}
\\
(a, A, b, B )=(1,2, \ldots, m)
} }
\sum_{
\substack{
\sigma \in  X \shuffle  Y^{\rm T}
\\
(j, X, p, Y )=(m{+}1,m{+}2, \ldots, n)
} }
 \! \! \! \! \!  (-1)^{|B|+|Y|}  {\bm \Omega}_{a,\rho,b} \,
x_{b,j} {\bm \Omega}_{j,\sigma,p}  \, x_{p, a} 
 \label{fusion2} \\
&=
\frac12
\sum_{\substack{
a,b  
 =1 \\ a\neq b}
}^{m} \,
  \sum_{\substack{
j, p 
 =m+1 \\ j\neq p}
}^{n} 
\sum_{
\substack{
\rho \in  A \shuffle  B^{\rm T}
\\
(a, A, b, B )=(1,2, \ldots, m)
} }
\sum_{
\substack{
\sigma \in  X \shuffle  Y^{\rm T}
\\
(j, X, p, Y )=(m{+}1,m{+}2, \ldots, n)
} }
 \! \! \! \! \!  (-1)^{|B|+|Y|}  {\bm \Omega}_{a,\rho,b} \,
x_{b,j} {\bm \Omega}_{j,\sigma,p}  \, x_{p, a} \,.
\nonumber
\end{align}
Hence, our final formula to break two cycles of arbitrary length is given by
  \begin{align}
  & ( 1{+}s_{12\cdots m}) ( 1{+}s_{m+1 \cdots n})  \CC_{(12\cdots m)}(\xi_1)\CC_{(m+1 \cdots, n)} (\xi_2)
\overset{\rm IBP}= 
\MM_{12\cdots m}(\xi_1)\MM_{m+1 \cdots, n}(\xi_2) \notag \\
&\qquad
-\MM_{12\cdots m} (\xi_1)
\sum_{
p  =m+2}^{n}
~
 \sum_{\substack{\sigma \in \{m+2,m+3,\cdots, p-1\}\\
\quad\shuffle \{n,n-1,\cdots, p+1 \}} }
\!\!\!\!\!\!\!\!\!\!\!\!\!
 (-1)^{n-m-p} 
{\bm \Omega}_{m+1,\sigma,p}  \sum_{k=1}^m  x_{p,k} 
\label{newcyc23pre22} \\
&\qquad -\MM_{m+1,m+2,\cdots, n}(\xi_2)
\sum_{
b  =2}^{m}
\!\!\!\!
\!\!\!\!
\!\!\!\!
 \sum_{\substack{\rho \in \{2,3,\cdots, b-1\}\\
\qquad\quad \shuffle \{m,m-1,\cdots, b+1 \}} }
\!\!\!\!\!\!\!\!\!\!\!\!\!\!\!\!\!\!\!\!
 (-1)^{m-b} 
{\bm \Omega}_{1,\rho,b}  \sum_{j=m+1}^n  x_{b,j} 
\nl
&\qquad  +
  \frac12
\sum_{\substack{
a,b  
 =1 \\ a\neq b}
}^{m} \,
  \sum_{\substack{
j, p 
 =m+1 \\ j\neq p}
}^{n} \sum_{
\substack{
\rho \in  A \shuffle  B^{\rm T}
\\
(a, A, b, B )=(1,2, \ldots, m)
} }
\sum_{
\substack{
\sigma \in  X \shuffle  Y^{\rm T}
\\
(j, X, p, Y )=(m{+}1,m{+}2, \ldots, n)
} }
\!\!\!\!
\!\!\!\!
\!\!\!\!
 (-1)^{|B|+|Y|}  {\bm \Omega}_{a,\rho,b} \,
x_{b,j} {\bm \Omega}_{j,\sigma,p}  \, x_{p, a}
\,,
\non
 \end{align}
where the $f$-$\Omega$ cycles of length $n$ in the last line can be broken by
isolating suitable components in the Laurent expansion of \eqref{rhorho}. 
As indicated by the equivalence relation $\overset{\rm IBP}=$, total Koba-Nielsen derivatives 
have been discarded on the right-hand side whose explicit form can be
found in appendix \ref{sec:2cyc.4}.
As before, the right-hand side of \eqref{newcyc23pre22} would take the same form
if the cycles had been broken in reversed order.

At length $m=2$, the sets $\rho$ in the last two lines of (\ref{newcyc23pre22}) are empty, which implies that ${\bm \Omega}_{1,\rho,b}$ in the third line becomes ${\Omega}_{12}(\eta_2)$, and ${\bm \Omega}_{a,\rho,b}$ in the last line becomes $\pm {\Omega}_{12}(\eta_2)$. Therefore, we can see how the specialization of \eqref{newcyc23pre22} to $m=2$ reproduces \eqref{newcyc23pref}. Applications 
 of \eqref{newcyc23pre22} to products $V_m(1,2,\cdots,m) V_{n-m}(m{+}1, \cdots, n)$ in
actual string integrands are straightforward, see \eqref{uniform31} for an example at
$(m,n)=(2,4)$.

\subsubsection{Example with two cycles of length three}

The simplest example of our general result (\ref{newcyc23pre22}) for
products of two cycles that has not been covered in section \ref{sec:2cyc.2}
is the IBP reduction of $\CC_{(123)} (\xi_1)\CC_{(456)}(\xi_2)$
at $n=6$ points,
 \begin{align}
 &(1{+}s_{123})(1{+}s_{456})\CC_{(123)} (\xi_1)\CC_{(456)}(\xi_2)
 \overset{\rm IBP}= 
  \MM_{123}(\xi_1)\MM_{456}(\xi_2)   \label{cyc33omega} \\
  &\quad - \MM_{123} (\xi_1) \big(
 {\bm \Omega_{456}} x_{6,123}-  {\bm \Omega}_{465} x_{5,123} \big)
-\big(
{\bm \Omega}_{123}  x_{3,456}
      - 
{\bm \Omega}_{132}  x_{2,456} 
\big)  \MM_{456}(\xi_2)
\notag \\
\nonumber
 &\quad  +{\bm\Omega}_{123} {\bm\Omega}_{456} \left(x_{1,4}
   x_{3,6}-x_{1,6} x_{3,4}\right)
  +{\bm\Omega}_{123} {\bm\Omega}_{465}
   \left(x_{1,5} x_{3,4}-x_{1,4}
   x_{3,5}\right) 
   \\ & \quad \nonumber 
   +{\bm\Omega}_{123} {\bm\Omega}_{546}
   \left(x_{1,6} x_{3,5}-x_{1,5}
   x_{3,6}\right)  +
   {\bm\Omega}_{132}
   {\bm\Omega}_{456} \left(x_{1,6} x_{2,4}-x_{1,4}
   x_{2,6}\right)
   \\ & \quad \nonumber 
   +{\bm\Omega}_{132} {\bm\Omega}_{546}
   \left(x_{1,5} x_{2,6}-x_{1,6}
   x_{2,5}\right)+{\bm\Omega}_{132} {\bm\Omega}_{465}
   \left(x_{1,4} x_{2,5}-x_{1,5}
   x_{2,4}\right)
      \\ & \quad \nonumber+{\bm\Omega}_{213} {\bm\Omega}_{456}
   \left(x_{2,6} x_{3,4}-x_{2,4}
   x_{3,6}\right)+{\bm\Omega}_{213} {\bm\Omega}_{465}
   \left(x_{2,4} x_{3,5}-x_{2,5}
   x_{3,4}\right)
     \\ & \quad \nonumber
     +{\bm\Omega}_{213} {\bm\Omega}_{546}
   \left(x_{2,5} x_{3,6}-x_{2,6} x_{3,5}\right)  \,.
 \end{align}

\subsection{Two $F$-cycles}
\label{sec:2cyc.5}

The procedure to break two meromorphic $F$-cycles in (\ref{defcf}) is similar to that of $\Omega$-cycles, with the additional consideration of terms involving $\ell{\cdot} k_b$ when applying \eqref{rhorhogf}. We shall first demonstrate this with a four-point example and
then state a general result for arbitrary lengths of the two cycles.

For a product of two length-two cycles,
$\CCF_{(12)}(\xi_1)= F_{12}(\eta_{2}{+}\xi_1)F_{21}(\xi_1)$
and $\CCF_{(34)}(\xi_2)=F_{34}(\eta_{4}{+}\xi_2)F_{43}(\xi_2)$
at $n=4$ points, we start by breaking the $F$-cycle $\CCF_{(12)}(\xi_1)$ via (\ref{2ptfinalF}),
\begin{align}
&(1{+}s_{12})\CCF_{(12)}(\xi_1)\CCF_{(34)} (\xi_2)
=  \MMF_{12}(\xi_1) \CCF_{(34)} (\xi_2)
\label{(12)(34)F1}\\
\nonumber
&\quad +  ( \ell {\cdot} k_2 - {\tilde  x}_{2,34})  F_{12}(\eta_2)  \CCF_{(34)} (\xi_2)
- {\tilde\nabla}_2  \big( F_{12}(\eta_2)  \CCF_{(34)} (\xi_2) \big)\,,\notag
   \end{align}
where $\tilde x_{i,j}$ is defined by (\ref{basebefore}). 
We then proceed to breaking the second cycle in two different ways,
depending on the attachment points $i=3,4$ of the products ${\tilde  x}_{2,i}\CCF_{(34)} (\xi_2) $,
   \begin{align}
(1{+}s_{12})&(1{+}s_{34})\CCF_{(12)}(\xi_1)\CCF_{(34)} (\xi_2)
=  \big(\MMF_{12}(\xi_1)
+F_{12}(\eta_2)  (\ell{\cdot} k_2-  {\tilde  x}_{2,3}) -{\tilde\nabla}_2  F_{12}(\eta_2) 
 \big) 
\notag \\
 &\quad\quad\quad\quad\quad\quad \quad \quad\quad \quad \quad \quad   \times
  \big(\MMF_{34}(\xi_2) 
  +F_{34}(\eta_4)  (\ell{\cdot} k_4-{\tilde  x}_{4,12})  -{\tilde\nabla}_4  F_{34}(\eta_4) \big) \nl
&\quad \quad\quad \quad\quad   - F_{12}(\eta_2)  {\tilde  x}_{2,4}   \big(\MMF_{34}(\xi_4) 
+ F_{34}(\eta_4) ( {\tilde  x}_{3,12}-   \ell{\cdot} k_3)
   +{\tilde\nabla}_3  F_{34}(\eta_4) \big)  \,. \label{(12)(34)F2}
   \end{align}
This can be rewritten as follows 
   \begin{align}
&(1{+}s_{12})(1{+}s_{34})\CCF_{(12)}(\xi_1)\CCF_{(34)} (\xi_2)
= \MMF_{12}(\xi_1)\MMF_{34}(\xi_2) \label{twopairsF}
\\
&\quad -\MMF_{12} (\xi_1)  {\tilde  x}_{4,12} F_{34}(\eta_{4}) -   {\tilde  x}_{2,34} F_{12}(\eta_{2}) \MMF_{34}(\xi_2)
   \nonumber \\
&\quad +\big(  {\tilde  x}_{2,3}{\tilde  x}_{4,1} -    {\tilde  x}_{2,4}{\tilde  x}_{3,1} \big) F_{12}(\eta_{2})F_{34}(\eta_{4})
\nl
&\quad +   \MMF_{12} (\xi_1)  F_{34}(\eta_{4}) \,\ell \!\cdot\! k_{4} +  \MMF_{34}  (\xi_2)  F_{12}(\eta_{2})   \,\ell \!\cdot\! k_{2} 
+ F_{12}(\eta_{2})   F_{34}(\eta_{4})   \,\ell\! \cdot\! k_{2}\,   \,\ell\! \cdot\!  k_{4} 
\nl
&\quad + F_{12}(\eta_{2})  F_{34}(\eta_{4}) \big(  {\tilde  x}_{2,4} \,\ell \!\cdot\! k_{3} 
- {\tilde  x}_{4,12}  \,\ell\! \cdot\!  k_{2} - {\tilde  x}_{2,3}   \,\ell\! \cdot\!   k_{4} \big)
\nl
&\quad + {\tilde\nabla}_4\big(   F_{12}(\eta_{2}) F_{34}(\eta_{4}) ( {\tilde  x}_{2,3}  - \ell\! \cdot\!  k_{2}) - \MMF_{12} (\xi_1)  F_{34}(\eta_{4})    \big)  \notag \\
&\quad + {\tilde\nabla}_2\big(     F_{12}(\eta_{2})  F_{34}(\eta_{2}) ( {\tilde  x}_{4,12}  -\ell\! \cdot\!  k_{4}) - \MMF_{34}  (\xi_2)  F_{12}(\eta_{2}) \big)
\nl
&\quad - {\tilde\nabla}_3 \big( F_{12}(\eta_{2})F_{34}(\eta_{4})    {\tilde  x}_{2,4}  \big)
+ {\tilde\nabla}_2 {\tilde\nabla}_4\big( F_{12}(\eta_{2})   F_{34}(\eta_{4}) \big) 
\,,
\non
\end{align} 
where the first three lines are free of $\ell$ and related to \eqref{ofof} 
by $\MMF_{ij}(\xi) \leftrightarrow \MM_{ij}(\xi) $ as well as
$F_{ij}(\eta)\leftrightarrow \Omega_{ij}(\eta)$ and 
${\tilde  x}_{i,j} \leftrightarrow x_{i,j}$ as expected. 
The symmetry of (\ref{twopairsF}) under the exchange
$(z_1,z_2,\eta_2,\xi_1) \leftrightarrow (z_3,z_4,\eta_4,\xi_2)$ of the cycles
is not fully manifest: The loop-momentum dependence in the 
fifth line differs from its image under the exchange of the cycles by a term
$\sim {\tilde x}_{2,4} \,\ell {\cdot} (k_{1}{+}k_{2}{+}k_{3}{+}k_{4})$. Following
the discussion below (\ref{chiKN}), translation invariance of the chiral 
Koba-Nielsen factor requires $\sum_{j=1}^n \ell {\cdot} k_j=0$ 
which establishes the expected exchange symmetry.

The total Koba-Nielsen derivatives in the last three lines of (\ref{twopairsF}) yield
boundary terms by the application of Stokes' theorem as in section \ref{sec:bdyibp}.
They cannot be discarded in a closed-string context since the
primitives $\sim F_{12}(\eta_{2})   F_{34}(\eta_{4})$ of the ${\tilde\nabla}_a$ in 
(\ref{twopairsF}) have B-cycle monodromies and therefore
yield a non-vanishing right-hand side of (\ref{eq:LoopMomentumStokesSecond}).

\subsubsection{Generalization to cycles of arbitrary length}

The additional $\ell$-dependence in the four-point example (\ref{twopairsF})
which is absent in its doubly-periodic counterpart (\ref{ofof}) 
can be easily generalized to cycles of arbitrary length. 
For the product of two arbitrary $F$-cycles $\CCF_{\ldots}(\xi)$
defined by (\ref{defcf}), we have
  \begin{align}
  \label{newcyc23pre22F}
  & ( 1{+}s_{12\cdots m}) ( 1{+}s_{m+1,\cdots n})  \CCF_{(12\cdots m)}(\xi_1)\CCF_{(m+1,\cdots, n)} (\xi_2) = \MMF_{12\cdots m}(\xi_1)\MMF_{m+1,\cdots, n}(\xi_2)
   \\
&\quad -\sum_{p  =m+2}^{n}
~
 \sum_{\substack{\sigma \in \{m+2,\cdots, p-1\}\\
\quad\shuffle \{n,\cdots, p+1 \}} }
\!\!\!\!\!\!\!\!\!
 (-1)^{n-m-p} 
 \left( \sum_{k=1}^m  {\tilde x}_{p,k} -\ell {\cdot} k_p + {\tilde\nabla}_p \right)
 \big({\bm F}_{m+1,\sigma,p}
 \MMF_{12\cdots m} (\xi_1) \big)
\nl
&\quad - \sum_{
b  =2}^{m}
\!\!\!\!
\!\!\!\!
 \sum_{\substack{\rho \in \{2,\cdots, b-1\}\\
\qquad\quad \shuffle \{m,\cdots, b+1 \}} }
\!\!\!\!\!\!\!\!\!\!\!\!\!
 (-1)^{m-b} 
\sum_{j=m+1}^n      \sum_{\substack{
p  =m+1
\\
p\neq j}}^{n}
\sum_{
\substack{
\sigma \in  X \shuffle  Y^{\rm T}
\\
(j, X, p, Y )=(m{+}1,m{+}2, \ldots, n)
} }
\!\!\!
\nonumber
\\
&\quad \quad \times  (-1)^{|Y|} 
(\ell\!\cdot\! k_p  -{\tilde\nabla}_p )
\left(
{\bm F}_{1,\rho,b}  {\tilde x}_{b,j}
{\bm F}_{j,\sigma,p} 
\right)
\nl\,\,
& \quad -\sum_{b  =2}^{m}
\!\!\!\!
\!\!\!\!
 \sum_{\substack{\rho \in \{2,\cdots, b-1\}\\
\qquad\quad \shuffle \{m,\cdots, b+1 \}} }
\!\!\!\!\!\!\!\!\!\!\!\!\!\!
 (-1)^{m-b} 
\Bigg( \sum_{j=m+1}^n  {\tilde x}_{b,j} -\ell{\cdot} k_b +{\tilde\nabla}_b \Bigg)
\big(
{\bm F}_{1,\rho,b}
\MMF_{m+1,m+2,\cdots, n}(\xi_2)
\big)
\nl
&\quad  -  \sum_{b  =2}^{m}
\!\!\!\!
\!\!\!\!
 \sum_{\substack{\rho \in \{2,\cdots, b-1\}\\
\qquad\quad \shuffle \{m,\cdots, b+1 \}} }
\!\!\!\!\!\!\!\!\!\!\!\!\!
 (-1)^{m-b} 
 \sum_{
p  =m+2}^{n}
 \sum_{\substack{\sigma \in \{m+2,\cdots, p-1\}\\
\quad\shuffle \{n,\cdots, p+1 \}} }
 \nonumber\\
&\quad\quad \times
 (-1)^{n-m-p} 
(\ell\!\cdot\! k_b -{\tilde\nabla}_b) 
\left[
\left( \sum_{k=1}^m  {\tilde x}_{p,k} -\ell{\cdot} k_p 
+{\tilde\nabla}_p \right)
\left({\bm F}_{1,\rho,b} {\bm F}_{m+1,\sigma,p} 
\right)
\right]
\nl\,\,
&\quad +\frac12
\sum_{\substack{a,b  
 =1 \\ a\neq b}
}^{m} \,
  \sum_{\substack{
j, p 
 =m+1 \\ j\neq p}
}^{n} \sum_{
\substack{
\rho \in  A \shuffle  B^{\rm T}
\\
(a, A, b, B )=(1,2, \ldots, m)
} }
\sum_{
\substack{
\sigma \in  X \shuffle  Y^{\rm T}
\\
(j, X, p, Y )=(m{+}1,m{+}2, \ldots, n)
} }
\!\!\!\!
\!\!\!\!
\!\!\!\!
 (-1)^{|B|+|Y|}  {\bm F}_{a,\rho,b} \,
{\tilde x}_{b,j} {\bm F}_{j,\sigma,p}  \, {\tilde x}_{p, a}
\,.
\non
 \end{align}
Apart from the last line, all terms not containing $\tilde\nabla_b$ on the right-hand side are free of $F$-cycles or combined cycles of $F_{ij}(\eta)$ and $g^{(1)}_{ij}$. Similar to the
doubly-periodic case in (\ref{newcyc23pre22}), the single-cycles ${\bm F}_{a,\rho,b} 
{\tilde x}_{b,j} {\bm F}_{j,\sigma,p}   {\tilde x}_{p, a}$
in the last line can be broken using \eqref{rhorhoF},
by isolating suitable terms in the Laurent expansion.

Following the substitution rules (\ref{subsrule}), any $ {\tilde\nabla}_b $ in
 (\ref{newcyc23pre22F}) can be anticipated from the total derivatives $\nabla_b$ 
 in the F-IBP reduction of two $\Omega$-cycles in appendix \ref{sec:2cyc.4}. Moreover, 
 the rules in (\ref{subsrule}) imply that the ${\tilde\nabla}_b $
always appear together with $\ell\!\cdot\! k_b$ with opposite signs. As discussed
below (\ref{twopairsF}), the total Koba-Nielsen derivatives ${\tilde\nabla}_b $
yield boundary terms by the B-cycle monodromies of the respective primitives.

\subsubsection{Example at $n=5$ points}

Specializing (\ref{newcyc23pre22F}) to $m=2$ and $n=5$ yields the
following meromorphic analogue of (\ref{cyc23}),
 \begin{align}
&  (1{+}s_{12}) (1{+}s_{345}) \CCF_{(12)} (\xi_1)\CCF_{(345)}(\xi_2)
=  \big(   {\rm RHS~ of ~ \eqref{cyc23}} \big|_{x\to {\tilde  x}, \,  \MM\to \MMF, \, \Omega\to F,\, {\bm \Omega}\to {\bm F} }\big)
  \label{cyc23F}   \\
&\quad + \MMF_{345}(\xi_2)
\big(\ell\! \cdot \!k_2\, -{\tilde\nabla}_2\big)  F_{12}(\eta_{2}) 
+
\MMF_{12}(\xi_1) \big(
( \ell\! \cdot \!k_5\, -{\tilde\nabla}_5){\bm F}_{345}
 -   (\ell\! \cdot \!k_4\,-{\tilde\nabla}_4)  {\bm F}_{354}\big) 
\nl
&\quad +( \ell\! \cdot \!k_2\, -{\tilde\nabla}_2 ) F_{12}(\eta_{2})\big(
  ( \ell\! \cdot \!k_5\,
      - {\tilde\nabla}_5-{\tilde x}_{5,12} ) {\bm F}_{345}  
-(  \ell\! \cdot \!k_4\,-
     {\tilde\nabla}_4-{\tilde x}_{4,12}){\bm F}_{354}
\big)  
\nonumber\\
&\quad+   F_{12}(\eta_{2})  
{\bm F}_{345} \big( 
      {\tilde x}_{2,5} 
(\ell\! \cdot \!k_3 -
     {\tilde\nabla}_3)
     - {\tilde x}_{2,3}  (\ell\! \cdot \!k_5 -
     {\tilde\nabla}_5)
\big) \non\\
& \quad
+   F_{12}(\eta_{2})  
{\bm F}_{354} \big( 
 {\tilde x}_{2,3}  (\ell\! \cdot \!k_4 -
     {\tilde\nabla}_4)
     - {\tilde x}_{2,4} 
(\ell\! \cdot \!k_3 -
     {\tilde\nabla}_3)
\big) \nonumber\\
 & \quad+   F_{12}(\eta_{2})  
{\bm F}_{435} \big( 
 {\tilde x}_{2,4}  (\ell\! \cdot \!k_5 -
     {\tilde\nabla}_5)
     - {\tilde x}_{2,5} 
(\ell\! \cdot \!k_4 -
     {\tilde\nabla}_4)
\big) 
\,, \non
 \end{align}
see \eqref{twopairsF} for its four-point counterpart.
  
\subsubsection{Example at $n=6$ points}

Specializing (\ref{newcyc23pre22F}) to $m=2$ and $n=6$ yields the
following meromorphic analogue of (\ref{cyc33omega}),
   \begin{align}
&  (1{+}s_{123}) (1{+}s_{456}) \CCF_{(123)}(\xi_1)\CCF_{(456)} (\xi_2)
=   \big(   {\rm RHS~ of ~ \eqref{cyc33omega}} \big|_{x\to {\tilde  x}, \,  \MM\to \MMF, \, {\bm \Omega}\to {\bm F}  }\big)\label{cyc33F} \\
&\quad+ \MMF_{123}(\xi_1) \big( \ell\! \cdot \!k_6\,   {\bm F}_{456}  - \ell\! \cdot \!k_5\,  {\bm F}_{465}
\big) +\big(\ell\! \cdot \!k_3\,  {\bm F}_{123} -    \ell\! \cdot \!k_2\, {\bm F}_{132}
\big) \MMF_{456}(\xi_2)
\nl
&\quad+\big(  \ell\! \cdot \!k_3\,  {\bm F}_{123}   -   \ell\! \cdot \!k_2\,  {\bm F}_{132}
\big) \big(
\ell\! \cdot \!k_6\, {\bm F}_{456} -   \ell\! \cdot \!k_5\,   {\bm F}_{465}
\big) 
 \nl
&\quad + \Big[ 
 \big(   {\tilde x}_{3,6}   \,\ell\! \cdot \!k_4\,    -{\tilde x}_{6,123}   \,\ell\! \cdot \!k_3\,  - {\tilde x}_{3,4}   \,\ell\! \cdot \!k_6    \big)  {\bm F}_{123} {\bm F}_{456} 
  +  \big(    {\tilde x}_{3,5}   \,\ell\! \cdot \!k_6\, - {\tilde x}_{3,6}   \,\ell\! \cdot \!k_5\,       \big) {\bm F}_{123} {\bm F}_{546} 
  \nl\,\,
&\quad
 \quad +   
  \big(   {\tilde x}_{5,123}   \,\ell\! \cdot \!k_3\, -  {\tilde x}_{3,5}   \,\ell\! \cdot \!k_4\, + {\tilde x}_{3,4}   \,\ell\! \cdot \!k_5\,       \big)
  {\bm F}_{123} {\bm F}_{465}   - \Big(2\leftrightarrow3\Big) \Big]
   \nonumber   \\
   &\quad+{\text{(total Koba-Nielsen derivatives)}}
 \,, \non
 \end{align}
where the total Koba-Nielsen derivatives can be reinstated by substituting $\ell {\cdot} k_i \rightarrow \ell {\cdot} k_i{-}\tilde\nabla_i$ acting from the left on the accompanying functions of $z_i$. Double derivatives due to bilinears in $\ell {\cdot} k_i$ (say $\tilde\nabla_3 \tilde\nabla_6 {\bm F}_{123}  
 {\bm F}_{456}  $ due to
$ \ell\! \cdot \!k_3\,  {\bm F}_{123}  
\ell\! \cdot \!k_6\, {\bm F}_{456} $) do not introduce any ordering ambiguities in view of $ \tilde\nabla_i \tilde\nabla_j-\tilde\nabla_j \tilde\nabla_i =0$ for any pair $i\neq j $.

 Similar to \eqref{twopairsF}, the first three lines on the right-hand side of   \eqref{cyc33F} are  manifestly symmetric under exchange of the cycles, i.e.\ under $(z_1,z_2,z_3,\eta_2,\eta_3,\xi_1) \leftrightarrow (z_4,z_5,z_6,\eta_5,\eta_6,\xi_2)$. The fourth and fifth 
 line in turn share this symmetry up to 
 \ba
  \,\ell \!\cdot\! (k_{1}{+}k_{2}{+}\cdots{+}k_{6}) 
  \big( 
 {\bm F}_{123} {\bm F}_{456} {\tilde x}_{3,6}-{\bm F}_{132} {\bm F}_{456} {\tilde x}_{2,6}-{\bm F}_{123} {\bm F}_{465} {\tilde x}_{3,5}+{\bm F}_{132} {\bm F}_{465} {\tilde x}_{2,5} 
 \big)\,,
  \ea
which vanishes by translation invariance of the chiral Koba-Nielsen factor. 
 
\subsection{Towards triple cycles}
\label{triplecyclenew}

We conclude this section with a glimpse of F-IBP reductions
of triple cycles, following the earlier approach of sequentially breaking 
the cycles and prioritizing the breaking of tadpoles.
Even though larger numbers of cycles do not introduce any conceptual 
challenges, the combinatorial complexity increases. In a companion 
paper \cite{companion}, we provide systematic methods for breaking three 
or more cycles as well as more general configurations of Kronecker-Eisenstein
series by introducing new terminologies
and applying combinatorial tools beyond the scope of
this work to simplify the expressions.

\subsubsection{Doubly-periodic cycles at six points}

By extending the techniques of section \ref{sec:2cyc.1}
to a third cycle of length two, we derive the following
six-point result:
\begin{align} 
&(1{+}s_{12}) (1{+}s_{34}) (1{+}s_{56})  \CC_{(12)} (\xi_1)\,\CC_{(34)} (\xi_2)\,\CC_{(56)} (\xi_3) 
\overset{\rm IBP}= 
\MM_{12}(\xi_1) \MM_{34} (\xi_2)\MM_{56}(\xi_3)  \label{triple6} \\
&-\MM_{12}(\xi_1)\MM_{34}(\xi_2) {\bm \Omega}_{56}  x_{6,1234} 
-\MM_{12} (\xi_1) \MM_{56}(\xi_3){\bm \Omega}_{34} x_{4,1256}
- \MM_{34} (\xi_2)\MM_{56}(\xi_3){\bm \Omega}_{12} x_{2,3456}
\nl
&+ \MM_{12}(\xi_1){\bm \Omega}_{34} {\bm \Omega}_{56}  (x_{4,12}x_{6,1234}+x_{4,5}x_{6,12}-x_{4,6}x_{5,12})
\nonumber
\\
&+ \MM_{34}(\xi_2){\bm \Omega}_{12} {\bm \Omega}_{56} (x_{2,34}x_{6,1234}+x_{2,5}x_{6,34}-x_{2,6}x_{5,34})
\nonumber
\\
&+ \MM_{56}(\xi_3){\bm \Omega}_{12} {\bm \Omega}_{34}  (x_{2,56}x_{4,1256}+x_{2,3}x_{4,56}-x_{2,4}x_{3,56})
\nonumber
\\
&+(1{+}s_{12}) \CC_{(12)} (\xi_1) {\bm \Omega}_{34} {\bm \Omega}_{56} (x_{4,5}x_{6,3}-x_{4,6}x_{5,3})\nonumber
\\
&+  (1{+}s_{34}) \CC_{(34)}(\xi_2) {\bm \Omega}_{12} {\bm \Omega}_{56} (x_{2,5}x_{6,1}-x_{2,6}x_{5,1})
\nonumber
\\
&+  (1{+}s_{56}) \CC_{(56)} (\xi_3){\bm \Omega}_{12} {\bm \Omega}_{34} (x_{2,3}x_{4,1}-x_{2,4}x_{3,1})
\nonumber
\\
&
+{\bm \Omega}_{12} {\bm \Omega}_{34}  {\bm \Omega}_{56} \big(
x_{1,4} x_{2,6} x_{3,5}+x_{1,5} x_{2,4} x_{3,6} -x_{1,6} x_{2,4} x_{3,5}
-x_{1,4} x_{2,5} x_{3,6}
\nonumber
\\
&\quad\quad\quad\quad\quad\; \, +x_{1,6} x_{2,3} x_{4,5}-x_{1,3} x_{2,6} x_{4,5}-x_{1,5} x_{2,3} x_{4,6}+x_{1,3} x_{2,5} x_{4,6}\big)\,.
   \nonumber
\end{align}
While the first five lines on the right-hand side are already in the
desired chain form, the remaining lines feature two types of
cycles of lower complexity:
\begin{itemize}
\item In the third to fifth line of (\ref{triple6}) from below, each term is
a product of a length-two cycle $\CC_{(ij)} (\xi)$ and a $f$-$\Omega$
cycle ${\bm \Omega}_{ab} x_{b,c} {\bm \Omega}_{cd} x_{d,a}$ of length four.
Their decomposition into the chain basis follows from Laurent expansion
of (\ref{cyc24}) in its bookkeeping variables.
\item The last two lines of (\ref{triple6}) feature single-cycles
${\bm \Omega}_{ab} x_{b,c} {\bm \Omega}_{cd} x_{d,e} 
{\bm \Omega}_{ef} x_{f,a}$ of length six whose F-IBP reduction
is determined by (\ref{rhorho}).
\end{itemize}
Hence, by importing results of earlier sections, the entire
right-hand side of (\ref{triple6}) can be reduced to expansion
coefficients of the conjectural chain basis ${\bm \Omega}_{1\rho(23456)} $
with $\rho \in S_5$.

\subsubsection{Doubly-periodic cycles at seven points}

The methods of deriving (\ref{triple6}) can be straightforwardly extended
to the following seven-point case, 

  \begin{align}
&(1{+}s_{12}) (1{+}s_{34}) (1{+}s_{567})  \CC_{(12)}(\xi_1) \,\CC_{(34)} (\xi_2)\,\CC_{(567)}  (\xi_3)
\overset{\rm IBP}= \MM_{12}(\xi_1)\MM_{34}(\xi_2)\MM_{567}(\xi_3) 
\notag \\
&\quad -\MM_{12}(\xi_1)\MM_{567}(\xi_3) {\bm \Omega}_{34}  x_{4,12567} 
- \MM_{34}(\xi_2)\MM_{567}(\xi_3){\bm \Omega}_{12}  x_{2,34567} 
\nl
&\quad - \MM_{12}(\xi_1)\MM_{34}(\xi_2) ( {\bm \Omega}_{567} x_{7,1234} -{\bm \Omega}_{576}  x_{6,1234} )
\\
&\quad+\Big[ \MM_{12}(\xi_1){\bm \Omega}_{34}  \Big(x_{4,12}({\bm \Omega}_{567}x_{7,1234}-{\bm \Omega}_{576}x_{6,1234})+x_{4,5}({\bm \Omega}_{567}x_{7,12}-{\bm \Omega}_{576}x_{6,12}) 
\nonumber
\\
&\quad
\quad+x_{4,6}({\bm \Omega}_{675}x_{5,12}-{\bm \Omega}_{657}x_{7,12}) 
+x_{4,7}({\bm \Omega}_{756}x_{6,12}-{\bm \Omega}_{765}x_{5,12}) 
\Big)+\Big(12\leftrightarrow 34\Big)\Big]
\nonumber
\\
&\quad+\MM_{567}(\xi_3){\bm \Omega}_{12} {\bm \Omega}_{34}(x_{2,567}x_{4,12567}+x_{2,3}x_{4,567}-x_{2,4}x_{3,567})
\nonumber
\\
&\quad+\Big[ (1{+}s_{34}) \CC_{(12)} (\xi_1){\bm \Omega}_{34} \big(
{\bm \Omega}_{567}(x_{4,5}x_{7,3}-x_{4,7}x_{5,3})
+{\bm \Omega}_{576}(x_{4,6}x_{5,3}-x_{4,5}x_{6,3})
\nonumber
\\
&\quad
\quad +{\bm \Omega}_{657}(x_{4,7}x_{6,3}-x_{4,6}x_{7,3})
\big)
+\Big(12\leftrightarrow 34\Big) \Big]
\nonumber
\\
&\quad
+ (1{+}s_{567}) \CC_{(567)} (\xi_3){\bm \Omega}_{12} {\bm \Omega}_{34} (x_{2,3}x_{4,1}-x_{2,4}x_{3,1})
\nonumber
\\
&\quad
+ {\bm \Omega}_{12} {\bm \Omega}_{34}   \Big[
\Big(x_{1,4} \left(x_{2,7} x_{3,5}-x_{2,5} x_{3,7}\right)+x_{1,7} \left(x_{2,3} x_{4,5}-x_{2,4}
   x_{3,5}\right)
   \nonumber
   \\
   &\quad
  \quad +x_{1,5} \left(x_{2,4} x_{3,7}-x_{2,3} x_{4,7}\right)+x_{1,3} \left(x_{2,5}
   x_{4,7}-x_{2,7} x_{4,5}\right)\Big) {\bm \Omega} _{567}+ {\rm cyc}(5,6,7)
   \Big]\,.
   \nonumber
\end{align}   
While the first six lines on the right-hand side are term-by-term in chain form, we have
\begin{itemize}
\item products of two cycles (lengths $2{+}5$ and $3{+}4$) in the third to fifth line from below
which can be broken via Laurent expansion of \eqref{newcyc23pre22};
\item single-cycles of length seven in the last two lines which can be broken via (\ref{rhorho}).
\end{itemize}
This example illustrates once more that a recursive
approach in the number of cycles is well adapted to the  
F-IBP reduction of multiple cycles.

\subsubsection{Meromorphic cycles at six points}

As a last case study of triple cycles, we shall spell out the
meromorphic analogue of the six-point F-IBP reduction in (\ref{triple6}) 
 \begin{align} 
&(1{+}s_{12}) (1{+}s_{34}) (1{+}s_{56})  \CCF_{(12)} (\xi_1)\,\CCF_{(34)}(\xi_2)\,\CCF_{(56)} (\xi_3) =   \big(   {\rm RHS~ of ~ \eqref{triple6}} \big|^{x\to {\tilde  x}, \, \CC\to\CCF,}_{ \MM\to \MMF, \, \Omega\to F  }\big)
\notag \\
 &\quad +\Big[
 \MMF_{12} (\xi_1) \MMF_{34} (\xi_2)  \ell \!\cdot\! k_{6}  { F}_{56} (\eta_6) 
 +  \MMF_{12}(\xi_1)  \ell \!\cdot\! k_{4} \ell \!\cdot\! k_{6}  { F}_{34} (\eta_4)  { F}_{56} (\eta_6)
 \label{cyc222F} \\
&\quad\quad + 
\MMF_{12}(\xi_1)  \big( \ell \!\cdot\! k_{5} {\tilde  x}_{4,6}  
 - \ell \!\cdot\! k_{4}     {\tilde  x}_{6,1234} 
   -\ell \!\cdot\! k_{6}   {\tilde  x}_{4,125}     \big)
  { F}_{34} (\eta_4)  { F}_{56} (\eta_6) + {\rm cyc}(12,34,56) \Big] 
\notag \\
&\quad+ { F}_{12} (\eta_2){ F}_{34} (\eta_4) { F}_{56}(\eta_6)\Big(\ell \!\cdot\! k_{2}  \ell \!\cdot\! k_{4}  \ell \!\cdot\! k_{6} 
-\ell \!\cdot\! k_{2}  \ell \!\cdot\! k_{4} \tilde x_{6,1234}
+ \ell \!\cdot\! k_{2}  \ell \!\cdot\! k_{5}  \tilde x_{4,6}
- \ell \!\cdot\! k_{2}  \ell \!\cdot\! k_{6} \tilde x_{4,125}
 \notag \\
&\quad \quad
+\ell \!\cdot\! k_{3}  \ell \!\cdot\! k_{6} \tilde x_{2,4}
+ \ell \!\cdot\! k_{4}  \ell \!\cdot\! k_{5} \tilde x_{2,6}
-\ell \!\cdot\! k_{4}  \ell \!\cdot\! k_{6} \tilde x_{2,35}
+  \ell \!\cdot\! k_{6} ( \tilde x_{2,3} \tilde x_{4,5}  - \tilde x_{2,4}  \tilde x_{3,5} +  \tilde x_{2,5}  \tilde x_{4,1256} ) \notag \\
 &\quad\quad
+  \ell \!\cdot\! k_{5} ( \tilde x_{2,4}  \tilde x_{3,6}  -\tilde x_{2,3} \tilde x_{4,6} - \tilde x_{2,6}  \tilde x_{4,1256}  )
+ \ell \!\cdot\! k_{4} ( \tilde x_{2,5}  \tilde x_{6,3}  - \tilde x_{2,6}  \tilde x_{5,3}  + \tilde x_{2,3} \tilde x_{6,1234} ) \notag \\
 &\quad\quad
+ \ell \!\cdot\! k_{3} ( \tilde x_{2,6}  \tilde x_{5,4}  - \tilde x_{2,5} \tilde x_{6,4}  - \tilde x_{2,4}  \tilde x_{6,1234} )
+ \ell \!\cdot\! k_{2} ( \tilde x_{4,5}  \tilde x_{6,12}  - \tilde x_{4,6}  \tilde x_{5,12} +\tilde x_{4,12}  \tilde x_{6,1234} )  \Big)
  \nonumber
   \\&\quad+{\text{(total Koba-Nielsen derivatives)}}
 \,. \non
 \end{align}
Both the second and the third line are manifestly symmetric under cyclic permutations of the three cycles, i.e.\ the associated groups of variables $(z_1,z_2,\eta_2,\xi_1)$, $(z_3,z_4,\eta_4,\xi_2)$, and $(z_5,z_6,\eta_6,\xi_3)$. Lines four through seven, by contrast, only share this symmetry in the cycles after imposing the corollary $\sum_{j=1}^6\ell {\cdot} k_j =0$ of translation invariance of the chiral Koba-Nielsen factor. Manifest permutation symmetry in $\CCF_{(12)}(\xi_1),\CCF_{(34)}(\xi_2),\CCF_{(56)}(\xi_3)$ can of course be enforced by averaging (\ref{cyc222F}) over permutations of the cycles. 

Similar to the discussion below (\ref{cyc33F}),
the total Koba-Nielsen derivatives in the last line can again be reinstated by
replacing $\ell {\cdot} k_i \rightarrow \ell {\cdot} k_i\,-\tilde\nabla_i$. This is unambiguous for any number of factors $\ell {\cdot} k_i$ since any pair of $\tilde\nabla_i,\tilde\nabla_j$ commutes.

\section{Conclusions and outlook \label{sec:conclusion}}

In this work, we have significantly advanced the integration-by-parts 
methodology for one-loop string integrals of Koba-Nielsen type. Specifically, we have reduced  
{\it cyclic products} of Kronecker-Eisenstein series and their coefficients $f^{(w)}(z_i{-}z_j,\tau)$
into conjectural bases of one-loop string integrals built from 
Kronecker-Eisenstein products of {\it chain topology}
\cite{Mafra:2019ddf,Mafra:2019xms, Gerken:2019cxz}.
Our results not only furnish strong validations of the chain bases in the references
but also provide explicit formulae for the basis decompositions of
one or two cycles of Kronecker-Eisenstein series of arbitrary length.
A companion paper \cite{companion} will 
\begin{itemize}
\item provide a {\tt Mathematica} implementation of our main formulae,
\item extend the recursive approach of this work to arbitrary numbers of Kronecker-Eisenstein cycles and identify the combinatorial structure of their integration-by-parts reduction,
\item address more general configurations of Kronecker-Eisenstein series and coefficients
besides cyclic products, to be represented via tadpoles, multibranch and even connected multiloop graphs.
\end{itemize} 
As a first motivation for the detailed integration-by-parts reductions in this work,
they can be applied to low-energy expansions of one-loop string amplitudes in bosonic, 
heterotic and supersymmetric theories. For the one-loop basis integrals of chain topology, 
differential equations in the modular parameter $\tau$ led to powerful expansion 
techniques for open strings \cite{Mafra:2019ddf,Mafra:2019xms} 
and for closed strings \cite{Gerken:2020yii}, supplemented by the {\tt Mathematica} package 
\cite{Gerken:2020aju}. The results of this work and \cite{companion}
allow to swiftly export these expansions of chain integrals to string 
amplitudes in their more basic representation involving cyclic products 
of Kronecker-Eisenstein coefficients. 
The most interesting applications should arise in heterotic string theories whose bosonic 
conformal-field-theory sector tends to yield numerous cyclic products 
but which at the same time offer a rewarding window into string dualities.

As a second motivation, the techniques for basis decompositions of
string integrals in this work pave the way for structural insights into one-loop
amplitudes in string and field theory. The basis decompositions of
one-loop string integrals unlocked in this work organize amplitudes
in various string theories into gauge-invariant kinematic functions
of external polarizations. For one-loop open-superstring amplitudes
with maximal supersymmetry, these kinematic functions admit a
field-theory interpretation which unravelled a surprising double-copy structure
\cite{Mafra:2017ioj, Mafra:2018qqe}. Our results give access to the analogous
one-loop kinematic functions in heterotic and bosonic theories
and therefore guide the quest for similar double-copy structures. 
At tree level, this line of investigations revealed an elegant web
of double-copy relations among different classes of open-string, closed-string 
and field-theory amplitudes \cite{Mafra:2011nv, Zfunctions, Carrasco:2016ldy, Azevedo:2018dgo}. 
Hence, this work is an opportunity to explore
loop-level echoes of this web of double copies.

This work additionally spawns several mathematical lines of follow-up research. 
The consideration of string integrals over all the punctures {\it at
fixed modular parameter $\tau$} initiated fruitful crosstalk with algebraic geometers
and number theorists through the appearance of elliptic multiple zeta
values \cite{Enriquez:Emzv, Broedel:2014vla} and modular graph forms \cite{DHoker:2015wxz, DHoker:2016mwo} in the low-energy expansion.
When keeping not only $\tau$ {\it but also some of the punctures $z_i$ fixed}, intermediate
steps of string-amplitude computations serve as
generating functions of elliptic polylogarithms \cite{BrownLev, Broedel:2014vla, Broedel:2017kkb} and their single-valued versions \cite{Ramakrish, DHoker:2018mys, Broedel:2019tlz, DHoker:2020hlp}.
Conjectural $n$-point integral bases of dimension $n!$ depending on one unintegrated puncture have been presented in \cite{Broedel:2019gba, Broedel:2020tmd} and generalized to an arbitrary number of unintegrated punctures in \cite{Kaderli:2022qeu}. 
This work offers concrete starting points to substantiate these bases through explicit
integration-by-parts relations, and the recursive methods
of the companion paper \cite{companion} may furthermore stimulate a general proof.

Moreover, the quest for bases of integrands under integration by parts is a 
common theme of string amplitudes and Feynman integrals in particle physics.
Integration by parts in the presence of an ubiquitous Koba-Nielsen factor
or its Feynman-integral counterparts \cite{Mastrolia:2018uzb, Frellesvig:2019kgj, Mizera:2019vvs, Frellesvig:2020qot, Caron-Huot:2021xqj, Caron-Huot:2021iev, Duhr:2023bku} is closest to the setting of the twisted de Rham theory, initiated by Aomoto \cite{Aomoto87} and beautifully communicated by Mizera to the physics community in \cite{Mizera:2017cqs, Mizera:2017rqa, Mizera:2019gea}. Finding a suitable framing in terms of
twisted de Rham theory may offer a particularly elegant way
to rigorously establish the Kronecker-Eisenstein chains as
a basis of genus-one string integrals. In particular, this kind of understanding
should offer a unified description of integration-by-parts reduction, 
monodromy relations \cite{Tourkine:2016bak, Hohenegger:2017kqy, Tourkine:2019ukp, Casali:2019ihm, Casali:2020knc} as well as relations between open and closed strings \cite{Broedel:2018izr, Gerken:2020xfv, Stieberger:2021daa, Stieberger:2022lss}
at genus one.
  
In fact, previous mathematical work \cite{ManoWatanabe2012,ghazouani2016moduli,goto2022intersection} identifies a twisted-cohomology setup where meromorphic Kronecker-Eisenstein series are proven to form a basis. The cohomology setup in these references enjoys striking parallels with the 
chiral-splitting approach
to string amplitudes \cite{DHoker:1988pdl, DHoker:1989cxq} but has so far only been developed for a single integrated puncture (with an arbitrary number of unintegrated ones). A generalization of their work to multiple integration variables could readily prove the conjectures addressed in the present work. Alternatively, one could take the Lie-algebraic toolkit of Felder and Varchenko \cite{Felder:1995iv} as a starting point to attempt an alternative derivation of the results in our work.

Finally, the conjectural integral bases and integration-by-parts techniques in this 
work call for generalizations to higher genus. Based on the recent
proposal for higher-genus analogues of the Kronecker-Eisenstein 
kernels \cite{DHoker:2023vax},
the most immediate question concerns their Fay identities and generating
functions of Koba-Nielsen integrals that close under moduli derivatives.
The mechanisms for integration by parts in this work including the role of Fay identities
and their coincident limit should offer essential guidance for several of the challenging
steps in developing a comprehensive framework for higher-genus string integrals. On the 
one hand, these generalizations of our results will feed into tools for concrete string-amplitude 
computations at higher genus. On the other hand, the study of suitable families of
Koba-Nielsen integrals will have valuable input for the construction of function spaces
of interest to particle physicists and mathematicians including higher-genus
incarnations of modular tensors, polylogarithms and multiple zeta values.

\acknowledgments
We would like to thank Rishabh Bhardwaj, Freddy Cachazo, Alex Edison, Max Guillen, Song He, Andrzej Pokraka, Lecheng Ren and in particular Filippo Balli for combinations of valuable discussions and collaboration on related topics. C.R.\ and O.S.\ are supported by the European Research Council under ERC-STG-804286 UNISCAMP. 
The research of Y.Z.\ is supported by the Knut and Alice Wallenberg Foundation under the grant KAW 2018.0116: From Scattering Amplitudes to Gravitational Waves.
The research of Y.Z.\ was also supported in part by a grant from the Gluskin Sheff/Onex Freeman Dyson Chair in Theoretical Physics and by Perimeter Institute. Research at Perimeter Institute is supported in part by the Government of Canada through the Department of Innovation, Science and Economic Development Canada and by the Province of Ontario through the Ministry of Colleges and Universities.


\appendix


\section{Chain decompositions of $V_m(1,2,\ldots,m)$ up to five points}
\label{singlecycle456}

This appendix is dedicated to the chain decomposition (\ref{vopen}) of the
elliptic $V_m(1,2,\ldots,m)$-functions in (\ref{1.1a}) at $m=4,5$ points
(see section \ref{sec:vmat3} for a detailed discussion at $m=3$). For simplicity, the
Koba-Nielsen factor is considered at the same multiplicity $n=m$: The
extra terms for $n>m$ can be straightforwardly reconstructed from
the second line of (\ref{vopen}).

\subsection{Four points}

The $(n=m=4)$-point instance of (\ref{vopen}) 
together with the tools in section \ref{sec:toolcoeff} lead to
\begin{align}
&\! \! \!\! \! \!\! \! \!\!(1{+}s_{1234}) V_4(1,2,3,4) 
\overset{\rm IBP}= \MM_{1234}\big|\big|_{\eta_2^0,\eta_3^0,\eta_{4}^0,\eta_{5}^0}
\notag\\
=\,\,& \hat{\text{G}}_2
 V_2(1,2,3,4)+{\rm G}_4(1{+}3s_{13}{+}3s_{24})  
 \nonumber
 \label{eqb1} \\
 &+\bigg[ 
 {\bm \Omega} _{1234} \left(\frac{s_{12}}{\eta _{234}} -\frac{s_{12}}{\eta _{23}} +\frac{2 \left(s_{14}{+}s_{24}{+}s_{34}\right)}{\eta _4}-\frac{s_{24}}{\eta _2}-\frac{s_{13}{+}s_{23}{+}s_{34}}{\eta _3}+\frac{s_{13}{+}s_{23}}{\eta _{34}}\right)
 \nl
 &\quad +{\bm \Omega} _{1243} \left(\frac{s_{12}{+}s_{23}}{\eta _2} -\frac{s_{12}}{\eta _{23}} +\frac{s_{14}{+}s_{24}{+}s_{34}}{\eta _4}-\frac{2 \left(s_{13}{+}s_{23}{+}s_{34}\right)}{\eta _3}-\frac{s_{14}{+}s_{24}}{\eta _{34}}\right)
 \nl
 &\quad +s_{13} {\bm \Omega} _{1324}\left(\frac{1}{\eta _{23}}+\frac{1}{\eta _{34}}-\frac{1}{\eta _{234}}- \frac{1}{\eta _3} \right)  +\big(2\leftrightarrow4\big)\bigg] \, \bigg|\bigg|_{\eta_2^0,\eta_3^0,\eta_4^0} \, ,
\end{align}
where the relabelling $2\leftrightarrow 4$ of the subscripts of $s_{ij},\eta_i,{\bm \Omega} _{1ijk}$
applies to the last three lines. As detailed below (\ref{(12)(34)}), the notation $\overset{\rm IBP}= $ in the first line indicates that total Koba-Nielsen derivatives $\nabla_i(\ldots)$ have been discarded in passing to the right-hand side. It remains to extract the coefficients of $\eta_2^0,\eta_3^0,\eta_4^0$ in ratios such as
\begin{align}
\frac{ {\bm \Omega} _{1234} }{\eta_2}\bigg|\bigg|_{\eta_2^0,\eta_3^0,\eta_4^0} &= 
f_{34}^{\text{(1)}} f_{23}^{\text{(1)}} f_{12}^{\text{(2)}}+2 f_{23}^{\text{(1)}} f_{12}^{\text{(3)}}-f_{23}^{\text{(1)}} f_{34}^{\text{(3)}}-f_{34}^{\text{(1)}} f_{23}^{\text{(3)}}+2 f_{34}^{\text{(1)}} f_{12}^{\text{(3)}} 
\label{likehere}\\
&\quad+f_{12}^{\text{(2)}} f_{23}^{\text{(2)}}-f_{34}^{\text{(2)}} f_{23}^{\text{(2)}}+f_{12}^{\text{(2)}} f_{34}^{\text{(2)}}-f_{23}^{\text{(4)}}+3 f_{12}^{\text{(4)}}-f_{34}^{\text{(4)}}\,,
 \notag \\
\frac{ {\bm \Omega} _{1324} }{\eta_{34}}\bigg|\bigg|_{\eta_2^0,\eta_3^0,\eta_4^0} &=
f_{24}^{\text{(1)}} f_{13}^{\text{(1)}} f_{23}^{\text{(2)}}-f_{23}^{\text{(1)}} f_{13}^{\text{(1)}} f_{24}^{\text{(2)}}-f_{23}^{\text{(1)}} f_{24}^{\text{(1)}} f_{13}^{\text{(2)}}-f_{13}^{\text{(1)}} f_{23}^{\text{(3)}}
+f_{13}^{\text{(1)}} f_{24}^{\text{(3)}}+f_{23}^{\text{(2)}} f_{24}^{\text{(2)}}
\notag \\
&\quad
+f_{24}^{\text{(1)}} ( f_{13}^{\text{(3)}} {-} f_{23}^{\text{(3)}} )
%
- f_{23}^{\text{(1)}} (f_{24}^{\text{(3)}}{+} f_{13}^{\text{(3)}} )
+ f_{13}^{\text{(2)}} (f_{23}^{\text{(2)}}{+}f_{24}^{\text{(2)}})
+f_{13}^{\text{(4)}}+f_{23}^{\text{(4)}}+f_{24}^{\text{(4)}}\,.
\notag
\end{align}
which is straightforward to implement via {\tt Mathematica}.

\subsection{Five points}

At $n=m=5$ points, (\ref{vopen}) together with the tools in section \ref{sec:toolcoeff} yield
\begin{align}
   &\! \! \! \! \! \!   (1{+}s_{12345})  V_5(1,2,3,4,5) \overset{\rm IBP}= 
  \hat  {\rm G}_2V_3(1,2,3,4,5)+{\rm G}_4 V_1(1,2,3,4,5)   \\
&+3 {\rm G}_4  \big[ (s_{14}{+}s_{24}{+}s_{35})f_{12}^{(1)} +s_{13}f_{13}^{(1)} +{\rm cyc}(1,2,3,4,5)\big]
\nl\,\,&+
\bigg[ 
{\bm \Omega} _{12345} \bigg(\frac{s_{12}}{\eta _{2345}} -\frac{s_{12}}{\eta _{234}} -\frac{s_{25}}{\eta _2}-\frac{s_{35}}{\eta _3}-\frac{s_{14}{+}s_{24}{+}s_{34}{+}s_{45}}{\eta _4}
\nl\,\,&\quad \ \ \ \ \
 +\frac{2 \left(s_{15}{+}s_{25}{+}s_{35}{+}s_{45}\right)}{\eta _5}
-\frac{s_{13}{+}s_{23}}{\eta _{34}}+\frac{s_{14}{+}s_{24}{+}s_{34}}{\eta _{45}}+\frac{s_{13}{+}s_{23}}{\eta _{345}}\bigg)
\nl\,\,&\quad
-({\bm \Omega} _{12534}+{\bm \Omega} _{15234}+{\bm \Omega} _{12354}) \bigg(-\frac{s_{12}}{\eta _{23}}+\frac{s_{12}}{\eta _{234}}-\frac{s_{24}}{\eta _2}-\frac{s_{13}{+}s_{23}{+}s_{34}{+}s_{35}}{\eta _3}
\nl\,\,&\quad \ \ \ \ \
+\frac{2 \left(s_{14}{+}s_{24}{+}s_{34}{+}s_{45}\right)}{\eta _4}-\frac{s_{15}{+}s_{25}{+}s_{45}}{\eta _5}+\frac{s_{13}{+}s_{23}{+}s_{35}}{\eta _{34}}+\frac{s_{15}{+}s_{25}}{\eta _{45}} \bigg)
\nl\,\,&\quad
+{\bm \Omega} _{12354}
\left(\frac{1}{\eta _5}+\frac{1}{\eta _{34}}-\frac{1}{\eta _{45}}-\frac{1}{\eta _3}\right) s_{35}
+{\bm \Omega} _{15234} 
\left(\frac{1}{\eta _{23}}+\frac{1}{\eta _{45}}-\frac{1}{\eta _{234}}-\frac{1}{\eta _5}\right) s_{25}
\nl\,\,&\quad
+\left(\frac{1}{\eta _{34}}+\frac{1}{\eta _{45}}-\frac{1}{\eta _{345}}-\frac{1}{\eta _4}\right) \left(s_{14}{+}s_{24}\right) \left({\bm \Omega} _{12435}
+{\bm \Omega} _{12453}\right)
\nl\,\,&\quad
+\left(\frac{1}{\eta _{234}}+\frac{1}{\eta _{345}}-\frac{1}{\eta _{2345}}-\frac{1}{\eta _{34}}\right) s_{13} \left({\bm \Omega} _{13245}+{\bm \Omega} _{13425}+{\bm \Omega} _{13452}\right)
\nl\,\,&\quad
+\left(\frac{1}{\eta _{234}}+\frac{1}{\eta _3} -\frac{1}{\eta _{23}}-\frac{1}{\eta _{34}}\right) s_{13} \left({\bm \Omega} _{13254}+{\bm \Omega} _{13524}+{\bm \Omega} _{13542}\right)
 -\Big(23\leftrightarrow 54\Big)\bigg]
\, \bigg|\bigg|_{\eta_2^0,\ldots ,\eta_5^0} 
\,,
\notag
\end{align}
where the simultaneous relabelling $2\leftrightarrow 5$ and $3\leftrightarrow 4$ 
applies to the last eight lines. The extraction of coefficients $\eta_2^0,\ldots ,\eta_5^0$ 
in the ratios $\frac{ {\bm \Omega} _{1ijkl} }{\eta_I}$
is again straightforward, see (\ref{likehere}) for examples at four points.

\section{Restoring total Koba-Nielsen derivatives in breaking double cycles}
\label{sec:2cyc.4}

In sections \ref{sec:2cyc.1} to \ref{sec:2cyc.3}, we explained how to break a product of
any two Kronecker-Eisenstein cycles. However, it is important to note that the main result \eqref{newcyc23pre22} is an equivalence relation as we omitted several total Koba-Nielsen derivatives. In this appendix, 
we reinstate these total Koba-Nielsen derivatives which not only feed into the discussions of
sections \ref{sec:2cyc.5} and \ref{triplecyclenew} but also pave the way for closed-string applications.

The exact version of \eqref{newcyc23pre22} including all total Koba-Nielsen derivatives reads
  \begin{align}
  & ( 1{+}s_{12\cdots m}) ( 1{+}s_{m+1\cdots n})  \CC_{(12\cdots m)}(\xi_1)\CC_{(m+1\cdots n)} (\xi_2)
=\,\,
\MM_{12\cdots m}(\xi_1)\MM_{m+1\cdots n}(\xi_2)
   \label{newcyc23pre22r} \\
&\quad- \sum_{p  =m+2}^{n}
~
 \sum_{\substack{\sigma \in \{m+2,\cdots, p-1\}\\
\quad\shuffle \{n,\cdots, p+1 \}} }
\!\!\!\!\!\!\!\!\!
 (-1)^{n-m-p} 
 \left( \sum_{k=1}^m  { x}_{p,k}  + {\nabla}_p \right)
 \big({\bm \Omega}_{m+1,\sigma,p} 
 \MM_{12\cdots m} (\xi_1)\big)
\nl
&\quad + \sum_{
b  =2}^{m}
\!\!\!\!
\!\!\!\!
 \sum_{\substack{\rho \in \{2,\cdots, b-1\}\\
\qquad\quad \shuffle \{m,\cdots, b+1 \}} }
\!\!\!\!\!\!\!\!\!\!\!\!\!
 (-1)^{m-b} 
\sum_{j=m+1}^n     \sum_{\substack{
p  =m+1
\\
p\neq j}}^{n}
\sum_{
\substack{
\sigma \in  X \shuffle  Y^{\rm T}
\\
(j, X, p, Y )=(m{+}1,m{+}2, \ldots, n)
} }
\!\!\!
\nonumber\\
&\qquad\qquad\qquad\qquad \times
 (-1)^{|Y|} 
{\nabla}_p 
\left( {\bm \Omega}_{1,\rho,b} x_{b,j} {\bm \Omega}_{j,\sigma,p} 
\right)
\nl\,\,
&\quad -
\sum_{
b  =2}^{m}
\!\!\!\!
\!\!\!\!
 \sum_{\substack{\rho \in \{2,\cdots, b-1\}\\
\qquad\quad \shuffle \{m,\cdots, b+1 \}} }
\!\!\!\!\!\!\!\!\!\!\!\!\!\!
 (-1)^{m-b} 
\Bigg( \sum_{j=m+1}^n  { x}_{b,j} +{\nabla}_b \Bigg)\big( {\bm \Omega}_{1,\rho,b} \MM_{m+1,m+2,\cdots, n}(\xi_2) \big)
\nl
&\quad +  \sum_{
b  =2}^{m}
\!\!\!\!
\!\!\!\!
 \sum_{\substack{\rho \in \{2,\cdots, b-1\}\\
\qquad\quad \shuffle \{m,\cdots, b+1 \}} }
\!\!\!\!\!\!\!\!\!\!\!\!\!
 (-1)^{m-b} 
\sum_{
p  =m+2}^{n}
 \sum_{\substack{\sigma \in \{m+2,\cdots, p-1\}\\
\quad\shuffle \{n,\cdots, p+1 \}} }
\!\!\!\!\!\!\!\!\!\!\!\!\!
 (-1)^{n-m-p}  \notag \\
&\qquad \qquad \qquad \qquad \times {\nabla}_b \left[ \left( \sum_{k=1}^m  {x}_{p,k}
+{\nabla}_p \right) \left(
 {\bm \Omega}_{1,\rho,b}  
{\bm \Omega}_{m+1,\sigma,p} \right)\right]
\nl\,\,
&\quad+\frac12
\sum_{\substack{
a,b  
 =1 \\ a\neq b}
}^{m} \,
  \sum_{\substack{
j, p 
 =m+1 \\ j\neq p}
}^{n} 
\sum_{
\substack{
\rho \in  A \shuffle  B^{\rm T}
\\
(a, A, b, B )=(1,2, \ldots, m)
} }
\sum_{
\substack{
\sigma \in  X \shuffle  Y^{\rm T}
\\
(j, X, p, Y )=(m{+}1,m{+}2, \ldots, n)
} }
\!\!\!\!
\!\!\!\!
\!\!\!\!
 (-1)^{|B|+|Y|}  {\bm \Omega}_{a,\rho,b} \,
{ x}_{b,j} {\bm \Omega}_{j,\sigma,p}  \, { x}_{p, a}\non \end{align}
and implies (\ref{newcyc23pre22F}) under the substitution rules (\ref{subsrule}).
At $n=4$ points with two cycles of length $m=2$, this
specializes to
\begin{align}
&(1{+}s_{12})(1{+}s_{34})\CC_{(12)}(\xi_1)\CC_{(34)} (\xi_2)
= \MM_{12}(\xi_1)\MM_{34}(\xi_2) \label{ofofr} \\
&\quad -\MM_{12} (\xi_1)  x_{4,12} \Omega_{34}(\eta_{4}) -   x_{2,34} \Omega_{12}(\eta_{2}) \MM_{34}(\xi_2)
   +\big( 
   x_{2,3}x_{4,1} -    x_{2,4}x_{3,1} 
\big) \Omega_{12}(\eta_{2})\Omega_{34}(\eta_{4})
\non
\nl
&\quad + {\nabla}_2 {\nabla}_4\big( \Omega_{12}(\eta_{2})   \Omega_{34}(\eta_{4}) \big) 
 - {\nabla}_4\big(  \MM_{12} (\xi_1)  \Omega_{34}(\eta_{4}) -  \Omega_{12}(\eta_{2}) \Omega_{34}(\eta_{4}) {  x}_{2,3}    \big) 
\nl&\quad 
-{\nabla}_2\big(  \MM_{34}  (\xi_2)  \Omega_{12}(\eta_{2}) -  \Omega_{12}(\eta_{2})  \Omega_{34}(\eta_{2}) {  x}_{4,12}   \big)
- {\nabla}_3 \big( 
\Omega_{12}(\eta_{2})
\Omega_{34}(\eta_{4})    {  x}_{2,4}  \big)
\,,
\non
\end{align} 
where the first two lines were already spelt out in \eqref{ofof},
and the last two lines are total Koba-Nielsen derivatives discarded
in section \ref{sec:2cyc.1}.

The tracking of total Koba-Nielsen derivatives in (\ref{newcyc23pre22r}) is essential
for applications to generating functions of closed-string integrands \eqref{stringint}.
As exemplified in section \ref{sec:conbas}, the factors of 
$\overline{\Omega(z_i{-}z_j, \eta, \tau)}$ in a closed-string
context -- to be collectively denoted by $\overline{\varphi}=\overline{\varphi(z_j,\tau)}$ 
in the rest of this appendix -- interfere with the $\nabla_b$ in the F-IBP manipulations
of $\Omega(z_i{-}z_j, \eta, \tau)$. Hence, the images of $\nabla_b$
in (\ref{newcyc23pre22r}) are key information for the chain decomposition of
products
\be
\CC_{(12\cdots m)}(\xi_1)\CC_{(m+1,m+2,\cdots, n)} (\xi_2) \overline{\varphi}
\ee
in presence of combinations of $\overline{\Omega(z_i{-}z_j, \eta, \tau)}$. The
extra terms from integration by parts of holomorphic derivatives are obtained from
\be 
\left(\nabla_b {\bm  \Omega}_{1,\rho,b} \right) \overline{\varphi} =
\nabla_b\left({\bm  \Omega}_{1,\rho,b} \overline{\varphi} \right)  -  {\bm  \Omega}_{1,\rho,b} \partial_i \overline{\varphi} \,,
\label{closed2cyc}
\ee
where ${\bm  \Omega}_{1,\rho,b}$ was chosen as a placeholder for any
$\nabla_b$-image on the right-hand side of (\ref{newcyc23pre22r}).
As long as $\overline{\varphi}$ is a product of $\overline{\Omega(z_i{-}z_j, \eta, \tau)}$,
the derivatives in the last term of (\ref{closed2cyc}) are easily evaluated via
\eqref{partialbar} or \eqref{antiholocyc}. If $\overline{\varphi}$ comprises 
individual Kronecker-Eisenstein coefficients, one can furthermore make use of 
$\partial_i \overline{ {f}^{(w)}_{ij} } =-\frac{\pi}{ {\rm Im} \,\tau}   \overline{ {f}^{(w-1)}_{ij} }$
with $w\geq 1$, see \eqref{partialbar}.


\bibliographystyle{JHEP}
\bibliography{citesIBP}{}

\providecommand{\href}[2]{#2}\begingroup\raggedright\begin{thebibliography}{10}

\bibitem{Aomoto87}
K.~Aomoto, \emph{{Gauss-Manin connection of integral of difference products}},
  \href{http://dx.doi.org/10.2969/jmsj/03920191}{\emph{J. Math. Soc. Japan}
  {\bf 39} (1987) 191--208}.

\bibitem{Mizera:2017cqs}
S.~Mizera, \emph{{Combinatorics and Topology of Kawai-Lewellen-Tye Relations}},
  \href{http://dx.doi.org/10.1007/JHEP08(2017)097}{\emph{JHEP} {\bf 08} (2017)
  097}, [\href{http://arxiv.org/abs/1706.08527}{{\tt 1706.08527}}].

\bibitem{Mizera:2017rqa}
S.~Mizera, \emph{{Scattering Amplitudes from Intersection Theory}},
  \href{http://dx.doi.org/10.1103/PhysRevLett.120.141602}{\emph{Phys. Rev.
  Lett.} {\bf 120} (2018) 141602}, [\href{http://arxiv.org/abs/1711.00469}{{\tt
  1711.00469}}].

\bibitem{Mizera:2019gea}
S.~Mizera, \emph{{Aspects of Scattering Amplitudes and Moduli Space
  Localization}}.
\newblock PhD thesis, Perimeter Inst. Theor. Phys., 2019.
\newblock \href{http://arxiv.org/abs/1906.02099}{{\tt 1906.02099}}.

\bibitem{Aomoto:1975Vanishing}
K.~Aomoto, \emph{{On vanishing of cohomology attached to certain many valued
  meromorphic functions}},
  \href{http://dx.doi.org/10.2969/jmsj/02720248}{\emph{Journal of the
  Mathematical Society of Japan} {\bf 27} (1975) 248 -- 255}.

\bibitem{Aomoto87ComplexSelberg}
K.~Aomoto, \emph{{On the complex Selberg integral}},
  \href{http://dx.doi.org/10.1093/qmath/38.4.385}{\emph{The Quarterly Journal
  of Mathematics} {\bf 38} (12, 1987) 385--399},
  [\href{http://arxiv.org/abs/https://academic.oup.com/qjmath/article-pdf/38/4/385/4383972/38-4-385.pdf}{{\tt
  https://academic.oup.com/qjmath/article-pdf/38/4/385/4383972/38-4-385.pdf}}].

\bibitem{KitaYoshida1994paperI}
M.~Kita and M.~Yoshida, \emph{Intersection theory for twisted cycles},
  \href{http://dx.doi.org/https://doi.org/10.1002/mana.19941660122}{\emph{Mathematische
  Nachrichten} {\bf 166} (1994) 287--304},
  [\href{http://arxiv.org/abs/https://onlinelibrary.wiley.com/doi/pdf/10.1002/mana.19941660122}{{\tt
  https://onlinelibrary.wiley.com/doi/pdf/10.1002/mana.19941660122}}].

\bibitem{Mimachi2004IntNumbersSelbergInt}
K.~Mimachi, K.~Ohara and M.~Yoshida, \emph{{Intersection numbers for loaded
  cycles associated with Selberg-type integrals}},
  \href{http://dx.doi.org/10.2748/tmj/1113246749}{\emph{Tohoku Mathematical
  Journal} {\bf 56} (2004) 531 -- 551}.

\bibitem{Mafra:2022wml}
C.~R. Mafra and O.~Schlotterer, \emph{{Tree-level amplitudes from the pure
  spinor superstring}},
  \href{http://dx.doi.org/10.1016/j.physrep.2023.04.001}{\emph{Phys. Rept.}
  {\bf 1020} (2023) 1--162}, [\href{http://arxiv.org/abs/2210.14241}{{\tt
  2210.14241}}].

\bibitem{Kleiss:1988ne}
R.~Kleiss and H.~Kuijf, \emph{{Multi-Gluon Cross-sections and Five Jet
  Production at Hadron Colliders}},
  \href{http://dx.doi.org/10.1016/0550-3213(89)90574-9}{\emph{Nucl. Phys.} {\bf
  B312} (1989) 616--644}.

\bibitem{BCJ}
Z.~Bern, J.~J.~M. Carrasco and H.~Johansson, \emph{{New Relations for
  Gauge-Theory Amplitudes}},
  \href{http://dx.doi.org/10.1103/PhysRevD.78.085011}{\emph{Phys. Rev.} {\bf
  D78} (2008) 085011}, [\href{http://arxiv.org/abs/0805.3993}{{\tt
  0805.3993}}].

\bibitem{Zfunctions}
J.~Broedel, O.~Schlotterer and S.~Stieberger, \emph{{Polylogarithms, Multiple
  Zeta Values and Superstring Amplitudes}},
  \href{http://dx.doi.org/10.1002/prop.201300019}{\emph{Fortsch. Phys.} {\bf
  61} (2013) 812--870}, [\href{http://arxiv.org/abs/1304.7267}{{\tt
  1304.7267}}].

\bibitem{Stieberger:2014hba}
S.~Stieberger and T.~R. Taylor, \emph{{Closed String Amplitudes as
  Single-Valued Open String Amplitudes}},
  \href{http://dx.doi.org/10.1016/j.nuclphysb.2014.02.005}{\emph{Nucl. Phys.}
  {\bf B881} (2014) 269--287}, [\href{http://arxiv.org/abs/1401.1218}{{\tt
  1401.1218}}].

\bibitem{Schlotterer:2016cxa}
O.~Schlotterer, \emph{{Amplitude relations in heterotic string theory and
  Einstein-Yang-Mills}},
  \href{http://dx.doi.org/10.1007/JHEP11(2016)074}{\emph{JHEP} {\bf 11} (2016)
  074}, [\href{http://arxiv.org/abs/1608.00130}{{\tt 1608.00130}}].

\bibitem{Mafra:2011nv}
C.~R. Mafra, O.~Schlotterer and S.~Stieberger, \emph{{Complete N-Point
  Superstring Disk Amplitude I. Pure Spinor Computation}},
  \href{http://dx.doi.org/10.1016/j.nuclphysb.2013.04.023}{\emph{Nucl. Phys.}
  {\bf B873} (2013) 419--460}, [\href{http://arxiv.org/abs/1106.2645}{{\tt
  1106.2645}}].

\bibitem{Carrasco:2016ldy}
J.~J.~M. Carrasco, C.~R. Mafra and O.~Schlotterer, \emph{{Abelian Z-theory:
  NLSM amplitudes and $\alpha$'-corrections from the open string}},
  \href{http://dx.doi.org/10.1007/JHEP06(2017)093}{\emph{JHEP} {\bf 06} (2017)
  093}, [\href{http://arxiv.org/abs/1608.02569}{{\tt 1608.02569}}].

\bibitem{Azevedo:2018dgo}
T.~Azevedo, M.~Chiodaroli, H.~Johansson and O.~Schlotterer, \emph{{Heterotic
  and bosonic string amplitudes via field theory}},
  \href{http://dx.doi.org/10.1007/JHEP10(2018)012}{\emph{JHEP} {\bf 10} (2018)
  012}, [\href{http://arxiv.org/abs/1803.05452}{{\tt 1803.05452}}].

\bibitem{Kawai:1985xq}
H.~Kawai, D.~C. Lewellen and S.~H.~H. Tye, \emph{{A Relation Between Tree
  Amplitudes of Closed and Open Strings}},
  \href{http://dx.doi.org/10.1016/0550-3213(86)90362-7}{\emph{Nucl. Phys.} {\bf
  B269} (1986) 1--23}.

\bibitem{Broedel:2013aza}
J.~Broedel, O.~Schlotterer, S.~Stieberger and T.~Terasoma, \emph{{All order
  $\alpha^{\prime}$-expansion of superstring trees from the Drinfeld
  associator}}, \href{http://dx.doi.org/10.1103/PhysRevD.89.066014}{\emph{Phys.
  Rev.} {\bf D89} (2014) 066014}, [\href{http://arxiv.org/abs/1304.7304}{{\tt
  1304.7304}}].

\bibitem{Kaderli:2019dny}
A.~Kaderli, \emph{{A note on the Drinfeld associator for genus-zero superstring
  amplitudes in twisted de Rham theory}},
  \href{http://dx.doi.org/10.1088/1751-8121/ab9462}{\emph{J. Phys. A} {\bf 53}
  (2020) 415401}, [\href{http://arxiv.org/abs/1912.09406}{{\tt 1912.09406}}].

\bibitem{Dolan:2007eh}
L.~Dolan and P.~Goddard, \emph{{Current Algebra on the Torus}},
  \href{http://dx.doi.org/10.1007/s00220-008-0542-1}{\emph{Commun. Math. Phys.}
  {\bf 285} (2009) 219--264}, [\href{http://arxiv.org/abs/0710.3743}{{\tt
  0710.3743}}].

\bibitem{Broedel:2014vla}
J.~Broedel, C.~R. Mafra, N.~Matthes and O.~Schlotterer, \emph{{Elliptic
  multiple zeta values and one-loop superstring amplitudes}},
  \href{http://dx.doi.org/10.1007/JHEP07(2015)112}{\emph{JHEP} {\bf 07} (2015)
  112}, [\href{http://arxiv.org/abs/1412.5535}{{\tt 1412.5535}}].

\bibitem{Gerken:2018jrq}
J.~E. Gerken, A.~Kleinschmidt and O.~Schlotterer, \emph{{Heterotic-string
  amplitudes at one loop: modular graph forms and relations to open strings}},
  \href{http://dx.doi.org/10.1007/JHEP01(2019)052}{\emph{JHEP} {\bf 01} (2019)
  052}, [\href{http://arxiv.org/abs/1811.02548}{{\tt 1811.02548}}].

\bibitem{Mafra:2019ddf}
C.~R. Mafra and O.~Schlotterer, \emph{{All-order alpha'-expansion of one-loop
  open-string integrals}},
  \href{http://dx.doi.org/10.1103/PhysRevLett.124.101603}{\emph{Phys. Rev.
  Lett.} {\bf 124} (2020) 101603}, [\href{http://arxiv.org/abs/1908.09848}{{\tt
  1908.09848}}].

\bibitem{Mafra:2019xms}
C.~R. Mafra and O.~Schlotterer, \emph{{One-loop open-string integrals from
  differential equations: all-order $\alpha$'-expansions at $n$ points}},
  \href{http://dx.doi.org/10.1007/JHEP03(2020)007}{\emph{JHEP} {\bf 03} (2020)
  007}, [\href{http://arxiv.org/abs/1908.10830}{{\tt 1908.10830}}].

\bibitem{Gerken:2019cxz}
J.~E. Gerken, A.~Kleinschmidt and O.~Schlotterer, \emph{{All-order differential
  equations for one-loop closed-string integrals and modular graph forms}},
  \href{http://dx.doi.org/10.1007/JHEP01(2020)064}{\emph{JHEP} {\bf 01} (2020)
  064}, [\href{http://arxiv.org/abs/1911.03476}{{\tt 1911.03476}}].

\bibitem{Enriquez:Emzv}
B.~Enriquez, \emph{Analogues elliptiques des nombres multiz\'etas},
  \href{http://dx.doi.org/10.24033/bsmf.2718}{\emph{Bull. Soc. Math. France}
  {\bf 144} (2016) 395--427}, [\href{http://arxiv.org/abs/1301.3042}{{\tt
  1301.3042}}].

\bibitem{Broedel:2019gba}
J.~Broedel and A.~Kaderli, \emph{{Amplitude recursions with an extra marked
  point}}, \href{http://dx.doi.org/10.4310/CNTP.2022.v16.n1.a3}{\emph{Commun.
  Num. Theor. Phys.} {\bf 16} (2022) 75--158},
  [\href{http://arxiv.org/abs/1912.09927}{{\tt 1912.09927}}].

\bibitem{Broedel:2020tmd}
J.~Broedel, A.~Kaderli and O.~Schlotterer, \emph{{Two dialects for KZB
  equations: generating one-loop open-string integrals}},
  \href{http://dx.doi.org/10.1007/JHEP12(2020)036}{\emph{JHEP} {\bf 12} (2020)
  036}, [\href{http://arxiv.org/abs/2007.03712}{{\tt 2007.03712}}].

\bibitem{Gerken:2020yii}
J.~E. Gerken, A.~Kleinschmidt and O.~Schlotterer, \emph{{Generating series of
  all modular graph forms from iterated Eisenstein integrals}},
  \href{http://dx.doi.org/10.1007/JHEP07(2020)190}{\emph{JHEP} {\bf 07} (2020)
  190}, [\href{http://arxiv.org/abs/2004.05156}{{\tt 2004.05156}}].

\bibitem{DHoker:2015wxz}
E.~D'Hoker, M.~B. Green, {\"O}.~G{\"u}rdogan and P.~Vanhove, \emph{Modular
  graph functions},
  \href{http://dx.doi.org/10.4310/CNTP.2017.v11.n1.a4}{\emph{Commun. Num.
  Theor. Phys.} {\bf 11} (2017) 165--218},
  [\href{http://arxiv.org/abs/1512.06779}{{\tt 1512.06779}}].

\bibitem{DHoker:2016mwo}
E.~D'Hoker and M.~B. Green, \emph{Identities between modular graph forms},
  \href{http://dx.doi.org/10.1016/j.jnt.2017.11.015}{\emph{J. Number Theory}
  {\bf 189} (2018) 25--80}, [\href{http://arxiv.org/abs/1603.00839}{{\tt
  1603.00839}}].

\bibitem{Dorigoni:2022npe}
D.~Dorigoni, M.~Doroudiani, J.~Drewitt, M.~Hidding, A.~Kleinschmidt, N.~Matthes
  et~al., \emph{{Modular graph forms from equivariant iterated Eisenstein
  integrals}}, \href{http://dx.doi.org/10.1007/JHEP12(2022)162}{\emph{JHEP}
  {\bf 12} (2022) 162}, [\href{http://arxiv.org/abs/2209.06772}{{\tt
  2209.06772}}].

\bibitem{Brown:2017qwo}
F.~Brown, \emph{{A class of non-holomorphic modular forms I}},
  \href{http://dx.doi.org/10.1007/s40687-018-0130-8}{\emph{Res. Math. Sci.}
  {\bf 5} (2018) 5:7}, [\href{http://arxiv.org/abs/1707.01230}{{\tt
  1707.01230}}].

\bibitem{Brown:2017qwo2}
F.~Brown, \emph{{A class of non-holomorphic modular forms II : equivariant
  iterated Eisenstein integrals}},
  \href{http://dx.doi.org/10.1017/fms.2020.24}{\emph{Forum~of~Mathematics,~Sigma}
  {\bf 8} (2020) 1}, [\href{http://arxiv.org/abs/1708.03354}{{\tt
  1708.03354}}].

\bibitem{Huang:2016tag}
Y.-t. Huang, O.~Schlotterer and C.~Wen, \emph{{Universality in string
  interactions}}, \href{http://dx.doi.org/10.1007/JHEP09(2016)155}{\emph{JHEP}
  {\bf 09} (2016) 155}, [\href{http://arxiv.org/abs/1602.01674}{{\tt
  1602.01674}}].

\bibitem{He:2018pol}
S.~He, F.~Teng and Y.~Zhang, \emph{{String amplitudes from field-theory
  amplitudes and vice versa}},
  \href{http://dx.doi.org/10.1103/PhysRevLett.122.211603}{\emph{Phys. Rev.
  Lett.} {\bf 122} (2019) 211603}, [\href{http://arxiv.org/abs/1812.03369}{{\tt
  1812.03369}}].

\bibitem{He:2019drm}
S.~He, F.~Teng and Y.~Zhang, \emph{{String Correlators: Recursive Expansion,
  Integration-by-Parts and Scattering Equations}},
  \href{http://dx.doi.org/10.1007/JHEP09(2019)085}{\emph{JHEP} {\bf 09} (2019)
  085}, [\href{http://arxiv.org/abs/1907.06041}{{\tt 1907.06041}}].

\bibitem{Mason:2013sva}
L.~Mason and D.~Skinner, \emph{{Ambitwistor strings and the scattering
  equations}}, \href{http://dx.doi.org/10.1007/JHEP07(2014)048}{\emph{JHEP}
  {\bf 07} (2014) 048}, [\href{http://arxiv.org/abs/1311.2564}{{\tt
  1311.2564}}].

\bibitem{Berkovits:2013xba}
N.~Berkovits, \emph{{Infinite Tension Limit of the Pure Spinor Superstring}},
  \href{http://dx.doi.org/10.1007/JHEP03(2014)017}{\emph{JHEP} {\bf 03} (2014)
  017}, [\href{http://arxiv.org/abs/1311.4156}{{\tt 1311.4156}}].

\bibitem{Hohm:2013jaa}
O.~Hohm, W.~Siegel and B.~Zwiebach, \emph{{Doubled $\alpha'$-geometry}},
  \href{http://dx.doi.org/10.1007/JHEP02(2014)065}{\emph{JHEP} {\bf 02} (2014)
  065}, [\href{http://arxiv.org/abs/1306.2970}{{\tt 1306.2970}}].

\bibitem{Huang:2016bdd}
Y.-t. Huang, W.~Siegel and E.~Y. Yuan, \emph{{Factorization of Chiral String
  Amplitudes}}, \href{http://dx.doi.org/10.1007/JHEP09(2016)101}{\emph{JHEP}
  {\bf 09} (2016) 101}, [\href{http://arxiv.org/abs/1603.02588}{{\tt
  1603.02588}}].

\bibitem{Gomez:2013wza}
H.~Gomez and E.~Y. Yuan, \emph{{N-point tree-level scattering amplitude in the
  new Berkovits` string}},
  \href{http://dx.doi.org/10.1007/JHEP04(2014)046}{\emph{JHEP} {\bf 04} (2014)
  046}, [\href{http://arxiv.org/abs/1312.5485}{{\tt 1312.5485}}].

\bibitem{He:2017spx}
S.~He, O.~Schlotterer and Y.~Zhang, \emph{{New BCJ representations for one-loop
  amplitudes in gauge theories and gravity}},
  \href{http://dx.doi.org/10.1016/j.nuclphysb.2018.03.003}{\emph{Nucl. Phys.}
  {\bf B930} (2018) 328--383}, [\href{http://arxiv.org/abs/1706.00640}{{\tt
  1706.00640}}].

\bibitem{Kalyanapuram:2021xow}
N.~Kalyanapuram, \emph{{Ambitwistor integrands from tensionless chiral
  superstring integrands}},
  \href{http://dx.doi.org/10.1007/JHEP10(2021)171}{\emph{JHEP} {\bf 10} (2021)
  171}, [\href{http://arxiv.org/abs/2103.07943}{{\tt 2103.07943}}].

\bibitem{Guillen:2021mwp}
M.~Guillen, H.~Johansson, R.~L. Jusinskas and O.~Schlotterer, \emph{{Scattering
  Massive String Resonances through Field-Theory Methods}},
  \href{http://dx.doi.org/10.1103/PhysRevLett.127.051601}{\emph{Phys. Rev.
  Lett.} {\bf 127} (2021) 051601}, [\href{http://arxiv.org/abs/2104.03314}{{\tt
  2104.03314}}].

\bibitem{DHoker:1988pdl}
E.~D'Hoker and D.~H. Phong, \emph{{The Geometry of String Perturbation
  Theory}}, \href{http://dx.doi.org/10.1103/RevModPhys.60.917}{\emph{Rev. Mod.
  Phys.} {\bf 60} (1988) 917}.

\bibitem{DHoker:1989cxq}
E.~D'Hoker and D.~H. Phong, \emph{{Conformal Scalar Fields and Chiral Splitting
  on Super Riemann Surfaces}},
  \href{http://dx.doi.org/10.1007/BF01218413}{\emph{Commun. Math. Phys.} {\bf
  125} (1989) 469}.

\bibitem{companion}
Y.~Zhang, \emph{{Advanced tools for basis decompositions of genus-one string
  integrals}},  \href{http://arxiv.org/abs/2403.18078}{{\tt 2403.18078}}.

\bibitem{Kronecker}
L.~Kronecker, \emph{{Zur Theorie der elliptischen Funktionen}},
  {\emph{Mathematische Werke} {\bf IV} (1881) 313--318}.

\bibitem{BrownLev}
F.~Brown and A.~Levin, \emph{{Multiple elliptic polylogarithms}},
  \href{http://arxiv.org/abs/1110.6917}{{\tt 1110.6917}}.

\bibitem{Broedel:2017kkb}
J.~Broedel, C.~Duhr, F.~Dulat and L.~Tancredi, \emph{{Elliptic polylogarithms
  and iterated integrals on elliptic curves. Part I: general formalism}},
  \href{http://dx.doi.org/10.1007/JHEP05(2018)093}{\emph{JHEP} {\bf 05} (2018)
  093}, [\href{http://arxiv.org/abs/1712.07089}{{\tt 1712.07089}}].

\bibitem{Berkovits:2022ivl}
N.~Berkovits, E.~D'Hoker, M.~B. Green, H.~Johansson and O.~Schlotterer,
  \emph{{Snowmass White Paper: String Perturbation Theory}},  in
  \emph{{Snowmass 2021}}, 3, 2022.
\newblock \href{http://arxiv.org/abs/2203.09099}{{\tt 2203.09099}}.

\bibitem{Bourjaily:2022bwx}
J.~L. Bourjaily et~al., \emph{{Functions Beyond Multiple Polylogarithms for
  Precision Collider Physics}},  in \emph{{Snowmass 2021}}, 3, 2022.
\newblock \href{http://arxiv.org/abs/2203.07088}{{\tt 2203.07088}}.

\bibitem{green1988superstring}
M.~B. Green, J.~Schwarz and E.~Witten, \emph{Superstring theory. vol. 2: Loop
  amplitudes, anomalies and phenomenology. {C}ambridge, {U}k: Univ. {P}r.(1987)
  596 {P}}, {\emph{Cambridge Monographs On Mathematical Physics} {\bf 197}
  (1988) 198}.

\bibitem{Minahan:1987ha}
J.~A. Minahan, \emph{{One Loop Amplitudes on Orbifolds and the Renormalization
  of Coupling Constants}},
  \href{http://dx.doi.org/10.1016/0550-3213(88)90303-3}{\emph{Nucl. Phys. B}
  {\bf 298} (1988) 36--74}.

\bibitem{Berg:2016wux}
M.~Berg, I.~Buchberger and O.~Schlotterer, \emph{{From maximal to minimal
  supersymmetry in string loop amplitudes}},
  \href{http://dx.doi.org/10.1007/JHEP04(2017)163}{\emph{JHEP} {\bf 04} (2017)
  163}, [\href{http://arxiv.org/abs/1603.05262}{{\tt 1603.05262}}].

\bibitem{Bianchi:2006nf}
M.~Bianchi and A.~V. Santini, \emph{{String predictions for near future
  colliders from one-loop scattering amplitudes around D-brane worlds}},
  \href{http://dx.doi.org/10.1088/1126-6708/2006/12/010}{\emph{JHEP} {\bf 12}
  (2006) 010}, [\href{http://arxiv.org/abs/hep-th/0607224}{{\tt
  hep-th/0607224}}].

\bibitem{Bianchi:2015vsa}
M.~Bianchi and D.~Consoli, \emph{{Simplifying one-loop amplitudes in
  superstring theory}},
  \href{http://dx.doi.org/10.1007/JHEP01(2016)043}{\emph{JHEP} {\bf 01} (2016)
  043}, [\href{http://arxiv.org/abs/1508.00421}{{\tt 1508.00421}}].

\bibitem{Green:1982sw}
M.~B. Green, J.~H. Schwarz and L.~Brink, \emph{{N=4 Yang-Mills and N=8
  Supergravity as Limits of String Theories}},
  \href{http://dx.doi.org/10.1016/0550-3213(82)90336-4}{\emph{Nucl. Phys.} {\bf
  B198} (1982) 474--492}.

\bibitem{Tsuchiya:1988va}
A.~Tsuchiya, \emph{{More on One Loop Massless Amplitudes of Superstring
  Theories}}, \href{http://dx.doi.org/10.1103/PhysRevD.39.1626}{\emph{Phys.
  Rev.} {\bf D39} (1989) 1626}.

\bibitem{Stieberger:2002wk}
S.~Stieberger and T.~R. Taylor, \emph{{NonAbelian Born-Infeld action and type
  1. - heterotic duality 2: Nonrenormalization theorems}},
  \href{http://dx.doi.org/10.1016/S0550-3213(02)00979-3}{\emph{Nucl. Phys.}
  {\bf B648} (2003) 3--34}, [\href{http://arxiv.org/abs/hep-th/0209064}{{\tt
  hep-th/0209064}}].

\bibitem{oeis431}
{\rm \url{https://oeis.org/A000431}}.

\bibitem{Tsuchiya:2017joo}
A.~G. Tsuchiya, \emph{{On new theta identities of fermion correlation functions
  on genus g Riemann surfaces}},  \href{http://arxiv.org/abs/1710.00206}{{\tt
  1710.00206}}.

\bibitem{Mafra:2018pll}
C.~R. Mafra and O.~Schlotterer, \emph{{Towards the n-point one-loop superstring
  amplitude. Part II. Worldsheet functions and their duality to kinematics}},
  \href{http://dx.doi.org/10.1007/JHEP08(2019)091}{\emph{JHEP} {\bf 08} (2019)
  091}, [\href{http://arxiv.org/abs/1812.10970}{{\tt 1812.10970}}].

\bibitem{Bern:2019prr}
Z.~Bern, J.~J. Carrasco, M.~Chiodaroli, H.~Johansson and R.~Roiban, \emph{{The
  Duality Between Color and Kinematics and its Applications}},
  \href{http://arxiv.org/abs/1909.01358}{{\tt 1909.01358}}.

\bibitem{Adamo:2022dcm}
T.~Adamo, J.~J.~M. Carrasco, M.~Carrillo-Gonz\'alez, M.~Chiodaroli, H.~Elvang,
  H.~Johansson et~al., \emph{{Snowmass White Paper: the Double Copy and its
  Applications}},  in \emph{{Snowmass 2021}}, 4, 2022.
\newblock \href{http://arxiv.org/abs/2204.06547}{{\tt 2204.06547}}.

\bibitem{Mafra:2017ioj}
C.~R. Mafra and O.~Schlotterer, \emph{{Double-Copy Structure of One-Loop
  Open-String Amplitudes}},
  \href{http://dx.doi.org/10.1103/PhysRevLett.121.011601}{\emph{Phys. Rev.
  Lett.} {\bf 121} (2018) 011601}, [\href{http://arxiv.org/abs/1711.09104}{{\tt
  1711.09104}}].

\bibitem{Mafra:2018qqe}
C.~R. Mafra and O.~Schlotterer, \emph{{Towards the n-point one-loop superstring
  amplitude. Part III. One-loop correlators and their double-copy structure}},
  \href{http://dx.doi.org/10.1007/JHEP08(2019)092}{\emph{JHEP} {\bf 08} (2019)
  092}, [\href{http://arxiv.org/abs/1812.10971}{{\tt 1812.10971}}].

\bibitem{DHoker:2020prr}
E.~D'Hoker, C.~R. Mafra, B.~Pioline and O.~Schlotterer, \emph{{Two-loop
  superstring five-point amplitudes. Part I. Construction via chiral splitting
  and pure spinors}},
  \href{http://dx.doi.org/10.1007/JHEP08(2020)135}{\emph{JHEP} {\bf 08} (2020)
  135}, [\href{http://arxiv.org/abs/2006.05270}{{\tt 2006.05270}}].

\bibitem{Geyer:2021oox}
Y.~Geyer, R.~Monteiro and R.~Stark-Much\~ao, \emph{{Superstring Loop Amplitudes
  from the Field Theory Limit}},
  \href{http://dx.doi.org/10.1103/PhysRevLett.127.211603}{\emph{Phys. Rev.
  Lett.} {\bf 127} (2021) 211603}, [\href{http://arxiv.org/abs/2106.03968}{{\tt
  2106.03968}}].

\bibitem{boundaryref}
F.~Balli, A.~Edison and O.~Schlotterer, \emph{{Pinching rules in chiral
  splitting and six-point supergravity amplitudes at one loop}}, {\emph{{\!\!
  work in progress\!\!}} }.

\bibitem{Gerken:2020aju}
J.~E. Gerken, \emph{{Basis Decompositions and a Mathematica Package for Modular
  Graph Forms}}, \href{http://dx.doi.org/10.1088/1751-8121/abbdf2}{\emph{J.
  Phys. A} {\bf 54} (2021) 195401},
  [\href{http://arxiv.org/abs/2007.05476}{{\tt 2007.05476}}].

\bibitem{Ramakrish}
D.~Zagier, \emph{The {B}loch-{W}igner-{R}amakrishnan polylogarithm function},
  {\emph{Math. Ann.} {\bf 286} (1990) 613}.

\bibitem{DHoker:2018mys}
E.~D'Hoker, M.~B. Green and B.~Pioline, \emph{{Asymptotics of the $D^8 R^4$
  genus-two string invariant}},
  \href{http://dx.doi.org/10.4310/CNTP.2019.v13.n2.a3}{\emph{Commun. Num.
  Theor. Phys.} {\bf 13} (2019) 351--462},
  [\href{http://arxiv.org/abs/1806.02691}{{\tt 1806.02691}}].

\bibitem{Broedel:2019tlz}
J.~Broedel and A.~Kaderli, \emph{{Functional relations for elliptic
  polylogarithms}}, \href{http://dx.doi.org/10.1088/1751-8121/ab81d7}{\emph{J.
  Phys. A} {\bf 53} (2020) 245201},
  [\href{http://arxiv.org/abs/1906.11857}{{\tt 1906.11857}}].

\bibitem{DHoker:2020hlp}
E.~D'Hoker, A.~Kleinschmidt and O.~Schlotterer, \emph{{Elliptic modular graph
  forms. Part I. Identities and generating series}},
  \href{http://dx.doi.org/10.1007/JHEP03(2021)151}{\emph{JHEP} {\bf 03} (2021)
  151}, [\href{http://arxiv.org/abs/2012.09198}{{\tt 2012.09198}}].

\bibitem{Kaderli:2022qeu}
A.~Kaderli and C.~Rodriguez, \emph{{Open-string integrals with multiple
  unintegrated punctures at genus one}},
  \href{http://dx.doi.org/10.1007/JHEP10(2022)159}{\emph{JHEP} {\bf 10} (2022)
  159}, [\href{http://arxiv.org/abs/2203.09649}{{\tt 2203.09649}}].

\bibitem{Mastrolia:2018uzb}
P.~Mastrolia and S.~Mizera, \emph{{Feynman Integrals and Intersection Theory}},
  \href{http://dx.doi.org/10.1007/JHEP02(2019)139}{\emph{JHEP} {\bf 02} (2019)
  139}, [\href{http://arxiv.org/abs/1810.03818}{{\tt 1810.03818}}].

\bibitem{Frellesvig:2019kgj}
H.~Frellesvig, F.~Gasparotto, S.~Laporta, M.~K. Mandal, P.~Mastrolia,
  L.~Mattiazzi et~al., \emph{{Decomposition of Feynman Integrals on the Maximal
  Cut by Intersection Numbers}},
  \href{http://dx.doi.org/10.1007/JHEP05(2019)153}{\emph{JHEP} {\bf 05} (2019)
  153}, [\href{http://arxiv.org/abs/1901.11510}{{\tt 1901.11510}}].

\bibitem{Mizera:2019vvs}
S.~Mizera and A.~Pokraka, \emph{{From Infinity to Four Dimensions: Higher
  Residue Pairings and Feynman Integrals}},
  \href{http://dx.doi.org/10.1007/JHEP02(2020)159}{\emph{JHEP} {\bf 02} (2020)
  159}, [\href{http://arxiv.org/abs/1910.11852}{{\tt 1910.11852}}].

\bibitem{Frellesvig:2020qot}
H.~Frellesvig, F.~Gasparotto, S.~Laporta, M.~K. Mandal, P.~Mastrolia,
  L.~Mattiazzi et~al., \emph{{Decomposition of Feynman Integrals by
  Multivariate Intersection Numbers}},
  \href{http://dx.doi.org/10.1007/JHEP03(2021)027}{\emph{JHEP} {\bf 03} (2021)
  027}, [\href{http://arxiv.org/abs/2008.04823}{{\tt 2008.04823}}].

\bibitem{Caron-Huot:2021xqj}
S.~Caron-Huot and A.~Pokraka, \emph{{Duals of Feynman integrals. Part I.
  Differential equations}},
  \href{http://dx.doi.org/10.1007/JHEP12(2021)045}{\emph{JHEP} {\bf 12} (2021)
  045}, [\href{http://arxiv.org/abs/2104.06898}{{\tt 2104.06898}}].

\bibitem{Caron-Huot:2021iev}
S.~Caron-Huot and A.~Pokraka, \emph{{Duals of Feynman Integrals. Part II.
  Generalized unitarity}},
  \href{http://dx.doi.org/10.1007/JHEP04(2022)078}{\emph{JHEP} {\bf 04} (2022)
  078}, [\href{http://arxiv.org/abs/2112.00055}{{\tt 2112.00055}}].

\bibitem{Duhr:2023bku}
C.~Duhr and F.~Porkert, \emph{{Feynman integrals in two dimensions and
  single-valued hypergeometric functions}},
  \href{http://arxiv.org/abs/2309.12772}{{\tt 2309.12772}}.

\bibitem{Tourkine:2016bak}
P.~Tourkine and P.~Vanhove, \emph{{Higher-loop amplitude monodromy relations in
  string and gauge theory}},
  \href{http://dx.doi.org/10.1103/PhysRevLett.117.211601}{\emph{Phys. Rev.
  Lett.} {\bf 117} (2016) 211601}, [\href{http://arxiv.org/abs/1608.01665}{{\tt
  1608.01665}}].

\bibitem{Hohenegger:2017kqy}
S.~Hohenegger and S.~Stieberger, \emph{{Monodromy Relations in Higher-Loop
  String Amplitudes}},
  \href{http://dx.doi.org/10.1016/j.nuclphysb.2017.09.020}{\emph{Nucl. Phys.}
  {\bf B925} (2017) 63--134}, [\href{http://arxiv.org/abs/1702.04963}{{\tt
  1702.04963}}].

\bibitem{Tourkine:2019ukp}
P.~Tourkine, \emph{{Integrands and loop momentum in string and field theory}},
  \href{http://dx.doi.org/10.1103/PhysRevD.102.026006}{\emph{Phys. Rev. D} {\bf
  102} (2020) 026006}, [\href{http://arxiv.org/abs/1901.02432}{{\tt
  1901.02432}}].

\bibitem{Casali:2019ihm}
E.~Casali, S.~Mizera and P.~Tourkine, \emph{{Monodromy relations from twisted
  homology}}, \href{http://dx.doi.org/10.1007/JHEP12(2019)087}{\emph{JHEP} {\bf
  12} (2019) 087}, [\href{http://arxiv.org/abs/1910.08514}{{\tt 1910.08514}}].

\bibitem{Casali:2020knc}
E.~Casali, S.~Mizera and P.~Tourkine, \emph{{Loop amplitudes monodromy
  relations and color-kinematics duality}},
  \href{http://dx.doi.org/10.1007/JHEP03(2021)048}{\emph{JHEP} {\bf 03} (2021)
  048}, [\href{http://arxiv.org/abs/2005.05329}{{\tt 2005.05329}}].

\bibitem{Broedel:2018izr}
J.~Broedel, O.~Schlotterer and F.~Zerbini, \emph{{From elliptic multiple zeta
  values to modular graph functions: open and closed strings at one loop}},
  \href{http://dx.doi.org/10.1007/JHEP01(2019)155}{\emph{JHEP} {\bf 01} (2019)
  155}, [\href{http://arxiv.org/abs/1803.00527}{{\tt 1803.00527}}].

\bibitem{Gerken:2020xfv}
J.~E. Gerken, A.~Kleinschmidt, C.~R. Mafra, O.~Schlotterer and B.~Verbeek,
  \emph{{Towards closed strings as single-valued open strings at genus one}},
  \href{http://dx.doi.org/10.1088/1751-8121/abe58b}{\emph{J. Phys. A} {\bf 55}
  (2022) 025401}, [\href{http://arxiv.org/abs/2010.10558}{{\tt 2010.10558}}].

\bibitem{Stieberger:2021daa}
S.~Stieberger, \emph{{Open \& Closed vs. Pure Open String One-Loop
  Amplitudes}},  \href{http://arxiv.org/abs/2105.06888}{{\tt 2105.06888}}.

\bibitem{Stieberger:2022lss}
S.~Stieberger, \emph{{A Relation between One-Loop Amplitudes of Closed and Open
  Strings (One-Loop KLT Relation)}},
  \href{http://arxiv.org/abs/2212.06816}{{\tt 2212.06816}}.

\bibitem{ManoWatanabe2012}
T.~Mano and H.~Watanabe, \emph{Twisted cohomology and homology groups
  associated to the {R}iemann-{W}irtinger integral}, {\emph{Proceedings of the
  American Mathematical Society} {\bf 140} (2012) 3867--3881},
  [\href{http://arxiv.org/abs/1804.00366}{{\tt 1804.00366}}].

\bibitem{ghazouani2016moduli}
S.~Ghazouani and L.~Pirio, \emph{Moduli spaces of flat tori and elliptic
  hypergeometric functions},  \href{http://arxiv.org/abs/1605.02356}{{\tt
  1605.02356}}.

\bibitem{goto2022intersection}
Y.~Goto, \emph{Intersection numbers of twisted homology and cohomology groups
  associated to the {R}iemann-{W}irtinger integral},
  \href{http://arxiv.org/abs/2206.03177}{{\tt 2206.03177}}.

\bibitem{Felder:1995iv}
G.~Felder and A.~Varchenko, \emph{{Integral representation of solutions of the
  elliptic Knizhnik-Zamolodchikov-Bernard equations}},
  \href{http://arxiv.org/abs/hep-th/9502165}{{\tt hep-th/9502165}}.

\bibitem{DHoker:2023vax}
E.~D'Hoker, M.~Hidding and O.~Schlotterer, \emph{{Constructing polylogarithms
  on higher-genus Riemann surfaces}},
  \href{http://arxiv.org/abs/2306.08644}{{\tt 2306.08644}}.

\end{thebibliography}\endgroup

%
%

\end{document}